%% file: main.tex
\documentclass[letterpaper,twocolumn,10pt]{article}

\let\OLDthebibliography\thebibliography
\renewcommand\thebibliography[1]{
  \OLDthebibliography{#1}
  \setlength{\parskip}{0pt}
  \setlength{\itemsep}{0pt plus 0.3ex}
}

\usepackage{usenix,epsfig,endnotes}
 \usepackage[utf8]{inputenc}%(only for the pdftex engine)
\usepackage{algorithm2e}
\usepackage{amsmath}
\usepackage{graphicx}
\usepackage{xcolor}

\usepackage{amsmath}

\usepackage{graphicx}
\usepackage{subcaption}

\usepackage{enumitem}

\usepackage{authblk}

 \newcommand{\opt}{\operatorname{opt}}
 \newcommand{\qual}{\operatorname{qual}}

 \newcommand{\high}{\operatorname{high}}
 \newcommand{\low}{\operatorname{low}}
 \newcommand{\mmid}{\operatorname{mid}}

\begin{document}

\author[ ]{Aneet Kumar Dutta\thanks{aneet.dutta@cispa.de}}
\author[ ]{Sebastian Brandt\thanks{brandt@cispa.de}}
\author[ ]{Mridula Singh\thanks{singh@cispa.de}}

\affil[ ]{CISPA Helmholtz Center for Information Security\\
Saarbrücken, Germany}

% copy the following lines to add more authors
% \and
% {\rm Name}\\
%Name Institution
 % end author

% \title{Anti-spoof: Secure and Scalable Location Estimation and Recovery}

\title{Location Estimation and Recovery using 5G Positioning: Thwarting GNSS Spoofing Attacks}

\maketitle

\begin{abstract}

The availability of cheap GNSS spoofers can prevent safe navigation and tracking of road users. It can lead to loss of assets, inaccurate fare estimation, enforcing the wrong speed limit, miscalculated toll tax, passengers reaching an incorrect location, etc. The techniques designed to prevent and detect spoofing by using cryptographic solutions or receivers capable of differentiating legitimate and attack signals are insufficient in detecting GNSS spoofing of road users. Recent studies, testbeds, and 3GPP standards are exploring the possibility of hybrid positioning, where GNSS data will be combined with the 5G-NR positioning to increase the security and accuracy of positioning. We design the \textit{Location Estimation and Recovery} (LER) systems to estimate the correct absolute position using the combination of GNSS and 5G positioning with other road users, where a subset of road users can be malicious and collude to prevent spoofing detection.
Our \textit{Location Verification Protocol} extends the understanding of Message Time of Arrival Codes (MTAC) to prevent attacks against malicious provers. The novel \textit{Recovery and Meta Protocol} uses road users' dynamic and unpredictable nature to detect GNSS spoofing. 
This protocol provides fast detection of GNSS spoofing with a very low rate of false positives and can be customized to a large family of settings.
Even in a (highly unrealistic) worst-case scenario where each user is malicious with a probability of as large as 0.3, our protocol detects GNSS spoofing with high probability after communication and ranging with at most 20 road users, with a false positive rate close to 0.
SUMO simulations for road traffic show that we can detect GNSS spoofing in 2.6 minutes since its start under moderate traffic conditions. 

\end{abstract}

% \section{Things to do}

% \begin{enumerate}
%     \item Find the probabilistic bound in passive protocol.
%     \item Simulation results 
%     \item Compare the two models
% \end{enumerate}

\input{intro}
\input{background}

\input{ProblemStatement}
\input{attacker_model}
\input{protocol}
\input{protocolAnalysis}

\input{evaluation}
\input{discussion}

\vspace{10pt}
\section{Conclusion}
\label{sec:conclusion}
With LER, we propose a protocol to detect GNSS spoofing and recover the correct location coordinates. We used a secure computing environment to prevent remote entities from colluding using Terrorist Fraud. We limited the effect of Distance Fraud by using multi-carrier modulated OFDM Symbols of wider subcarrier bandwidth. We consider the worst-case scenario, where each node is malicious with a probability of 0.3, and show that we can detect it within $2.62$ to $4.05$ minutes depending upon traffic condition Using sumo simulations, we confirm that moderate road traffic is sufficient to detect and recover from GNSS spoofing. 
\clearpage

\bibliographystyle{plain}
\bibliography{sample}

\appendix

\input{appendix}
\end{document}

%% file: intro.tex
\section{Introduction}
\label{sec:into}

Location information is the most critical and essential for a wide variety of navigation and tracking applications. For example, law enforcement officials use ankle bracelets to track the location of an offender on parole~\cite{Parole}, logistics
and supply chain management companies that handle high-value commodities monitor the locations of every vehicle in their fleet to ensure secure transportation~\cite{fleet_tracking_1,fleet_tracking_2}, ride-hailing applications use location information for assigning drivers to trips and tracking them for billing~\cite{CarRentalSpoofing}, and nowadays hardly we can find anyone who is not dependent on Google Maps for navigation. Furthermore, public transport locations are continuously monitored to ensure smooth and timely operation of services~\cite{publicTransportGPS}. Dependency on location information is further increasing with the development of self-driving vehicles~\cite{tesla_hack}. 

All the applications stated above use Global Navigation Satellite Systems (GNSS) such as GPS, Galileo, and GLONASS. A GNSS receiver uses the time-of-arrival of the signal from four or more satellites to ascertain its precise location coordinates. 
It is well known that  GNSS is vulnerable to signal spoofing attacks. By using low-cost hardware (<50\$), an external attacker can generate a spoofing signal to change the estimated position at the GNSS receiver~\cite{GPS_spoofing_HackRF,tesla_gallilio}. These affordable attack devices can hijack vehicle navigation systems while evading detection by law enforcement~\cite{GNSS_spoofing_road}.

\begin{figure}
    \centering
    \includegraphics[width=8cm]{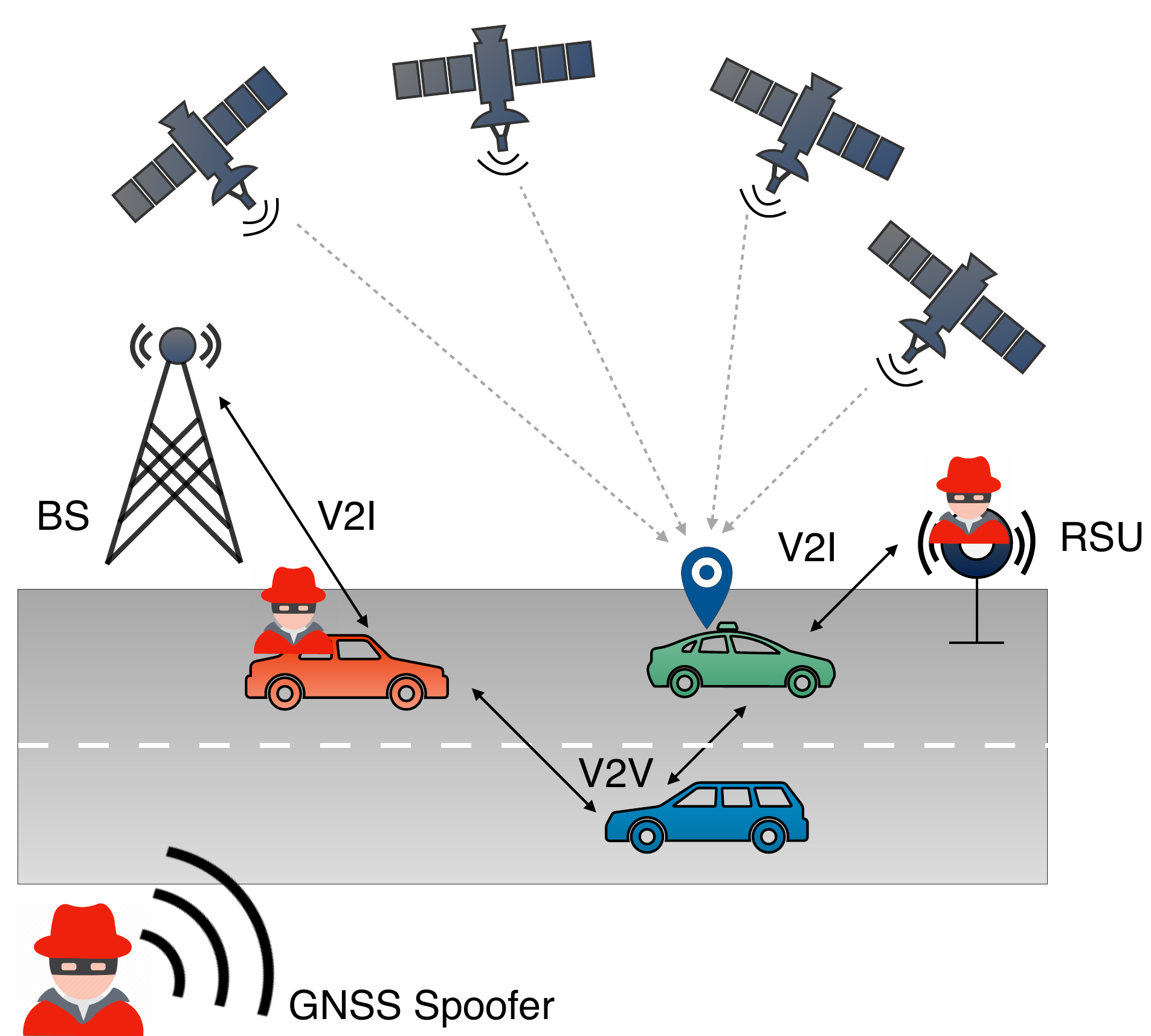}
    \caption{Standards are exploring the possibility of Hybrid positioning by combining GNSS positioning with V2X positioning, which would require ranging of road users with BS, RSUs, and other vehicles. In our design, we consider the presence of GNSS spoofer and road user entities that may collude with the spoofer.  }
    \label{fig:intro}
\end{figure}

The countermeasures to prevent GNSS spoofing by using cryptographic solutions or leveraging physical-layer signal properties are ineffective in securing vehicular navigation~\cite{Humphreys_GPS_sec,Kuhn_asymmetric_crypto,Akos2012WhosAO}.  
The recently launched cryptographic authentication techniques like Galileo's Open Service Navigation Message Authentication (OSNMA)~\cite{osnma}  authenticates the navigation message contents based on the TESLA protocol~\cite{TESLA_perrig} and one-way hash functions. Since in GNSS, the user's location is computed based on both the navigation message contents \emph{and} its time-of-arrival, such localization is still vulnerable to signal relay/replay attacks. Techniques that use physical-layer signal properties are ineffective against strong adversaries capable of completely overshadowing legitimate signals and stealthy attackers, e.g., seamless takeover attacks~\cite{Aanjhan_Spree}.

On the other hand, 3GPP, the standard organization responsible for developing the 5G New Radio (5G-NR) architecture, is exploring the possibility of hybrid positioning using GNSS and 5G-NR. Development of many hybrid positioning testbeds and testing is happening in parallel~\cite{5G_positioning_Chalmers,HybridPositioning_ESA}.
The positioning will be enabled using basestations and roadside units, but a strong focus is on enabling vehicle-to-vehicle ranging in order to reduce the load on the infrastructure. While positioning techniques currently added to standards are not designed to provide secure positioning, the research community is actively working to enable secure relative positioning~\cite{vrange, 5G_positioning_Chalmers}.

With this work, we extend the understanding of secure hybrid positioning using the combination of GNSS and 5G. We consider the GNSS spoofing scenario and assume that a subset of the 5G nodes can be malicious (i.e., fake basestation, malicious road users), as illustrated in Figure~\ref{fig:intro}. The goal is to detect and recover from the GNSS spoofing by performing opportunistic distance estimation with other road users and 5G infrastructure. To ensure detection and recovery from the GNSS spoofing, we propose a Location Estimation and Recovery (LER), a hybrid positioning system that combines GNSS and 5G. We make the following contributions: 
\begin{itemize}[noitemsep]
    \item We design a Location Verification Protocol to ensure secure distance measurement among road users and with the infrastructure. We extend the concept of Message Time of Arrival Codes (MTACs), a security primitive, to limit spurious distance measurements. 
    \item Our novel recovery and meta-protocol interprets the outcome of distance measurements to detect and recover from the GNSS spoofing while minimizing false alarms. 
    \item We use SUMO simulations to provide optimal time between the start and detection of the GNSS spoofing. 
\end{itemize}

Using MTACs with multicarrier modulation at the physical layer and trusted components for data processing S-LER protocol limits the distance manipulation caused by malicious users/devices and prevents them from colluding. Analysis using SUMO shows that by using a combination of location verification protocol, recovery and meta protocol, we can detect GNSS spoofing within $2.6$ to $4.05$ minutes in most traffic conditions, even when each instance of distance measurement is malicious with probability $0.3$. These contributions collectively ensure high security guarantees against location manipulation attacks, ensuring safe navigation and tracking of road users.

The remainder of this paper is organized as follows. Sections \ref{sec:background} and \ref{sec:threatmodel} provide background and detail the threat model. The LER protocol is explained in Section~\ref{sec:protocol}, and analyzed in Section~\ref{sec:analysis}. Section~\ref{sec:recoveval} complements with further evaluation. We discuss future work in Section~\ref{sec:discussion} and conclude in Section~\ref{sec:conclusion}

%% file: background.tex
\section{Background and Related work}
\label{sec:background}

\subsection{GNSS}
GNSS is a one-way ranging system. The receivers locate themselves by calculating the time-of-arrival (ToA) of the signals originating from four or more satellites. Due to the availability of cheap software-defined radios, the instances of GNSS spoofing are increasing rapidly. These attacks can be performed by manipulating the data transmitted by the satellites or their arrival time at the receiver~\cite{gps_spoof_1}. Several countermeasures leveraging spatial diversity, inertial sensors, opportunistic signals, and delayed key disclosure have been studied to increase resilience against GNSS spoofing~\cite{rabinowitz_capabilities_2000,GPS_harshad,Nils2011requirements}. The techniques to secure against GNSS spoofing include advanced signal processing~\cite{Aanjhan_Spree},  Cryptographic method~\cite{osnma,chimera}, correlating with other sensors, such as inertial navigation system (INS)~\cite{GPS_aanjhan}. With Open Service Navigation Message Authentication (OSNMA), the European Galileo satellite system introduced an anti-spoofing service directly on a civil GNSS signal.
Similarly, GPS is expected to use Chips-Message Robust Authentication (Chimera). However, adding authentication does not increase the security of the GNSS, as an adversary can spoof receivers to any location independent of the cryptographic primitive implemented by manipulating the arrival time of the signals~\cite{maryam_GNSS}. Moreover, even advanced signal processing cannot detect an attacker that completely overshadows the legitimate signal~\cite{Aanjhan_Spree}. While attempts have been made to combine inertial navigation sensors and GNSS to detect GNSS spoofing instances in vehicular settings, the build-up of errors in inertial measurement prevents the detection of GNSS spoofing. The existing approaches to detect GNSS spoofing have shortcomings and cannot be trusted to provide secure localization and tracking. 

\subsection{5G Positioning}

3GPP standards and  5GAA Automotive Association are exploring the possibility of designing terrestrial positioning infrastructure using 5G-NR~\cite{5AA_V2X_positioning,3GPP,3GPP_38.855}. The availability of larger bandwidth makes 5G a perfect fit for high-accuracy positioning. In the transportation sector, road users (cyclists, pedestrians, and vehicles) are expected to use both absolute and relative positioning with assistance from the roadside units (RSU) and base stations (BS). A significant amount of mobility is expected in Cellular Vehicle to Everything (C-V2X) positioning, e.g. vehicle driving by an RSU, a stationary vehicle trying to position with respect to an anchor vehicle that is driving past, etc. These systems are expected to support traffic management and collision prevention. Several field tests are already exploring capabilities of 5G positioning~\cite{ESA_5G_GPS,testbed1}. 

The 5G standard has currently adopted Observed Time Difference of Arrival (OTDoA) using Positioning Reference Signals (PRS) for distance measurement~\cite{3gpp_rel16}, which are vulnerable to distance manipulation by overshadowing attack~\cite{vrange}. An external adversary can shorten and enlarge the measured distance. Therefore, PRS cannot be trusted to provide secure positioning. The possibility of the two-way ranging is already explored in Release-16. 3GPP has plans to enable secure positioning for C-V2X in Release-18, and it will consider different physical and logical layer designs. The techniques included in the standard so far cannot be trusted to provide secure positioning.

\subsection{Secure Distance Estimation}
The two-way ToF-based ranging systems have emerged as an approach to achieve high precision secure positioning~\cite{LRP_HRP_comparision,Nils_wisec2017}. However, not all ToF-based positioning systems provide secure positioning. The security of wireless positioning depends on logical and physical layer designs and integrity checks at the receiver. Most research on enabling secure positioning focuses on upper bounding measured distance using logical layer cryptographic protocols known as distance bounding protocols~\cite{Brands1994,DistanceHijacking,DB_10,DB_EPFL,DB_Ariadne_forRouting,DB_Multi,kasper_DB,GDB_Capkun} and then using them for verifiable multilateration~\cite{VerifiableMultilateration}. The logical layer distance bounding protocols fall short if an adversary is capable of performing physical layer attacks~\cite{cicadaEPFL,edlcEPFl,Flury_reductionattack,Patrick_HRP,HRP_singh_wisec}.
An attacker can manipulate distance measurement if the notion of ToA can be manipulated at the receiver. For example, an Early Detect and Late Commit (ED/LC) attacker reduces the measured distance by preemptively injecting a non-committal waveform that triggers an early signal detection~\cite{cicadaEPFL,edlcEPFl,Flury_reductionattack}. The goal is to cause the receiver to register an earlier ToA, even if the attacker cannot generate the full symbol in advance due to a lack of knowledge about the signal content. An attacker can delay the ToA estimation by overshadowing attacks.

A class of cryptographic primitives, MTACs, have been designed to check the integrity of the ToA manipulation against physical layer attackers. In a similar way that Message Authentication Codes protect message integrity, MTACs preserve the integrity of message arrival times~\cite{MTAC}. Existing MTAC designs like V-Range~\cite{vrange} and UWB-ED~\cite{UWB-ED} provide security against distance reduction and enlargement attacks at the physical layer against an external attacker (i.e., Mafia Fraud). Providing security against Mafia Fraud is sufficient for use cases like Passive Keyless Entry and Start systems, where both car and key are part of the same system. However, in collaborative and opportunistic ranging, where the road user (used for the distance estimation) and GNSS spoofer can be the same entity/user, we need to operate under a broader attacker model where the device involved in the distance measurement can be malicious and collude with other attackers. 

\begin{figure}
    \centering
    \includegraphics[width=6.5cm]{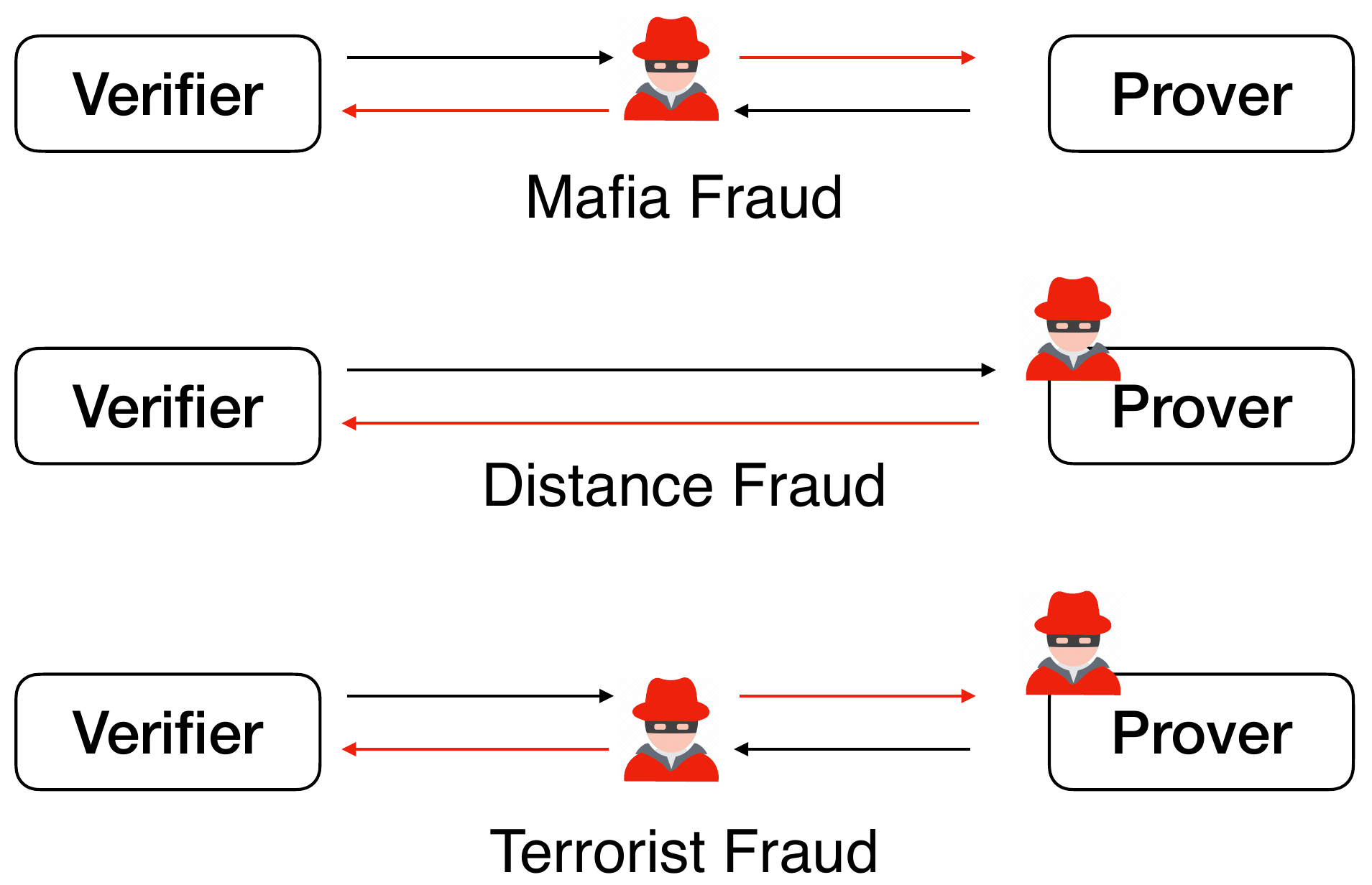}
    \caption{There exists the possibility of different distance manipulation frauds. A malicious user device may exist in the communication region and perform Distance Fraud or use a proxy present in the communication to perform Terrorist Fraud or Distance Hijacking to prove its presence in the communication region. }
    \label{fig:db}
\end{figure}

%% file: attacker_model.tex
\section{Threat Model}
\label{sec:threatmodel}

First and foremost, we consider the possibility of GNSS spoofing, where an attacker can simultaneously manipulate the GNSS of one or more road users. The attacker may have multiple antennas and perform beamforming to manipulate the perceived location of the receivers to an arbitrary location.

Second, we consider logical and physical layer attacks on the distance the road user measures with other users and infrastructure. We consider the possibility of the external attacker (i.e., Mafia Fraud), where this attacker may want to influence the distance measurement between two legitimate users in order to manipulate the outcome of the hybrid positioning. For these attacks, we consider the possibility of Dolev-Yao’s attacker~\cite{DolevYao}. Malicious entities can listen to any communication that happen in the communication region, as well as communicate with the other malicious nodes outside the communication region. Within the communication region, they can also inject signals, and through such injection it can block/modify the authentic signals. If successful, this injection can lead to jamming, signal annihilation, and/or content modification. 

We also consider that a malicious entity can be actively involved in distance measurements (i.e., Distance Fraud), such as fake basestations and malicious road users. These attackers can perform GNSS spoofing and distance manipulation simultaneously. While we consider that some of the entities can be owned by an adversary, the application responsible for executing our protocol will run inside a secure integrity-protected processing environment, consisting of memory and storage capabilities~\cite{tee}. We assume that attacker have limited access to this secure environment. However, the transceiver module used for receiving and transmitting signal for the distance measurement can be a shared resource controlled, as a device many use it for the data communication not related to our application. We also consider that attacker can manipulate internal clock at the receiver in order to manipulate the processing time computed inside the trusted environment. 

Moreover, we assume that the malicious users can collude with each other or use legitimate users as a proxy for distance measurement (i.e., Distance Hijacking and Terrorist Fraud).

%% file: protocol.tex
\section{Methodology}
\label{sec:protocol}

\subsection{Overview}

We design the Location estimation and Recovery (LER) system,  a hybrid positioning system that combines GNSS and 5G positioning to detect spoofing and recover correct location coordinates. The positioning will be mostly carried out between road users using minimal support from the 5G infrastructure. With this protocol, we want to answer the following question: \textit{can we use distance measurements between road users to detect GNSS spoofing? How many such distance measurement instances with other road users will be enough to detect and recover from the GNSS spoofing?} 

Since 5G positioning is currently vulnerable to distance manipulation attacks, we cannot directly use them to ensure secure estimation. Therefore, we design a Location Verification Protocol based on V-Range~\cite{vrange}. The V-Range is an MTAC based on Orthogonal Frequency Division Multiplexing (OFDM) and is compatible with the 5G. Since MTAC only provides security against Mafia Fraud, we extended its design to prevent Distance Fraud, Terrorist Fraud, and Distance Hijacking.
We use wider subcarrier bandwidth OFDM symbols to reduce the effect of Distance Fraud. The challenge and response for the distance bounding will also be computed inside a trusted environment, and the external part, which may be under the attacker's control, will not have access to the keys used for computing them. The Location Verification Protocol ensures protection against Mafia Fraud, Distance Hijacking, and Terrorist Fraud and limits the effect of Distance Fraud. Every instance of the Location Verification Protocol between two devices gives a binary output based on whether GNSS coordinates match with distance measurement.

The recovery protocol is the algorithm that interprets the output of the S-LER protocol such that it guarantees GNSS spoofing detection with minimal false positives. The meta-protocol optimizes the parameters for successful GNSS spoofing detection within a fixed number of verification and also minimizes the false positive rate. The algorithm considers the presence of malicious nodes in the verification process and interprets the verification output accordingly. The final weight function obtained captures the temporal and dynamic nature of the mobile entities.

%The use of MTACs in this scenario will face the following two problems. First, two-way ranging increases the latency of the systems and is not considered scalable. Therefore, it may not be possible to perform ranging with every road user in the communication region. Second, MTACs are secure against an external adversary. However, V2X is a special case, RSU may not always exist, and another road user would act as an anchor for positioning; some of them may be malicious and collude with the GNSS spoofer. Therefore, we need to come up with a design that prevents all known distance manipulation frauds, including Distance Fraud, Terrorist Fraud, and Distance Hijacking.

\subsection{Location Verification Protocol}
The Location Verification Protocol enhance MTACs to detect all known distance manipulation attacks by enhancing security at the physical layer as well as on the computation of responses. It is responsible for executing distance bounding with the nearby road users. Each instance of this protocol output the success or failure in matching GNSS coordinates with the distance estimate, which we then feed to Recovery and Meta Protocol to detect and recover from GNSS spoofing. In the following we provide details of the protocol and involved entities.

% uses distance bounding between road users to detect GNSS spoofing and recovery actual location coordinates. We discuss the role of different entities involved in the process of distance bounding protocol. The output of this protocol is the success or failure in matching GNSS coordinates with the distance estimate. We feed output of

% and the algorithm that takes distance bounding success and failure as an input to decide if the GNSS is actually spoofed and recover from it.
% The following is the list of entities and their roles.

\begin{figure}[t]
  \centering
  \includegraphics[width=0.7\linewidth]{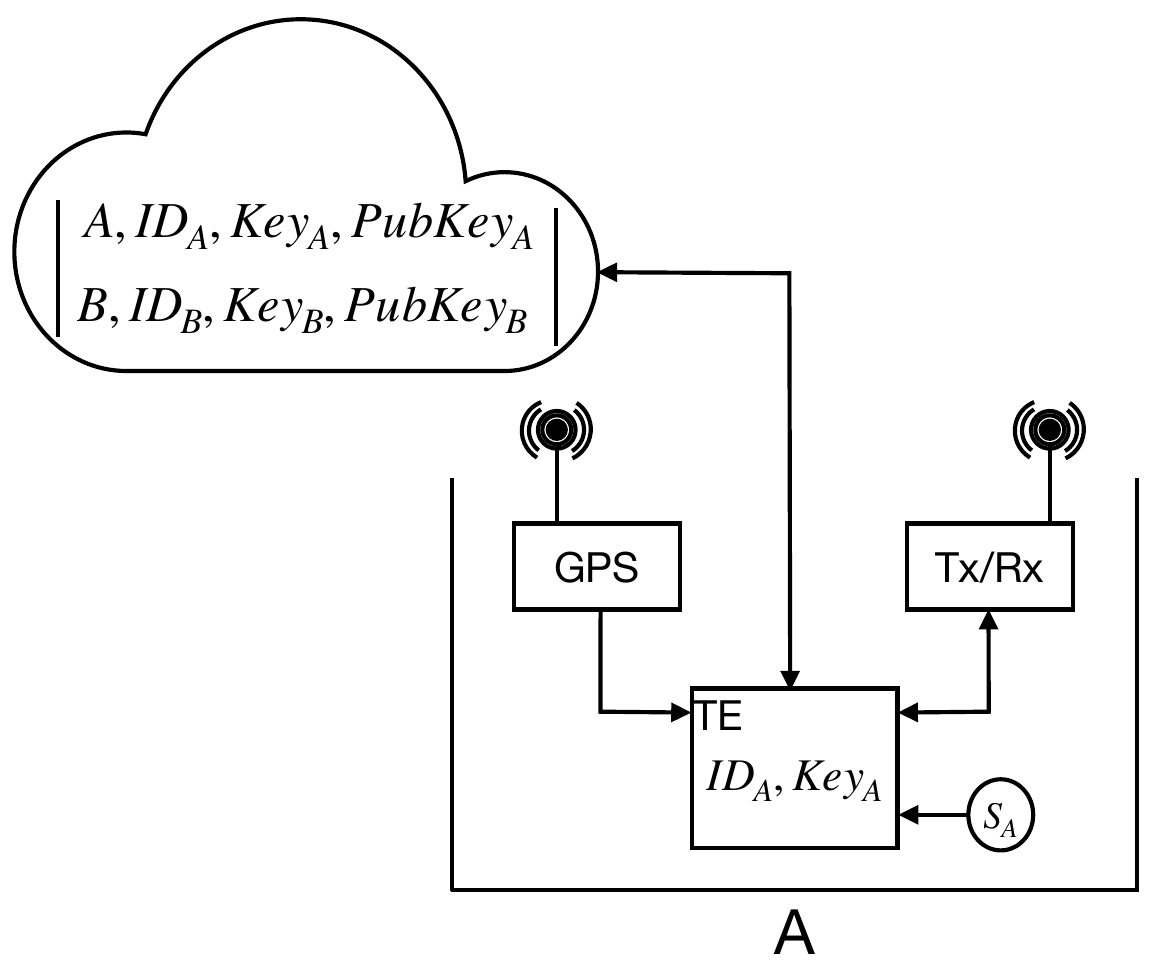}
  \caption{Key distribution, using key server and trusted environment of the user's device.
  }
  \label{fig:keyDist}
\end{figure}

% \begin{figure}[t]
%   \centering
%   \includegraphics[width=1\linewidth]{images/Error_region}
%   \caption{In this scenario, the entities $A$, $B$, and $C$ alternatively take the role of the verifier and the prover. A passive verifier D can passively validate its GNSS coordinates without transmitting any signal. 
%   }
%   \label{fig:error_region}
% \end{figure}

% \begin{itemize}
% \item 

\begin{figure*}[t]
  \centering
  \includegraphics[width=0.76\textwidth]{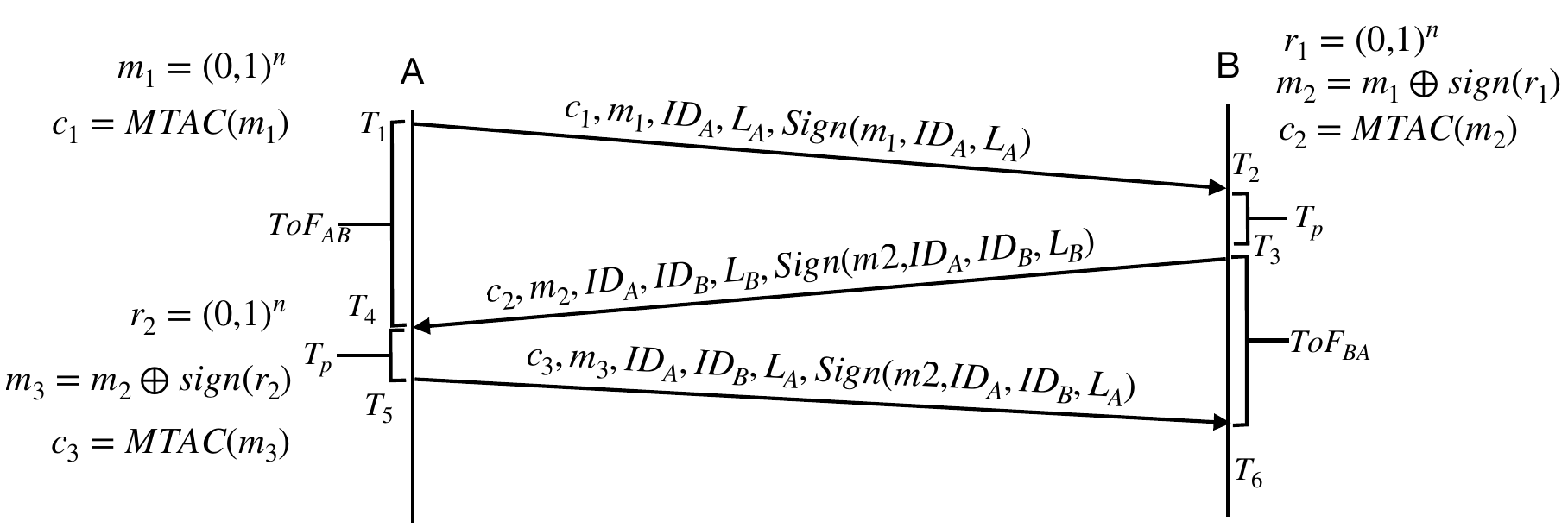}
  \caption{ Distance bounding protocol with MTAC.}
  \label{fig:protocol}
\end{figure*}

\noindent\textbf{Key Server.} It is responsible for providing $(ID,key)$ pair to all road users. As shown in the Figure~\ref{fig:keyDist}, we consider that a protocol will be executed between the key server and trusted environment at the user device to obtain a new identity and public key. The device can use the long-term shared key $(S_k)$ for accessing the server, establishing its identity and obtaining a new pair of $(ID,key)$. Assigned $(ID,Key)$ pair is valid only for the the $T_k$ duration and a new pair is assigned after this time. We also consider that temporary $(ID,Key)$ pair is only available inside the trusted environment(TE) and it is inaccessible by any unauthorized code or user. We use TE as a concept, where keys will be stored and used for generating challenge and response at the user's device. 
We consider TE as the root-of-trust and fully trusted. Also, the adversary cannot manipulate $(Key, ID)$ mapping stored at the server or manipulate it during a query because of the already established public key infrastructure (PKI) in place. However, an adversary can also query the server and acquire the public key $PubKey$ of any user. One of the important requirements is that each user should be assigned only one ID.  Such would be required for many C-V2X applications, where users must rely on other road users for information propagation. We do not provide exact details of the protocol that would be executed between device and key server, as many other applications demand for such architecture. For example, 3GPP is designing the SEAL architecture for key distribution in the C-V2X scenarios ~\cite{3gpp_seal}.

%https://www.allaboutcircuits.com/industry-articles/trusted-execution-environments-tees-in-connected-cars/

\vspace{10pt}
\noindent \textbf{Verifier and Prover} 
Any road user, RSU, or BS can assume the role of verifier and prover. The verifier initiates the ranging process with the prover device present in the communication region. The prover responds to the ranging message after processing time $T_p$. The verifier and prover can query the key server to access the public key of the prover and verifier, respectively. They both announce their IDs and all the messages transmitted during the protocol execution are tied to them. The integrity of messages can be established by accessing  $PubKey$ associated with these IDs from the key server. In these devices, while time-of-arrival estimation occurs at the transceiver, the received messages are passed to the TE for data-level integrity checks and response computation. 

% \item \textbf{.} The Prover announces its ID and shares its GNSS coordinates with the co-located vehicles. The Prover responds to a ranging message after processing time $(T_p)$. The Prover can query the server to access the Verifier's public key.  

% \end{itemize}

While the Key server is a trusted entity, the verifier and prover can be malicious and collude with the GNSS spoofer. 
They can share incorrect GNSS coordinates and IDs. For the ranging process, we use a combination of Message Time of Arrival Codes (MTACs) and Digital Signatures (Sign) to check the integrity of arrival time and data, respectively. Each road user can select their role as verifier or prover. We discuss and analyze the security of these design choices in Section~\ref{sec:analysis}.

\begin{figure}[t]
  \centering
  \includegraphics[width=1\linewidth]{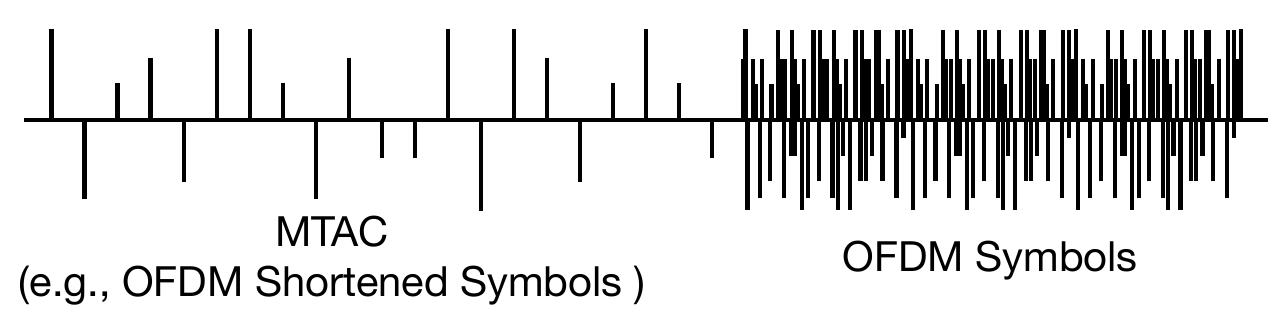}
  \caption{ Physical Layer of the messages exchanges for the distance bounding. 
  }
  \label{fig:phyLayer}
\end{figure}

We execute a distance bounding protocol to measure the distance between entities, including road users, RSUs, and BSs. The two entities, A and B, take up the role of the verifier and the prover alternatively. As shown in Figure~\ref{fig:phyLayer}, the following are the protocol steps.

\begin{enumerate}[noitemsep]
    \item  The entity $A$ takes the verifier role and establishes communication with another entity $B$. The $A$ transmits the physical layer code $c_1=MTAC(m_1)$ at time $T_1$, where $m_1$ is randomly generated by the user $A$. Along with the code $c_1$,  the packet sent by the A contains message $m_1$, its location coordinates $L_A$, $ID_A$, and the digital signature of this data generated inside TE. 

    \item The $B$ takes the role of the prover and estimates the arrival time $T_2$ of $c_1'$\footnote{ The received code $c'$ may not be identical to $c$ due to channel noise. } at the transceiver. The transceiver forwards the code and data to TE for integrity checks and response generation. The $B$ transmits the response after processing time $T_p$ at time $T_3$. The response contain code $c_2=MTAC(m_2)$, location coordinates $L_{B}$, $ID_{A}$, $ID_{B}$, and digital signature of this data. 

    \item The $A$ estimate arrival time of code $c_2'$ as $T_4$ and forward data to TE for processing. After $T_p$ duration, at time $T_5$, $A$ send an code $c_3=MTAC(m_3)$, its GNSS coordinates $L_{A}$,  $ID_A$, $ID_{B}$ and digital signature of this data\footnote{The receiver did not need to recalculate the values of $m_2$ and $m_3$ as they are directly sent to the receiver and as well as the signed version of message $m_2$ and $m_3$ is sent. The receiver retrieves the $sign(r_i)$ by $m1 \oplus m2$ which will be present to him.\\
    $m_2=m_1 \oplus sign(r_1)$\\
    $m_1 \oplus m_2=(m_1 \oplus m_1) \oplus sign(r_1)$\\
    $m_1 \oplus m_2= 0 \oplus sign(r_1)$ (Since, $A \oplus A=0$)\\
    $m_1 \oplus m_2= sign(r_1)$}

    \item  The $B$ estimate arrival time of code $c_3'$ as $T_6$, and forward data to TE for the further processing. 

   % \item All other entities in the passive Verifier mode listen to the communication between $A$ and $B$ and store this information for further processing. \textcolor{red}{It estimate} time-of-arrival of codes $c_1$, $c_2$ and $c_3$, as $T'_1$, $T'_2$ and $T'_3$, respectively. 

    \item In the next iteration, the roles may change between the verifier and the prover.%where some of the passive entities take up the role of the verifier and the prover.   

\end{enumerate}

The message in this protocol contains MTAC, location coordinates, and other information needed for the integrity checks. As shown in Figure~\ref{fig:phyLayer}, shortened OFDM (V-Range) symbols can be used as MTAC, and the remaining data is transmitted as OFDM symbols. After the successful exchange, users can process these messages, even if users $A$ and $B$ are no longer in the communication range. In our protocol, it is important to check the integrity of the arrival time of the signal as well as data integrity. The following checks are needed to declare the execution of this protocol successful. 

\vspace{10pt}
\noindent\textbf{MTAC Verification.} This check is performed as soon as the signal arrives at the receiver. It is imperative to perform this check for secure time-of-arrival estimation. The exact procedure to perform these checks depends on the MTAC design for transmitting code $c_i$~\cite{MTAC,vrange}. The MTAC verification is done in two steps. First, we must perform a signal integrity check to ensure the code is not maliciously tampered with during the transmission. As shown above, received MTAC samples $c_i'$ are affected by channel noise and, consequently, are not identical to $c_i$. The difference in the $c_i$ and $c_i'$ depends on the modulation and channel conditions. Therefore, each receiver may have a different version of the MTAC $c_i$ due to different channel conditions. The receiver needs to assess the samples' power and variance in $c_i'$ to perform this check. Once sanity at the physical layer is established, the receiver checks for the bit error. The receiver demodulate code $c'_i$ to message $m'_i$ and checks for the bit error by comparing it with the message $m_i$. Next, we check for data integrity if signal integrity is established and bit errors are acceptable.

\vspace{10pt}
\noindent
\textbf{Data Validation.} Checking the integrity of the messages exchanged between users is crucial. It determines if messages, GNSS coordinates, and ID information exchanged between the users can be trusted. Each user independently queries the key server to acquire public keys of the ID associated with other users. Users A and B need access to the public keys of B and A, respectively. If the integrity of the messages cannot be established, then the ranging process is marked unsuccessful. This validation ensures that the data is generated within the time $T_k$ and when location coordinates of entities with $ID_A$ and $ID_B$ are $L_A$ and $L_B$, respectively.  If multiple ranging are performed with the device with the same ID, then only the first ranging is used, and any further ranging instances are discarded.

\vspace{10pt}
\noindent
\textbf{Distance Estimate Validation.} Using GNSS coordinates and estimated arrival time of messages, each user can validate their location coordinates. Using timestamp $T_1$ and $T_4$, the entity $A$ estimate time-of-flight $ToF_{AB}$ and compare it with the location coordinates. If $|(ToF_{AB}/2) - (D_{AB}/c)| \le \delta $, where $D_{AB}$ is the Euclidean distance between the GNSS coordinates of $A$ and $B$, \textit{c} is the speed of signal propagation, and $\delta$ error allowed in the distance measurement due to channel conditions, then entity $A$ accept this measurement. Similarly, if $|ToF_{BA} - (D_{BA}/c)| \le \delta $, then entity $B$ accepts this measurement.

\subsection{Recovery and Meta Protocol}

In this section, we describe the algorithm that uses the output of the verification phase (success/failure) with the neighboring entities to detect GNSS spoofing with minimal false positives in the presence of malicious entities in the network. This algorithm---called the \emph{recovery protocol}---decides when to go into recovery mode based on the information obtained in the verification phase. 
The two objectives of the recovery protocol are to detect GNSS spoofing as fast as possible when it occurred, and to minimize false positives (i.e., going into recovery mode despite the GNSS not being spoofed).
The recovery protocol depends on several parameters that allow the user to fine-tune the protocol to the respective setting (i.e., the assumed percentage of malicious vehicles) and preferences (e.g., the trade-off between speed of spoofing detection and false positives).

To find the optimal parameters in the recovery protocol for any combination of setting and preferences chosen by the user, we provide a meta-protocol for adapting the parameters of the recovery protocol to the chosen setting/preferences (pseudocode in the appendix Section~\ref{alg:one}).
In the meta-protocol, the user can select values for three parameters $p$, $E$ and $T$ and is provided with a recovery protocol fine-tuned for this specific parameter combination.
The first parameter $p$ describes the (assumed) percentage of malicious entities in the verification network. %\snote{is it clear what is meant by that?} \snote{do we say earlier that we assume a setting where each vehicle is malicious with the chosen probability independently of each other?}
The second parameter $E$ describes the number of distance measurements after which GNSS spoofing is detected in expectation, and the third parameter $T$
is the number of (most recently) encountered vehicles that the recovery protocol bases its decision on whether to go into recovery mode.
%is the total number of distance measurements after which we are guaranteed to detect GNSS spoofing
%% \snote{check this whole section w.r.t. whether one should say vehicles here or measurements or ...}
%, measured from the point in time when the GNSS starts to provide false coordinates.
In particular, the user can decide how long the protocol is allowed to take to detect GNSS spoofing (which in turn affects the false positives rate)---time here is measured in number of encountered vehicles, but this can be translated to time in seconds via estimations about traffic density.
We first describe the recovery protocol and then the meta-protocol for parameter optimization.

\subsubsection{The Recovery Protocol}\label{sec:recproc}
The recovery protocol uses a sliding window mechanism with a sliding window of fixed size $T$. %\snote{should we mention that this parameter provides a trade-off between on-the-fly computation time and optimizability of the protocol, i.e., we can get a slightly better protocol (in terms of quality) if we take a larger T? And that, in principle, this parameter can be chosen by the user, but it only affects both quality and computation time in a more or less negligibly way if the parameter is chosen somewhat reasonably?}
The sliding window keeps track of the most recently encountered vehicles; each time a new distance measurement enters the sliding window, the least recently encountered one is dropped from the window, catering to the fact that more recently obtained information is more valuable.
Also, each time a new measurement enters the sliding window, a computation takes place based on which the decision is taken whether to go into recovery mode or not. %\snote{resp.\ whether to go into alert phase or exactly whatever else we want here; insert the right thing here and elaborate on the things that happen when we go into recovery mode}
The main two ingredients for this computation are
\begin{enumerate}[noitemsep]
    \item a weight function $w$ that assigns to each of the $T$ vehicles in the sliding window an individual weight, where more recently encountered vehicles are given a larger weight (again representing the fact that more recently obtained information is more valuable), and
    \item a threshold parameter $t$ (which represents the trade-off between the speed of spoofing detection and the minimization of false positives).
\end{enumerate}
The first step in the computation is, roughly speaking, to calculate the weighted sum of the failed verification attempts.
More precisely, we calculate the sum
\begin{equation}\label{eq:sum}
    D=\sum_{i=1}^{T}(w_i \cdot v_i),
\end{equation}
where $w_i$ is the weight assigned to the $i$th recently encountered vehicle and $v_i$ is a binary variable that has value $1$ if and only if the coordinates received from the $i$th recently encountered vehicle do not match the coordinates provided by the GNSS. %\snote{Is this "matching coordinates" way of phrasing things correct? Please replace by something more correct if applicable.}
In a second step, we compare the obtained sum $D$ with our threshold parameter $t$: if $D > t$, then we go into recovery mode, if $D \leq t$, then we do not.
% \anote{Recovery protocol?}
% \msnote{Find right place for this, and maybe re-write....Is this enough? 
In the recovery mode, the entity turns to the fully active mode and executes the LER protocol more aggressively. It also increases the frequency of the distance measurement with the RSUs and BSs, as these entities, if available, have a smaller probability of being malicious. By increasing the frequency of protocol execution, the entity forces the attacker to introduce more colluding malicious entities in the communication region.

\subsubsection{The Meta-Protocol}
In this section, we describe how we optimize the parameters for the recovery protocol (in particular, the values of the weight function $w$ and the threshold value $t$), by means of the meta-protocol.
The purpose of the meta-protocol is to obtain a fine-tuned recovery protocol for different settings (characterized by the percentage $p$ of malicious vehicles) and preferences (characterized by the expected number $E$ of vehicles after which GNSS spoofing is detected and total window size $T$).
Accordingly, the three values $p$, $E$  and $T$ are inputs to the meta-protocol, to be specified by the user.
As in the recovery protocol, the parameter $T$ describes the size of the sliding window.
Additionally, we will make use of a parameter $N$ that describes how often certain experiments are repeated in the meta-protocol.
Again, $N$ can be chosen by the user, where higher values for $N$ provide a better optimization at the expense of an increased runtime of the meta-protocol.

The objective of the meta-protocol is to find a good weight function $w$ and threshold parameter $t$ that optimize the recovery protocol for the user-specified values of $p$, $E$ and $T$.
Due to the nature of the expression for $D$ (s.\ Equation~\ref{eq:sum}) in the recovery protocol (which is clean, but too complex to allow for a direct deduction of the optimal parameters as closed-form expressions), searching for optimal parameters via a gradient descent approach seems like a good choice.
This is supported by the fact that more recently encountered vehicles will be given larger weight in the recovery protocol (providing a monotonic weight function), which makes it plausible that all local minima in the parameter optimization problem are also global minima.
Moreover, the experimental results provided in Section~\ref{sec:recoveval} confirm that the parameters found by our approach are indeed (close to) optimal.
We observe that due to the discrete nature of the recovery protocol, there are entire ranges of parameters (more specifically of the weights of the weight function) that provide exactly the same guarantees, making our optimization task more complicated.
We overcome this challenge with an approach based on suitably shrinking intervals.

On a high level, the approach of the meta-protocol is the following:
We start with some arbitrary function-threshold pair $(w,t)$ and then optimize the threshold parameter $t$ to yield the desired expected number of steps ($E$) for detecting GNSS spoofing when running the recovery protocol with parameters $w$ and $t$.
Next, we modify the arbitrarily chosen weight function values index-wise.
We increase the weight at a particular index to
%its highest possible value (we have set an upper bound on the highest value to be $2*t$, where $t$ is the optimal threshold we found for the arbitrarily chosen function $w$)
a much larger value ``$\high$'' and compute the optimal threshold $t_{\high}$ and false positives rate $\qual_{\high}$ of the obtained function.
The same procedure is followed by setting the value at the lowest possible value ($\low = 0$) and for a middle value $\mmid$ calculated by $(\low + \high)/2$.
Then we compare the false positive rates $\qual_{\low}$, $\qual_{\mmid}$, and $\qual_{\high}$, and identify intervals of weight values at the chosen index for which the false positives rates are lowest.
Then we adjust the $\high$ and $\low$ values accordingly and repeat this method in a recursive manner until after repeated adjustment of the intervals the interval size becomes very small.
The motivation for this approach is to identify the smallest and the largest weight at that particular index that will guarantee to detect GNSS spoofing after $E$ verifications in expectation and have the lowest false positives rates. Once we identify the smallest and largest for a particular index, the same process is repeated for every other index.

 The final $(w_{\opt}, t_{\opt})$ pair lies within the interval and we always guarantee a monotonic non-increasing weight function to exist. The significance of monotonic non-increasing weight function lies in the fact that the participating entities are dynamic in nature and we are verifying the locations with respect to it. To capture this dynamic and temporal nature of real-world traffic we always a choose a monotonic non-increasing function that is guaranteed to lie in the interval which guarantees GNSS spoofing with minimal false positive rate.  
%Next, we do exactly the same by \textcolor{red}{setting the value of the  particular index to the maximum value possible(we have set an upper bound on the highest value to be $2*t_{\opt}$) and also setting the value to the lowest value i.e., $0$. Now, we calculate the $t_{\opt}$ value for the lowest setting and measure the $\qual$ value(measures true negative rates) for the lowest setting. The same set is }slightly perturbed function-threshold pair $(w',t')$ and then compare the two pairs $(w,t)$ and $(w',t')$ w.r.t.\ which combination will produce fewer false positives in the recovery protocol.
%Then we will recurse, by keeping the pair that produced fewer false positives, again perturbing it slightly to obtain a new pair to compare it to, and so on, until we obtain a pair $(w_{\opt}, t_{\opt})$ that produces very few false positives (while still guaranteeing detection of GNSS spoofing after $E$ steps in expectation).
More precisely, the meta-protocol proceeds as follows (pseudocode in appendix~\ref{alg:meta}).

\begin{enumerate}[noitemsep]
    \item\label{step:start} Choose an arbitrary weight function $w$ and threshold $t$.
    \item\label{step:spoofed} Run the following experiment N times, each time storing the number of verification steps it takes to go into recovery mode.
    \begin{enumerate}
        \item Assume that the GNSS is spoofed and provides false coordinates starting at time $0$. Throw a biased coin that shows heads with probability $1-p$ for each vehicle encountered after time $0$.\footnote{Technically, for each such vehicle, we will only throw the respective coin whenever the vehicle is considered in the subsequent calculations, preventing us from throwing coins prematurely.} Whenever we throw heads, the respective vehicle is assumed to be non-malicious, conforming to the expected percentage $p$ of malicious vehicles.
        \item Run the recovery protocol with parameters $w, t$ on the sequence of vehicles\footnote{This (imaginary) sequence of vehicles is simply characterized by a value for each vehicle that describes whether the vehicle is malicious or not, determined by the respective coin throw. } until the protocol decides to go into recovery mode. Here, each non-malicious vehicle $i$ provides a value of $v_i = 1$ in the sum~(\ref{eq:sum}), while each malicious vehicle $i$ provides a value of $v_i = 0$. This mirrors the fact that the coordinates received from the non-malicious vehicles are exactly those that disagree with the coordinates provided by the (spoofed) GNSS.\footnote{For malicious vehicles, we can assume in a worst-case fashion that they provide exactly the same coordinates as the spoofed GNSS. This only makes our results stronger as such perfect coordination on the side of the attacker might be hard to achieve.}
    \end{enumerate}
        \item\label{step:average} Compute the average number of verification steps across all $N$ experiments. If this number is larger than $E$, then update $t$ to a slightly lower value; if it is smaller than $E$, then increase it slightly.
        \item\label{step:getE} Iterate Steps~\ref{step:spoofed} and~\ref{step:average}. The updates performed in Step~\ref{step:average} ensure that the pair $(w,t)$ will get closer and closer to a function-threshold pair that needs exactly $E$ steps in expectation to reach recovery mode. Consequently, when the average number of verification steps in Step~\ref{step:spoofed} is precisely $E$, then we have found the optimal threshold value $t_{\opt} = t$ for our fixed weight function $w$; in this case terminate the iteration of Steps~\ref{step:spoofed} and~\ref{step:average}.
        \item\label{step:qual} Next, we assign a value $\qual$ to the obtained pair $(w, t_{\opt})$ that measures the false positives rate. To this end, we consider the non-spoofed case (in which the coordinates provided by the malicious vehicles disagree with the coordinates provided by the GNSS). For each vehicle in a window of size $T$, toss a biased coin that shows heads with probability $p$ (marking it as a malicious vehicle that will provide a value of $v_i = 1$ in the recovery protocol), and compute whether the recovery protocol, when executed on this window, would go into recovery mode. Repeat this experiment $N$ times. The $\qual$ value for the pair $(w, t_{\opt})$ is defined as the number of experiments in which the recovery protocol does not go into recovery mode divided by $N$, i.e., the fraction of true negatives.
        \item Set the $\high$ value to the $2 \cdot t_{\opt}$ of the first arbitrary function $w$ chosen and the $\low$ value to 0.
     
     \item For each index starting from $0$ to $T$ the following steps are taken with the starting $\low$ and $\high$ value.

         \begin{enumerate}
                \item \label{step:recursion} Three new weight functions $w_{\low}$, $w_{\mmid}$ and $w_{\high}$ are obtained, by setting the weight at the chosen index to $\low$, $\mmid$, and $\high$, respectively, where $\mmid=(\low + \high)/2$.
                \item For each of the three modified weight functions, go to step \ref{step:spoofed}  and obtain the optimal thresholds $t_{\low}$, $t_{\mmid}$, and $t_{\high}$.
                \item Now, for the three weight functions and their corresponding optimal thresholds go to ~\ref{step:qual} and obtain the false positive rates $qual_{\low}$, $qual_{\mmid}$ and $qual_{\high}$.
                \item Then, based on the false positive rates we obtained, we carry out the recursion in following manner:
                \begin{enumerate}
                    \item If $\qual_{\mmid}$ is less than $\qual_{\low}$ set $\low=\low$ and $\high=\mmid$ and go to step \ref{step:recursion}.
                    \item if $\qual_{\mmid}$ is less than $\qual_{\high}$ set $\low=\mmid$ and $\high=\high$ and go to step \ref{step:recursion}.
                    \item Else, we consider the entire interval and split it into two parts i.e.,$[\low, \mmid]$ and $[\mmid, \high]$ and go to \ref{step:case2} for each of the intervals separately. 
                    \begin{enumerate}
                        \item \label{step:case2} If $\qual_{\mmid}$ is less than $\qual_{\high}$ set $\low=\mmid$ and go to \ref{step:case2}.
                        \item If  $\qual_{\mmid}$ is equal to $\qual_{\high}$ set $\high=\mmid$ and go to ~\ref{step:case2}.
                        \item if $\qual_{\mmid}$ is higher than $\qual_{\high}$ set $\mmid=\high$ and go to ~\ref{step:case2}.
                        \item This will continue until both interval sizes become very small ($\leq 0.00001$). Then we calculate the $\qual$ value for the four points at the boundaries of the two intervals and return the two (consecutive)points where the $\qual$ value is highest. These two points will comprise the lower bound and the higher bound for the weight at that particular index.
                    \end{enumerate}
                \end{enumerate}
         \item Go to the next index and continue from ~\ref{step:recursion}.
         \end{enumerate}
     
      \item Finally, we make use of the intuition that more recent vehicles should feature with a higher weight in the recovery protocol as follows. Turn the weight function obtained at the end of Step ~\ref{step:recursion} into a monotonically non-increasing function that will lie within the lower and higher bound obtained. 
   % \item As a further post-processing step, one can also decrease the window size $T$ if some values of the obtained weight function are $0$. In this case, we reduce the window size $T$ to the number of nonzero values, truncate the weight function in the natural way to this smaller window, and run the overall algorithm again. This can be iterated in case the repetition of the overall algorithm results in further $0$-values of the weight function.
\end{enumerate}
     
   %  Change the weight function $w$ to a new weight functions $w_{\lower}$, $w_{mid}$ by
     
   %  via some random perturbation (and choose a random initial threshold value $t'$). Repeat Steps~\ref{step:spoofed}--\ref{step:qual} for the pair $(w',t')$. Compare the $\qual$ values of $(w,t_{\opt})$ and $(w',t'_{\opt})$ and keep the pair with higher $\qual$ value. Iterate by perturbing the function in the kept pair randomly (and choosing a random initial threshold value), executing Steps~\ref{step:spoofed}--\ref{step:qual}, and so on. Stop when the $\qual$ value no longer increases.

\begin{figure}[t]
\centering
      \includegraphics[width=0.8\linewidth]{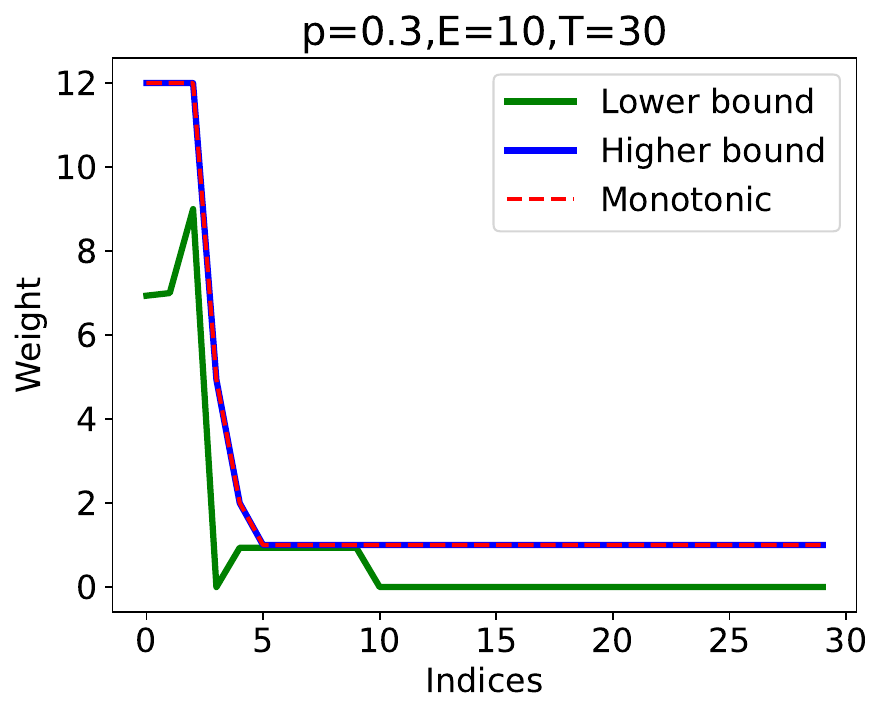}
      \caption{Optimized weight function achieved after executing the meta-protocol}
      \label{fig:optimization}
    \end{figure}

To briefly illustrate the above protocol, consider the parameter choice $E=10$, $T=30$ and $p=0.3$. Figure \ref{fig:optimization} shows the lower bound and upper bound for the weights at each index. Figure \ref{fig:optimization} also demonstrates the monotonic non-increasing weight function that exists within the bounds. The $qual$ value achieved in the mentioned setting is $0.9927$. The optimal threshold for the mentioned setting is $47.9375$.

%% file: protocolAnalysis.tex
\section{Security Analysis of Location Verification Protocol}
\label{sec:analysis}

The verification process uses MTAC for GNSS spoofing detection and recovery. The key server is responsible for providing keys for the verification phase of the MTAC and data exchanged during the protocol. In this section, we analyze this protocol in the presence of an adversary. 
We assume the attacker can spoof the GNSS of the target device to any location of her choice. We also consider that the attacker may have a subset of vehicles under her control, where she can influence the distance measurement. An attacker would manipulate messages or their arrival time to perform attacks. There could be the following attack scenarios for location spoofing: the attacker starts with the cold start, where the attacker manipulates GNSS coordinates when the device is static so that it already has the wrong GNSS coordinates when it starts moving. Otherwise, she starts GNSS spoofing when the entity is already moving on the road and already has some confidence in its location. 

The road users continuously measure distance with the nearby node and determine if this distance measurement matches the GNSS coordinates. 
Therefore, an attacker must pass these checks numerous times to perform a successful GNSS spoofing attack. An attacker cannot use a single malicious node to perform the attack repeatedly. If a node uses the same ID and key mapping for performing ranging in the duration $T_k$, then only the first instance of the ranging is considered. In the time duration $T_k$, during which the $ID,key$ pair stays the same, the receiver only considers these nodes once, and they will be again accepted after time duration $T_k$. Therefore, selecting the same node again does not help the attacker. Therefore, the attacker needs to either influence the distance measurement between legitimate nodes (i.e., Mafia Fraud) or increase the number of malicious nodes (i.e., Distance Fraud), or collude with the entities outside the communication range (i.e., Terrorist Fraud). In the following, we analyze the security of the location verification  protocol under different distance measurement frauds. While these frauds are mostly named in the literature in regard to the upper bound of the measured distance, we are considering them for both the upper and lower bound of the measured distance

% We have this constraint so that any entity should not be able to influence the device's confidence in its current GNSS coordinates. This assumption also includes the possibility of attackers preventing communication with other legitimate devices. 
% For example, an attacker can prevent the legitimate entity from communicating by jamming the signal they transmit and selecting their own node to execute this protocol

%\textcolor{red}{We assume processing time $T_p$ to be in the order of $ns$ to a few $\mu$s, which is insufficient to relay this information to another entity outside the communication region to generate a valid digital signature. }

% \subsection{Properties of Malicious Nodes} 
%A spoofed location coordinate can be accepted by the target node, if an attacker can also control the arrival time of the MTACs. For the MATCs that provide security against Mafia Fraud, for both distance reduction and distance enlargement attack, an external attacker cannot manipulate ToF measurement. As performing such attack would lead to failure during the MTAC Verification check at the receiver. Therefore, an attacker need to actively participate in the verification process. 

%We use V-Range as an example MTAC. However, any MTAC design that protect against both distance reduction and enlargement attacks can be used. 

\vspace{10pt}
\noindent\textbf{Mafia Fraud.} 
Location verification protocol protects against logical and physical layer attacks against an external adversary. For the distance measurement, we use freshly generated nonces and transmit them as MTACs at the physical layer. MTACs, such as V-Range, provide security against distance reduction and enlargement attacks. As shown in Figure~\ref{fig:phyLayer}, if the receiver uses the first part where MTAC is transmitted for the ToA estimation and the remaining part of the signal to validate the data, then the attacker would need to manipulate MTAC during transmission. The physical layer attacks, such as  ED/LC,  Cicada, signal annihilation, and Overshadowing to advance or delay the signal's arrival time has success probability below $10^{-4}$~\cite{vrange}. 

\vspace{10pt}
\noindent\textbf{Distance Fraud.} 
We consider that an attacker owns/colludes with a subset of entities. A malicious node in the communication range can transmit incorrect GNSS coordinates as well as compromise any distance measured with this nodes. GNSS coordinates can always be manipulated by spoofing, and distance measured with other entities can be enlarged, for example, by adding a delay circuit between the transmitter and the antenna. Since the legitimate signal has not yet left the device, signal annihilation or overshadowing would not be required. 

The possibility of distance shortening by distance fraud depends on the architecture. Suppose the transceiver is a trusted peripheral, accessible only to the trusted environment. In that case, an attacker cannot advance the arrival time of the message, as it would need to advance the arrival time of the message, as in the Mafia Fraud. The stronger possibility is that transceivers would be a shared resource, as the same hardware could be used for the ranging and positioning in 5G~\cite{Valero2022}. This scenario allows for the distance-shortening attack by distance fraud. For example, an attacker can predict some of the bits of the packet and forward it to the trusted environment for processing, advancing the generation of the response. 

\begin{figure}[t]
  \centering
  \includegraphics[width=1\linewidth]{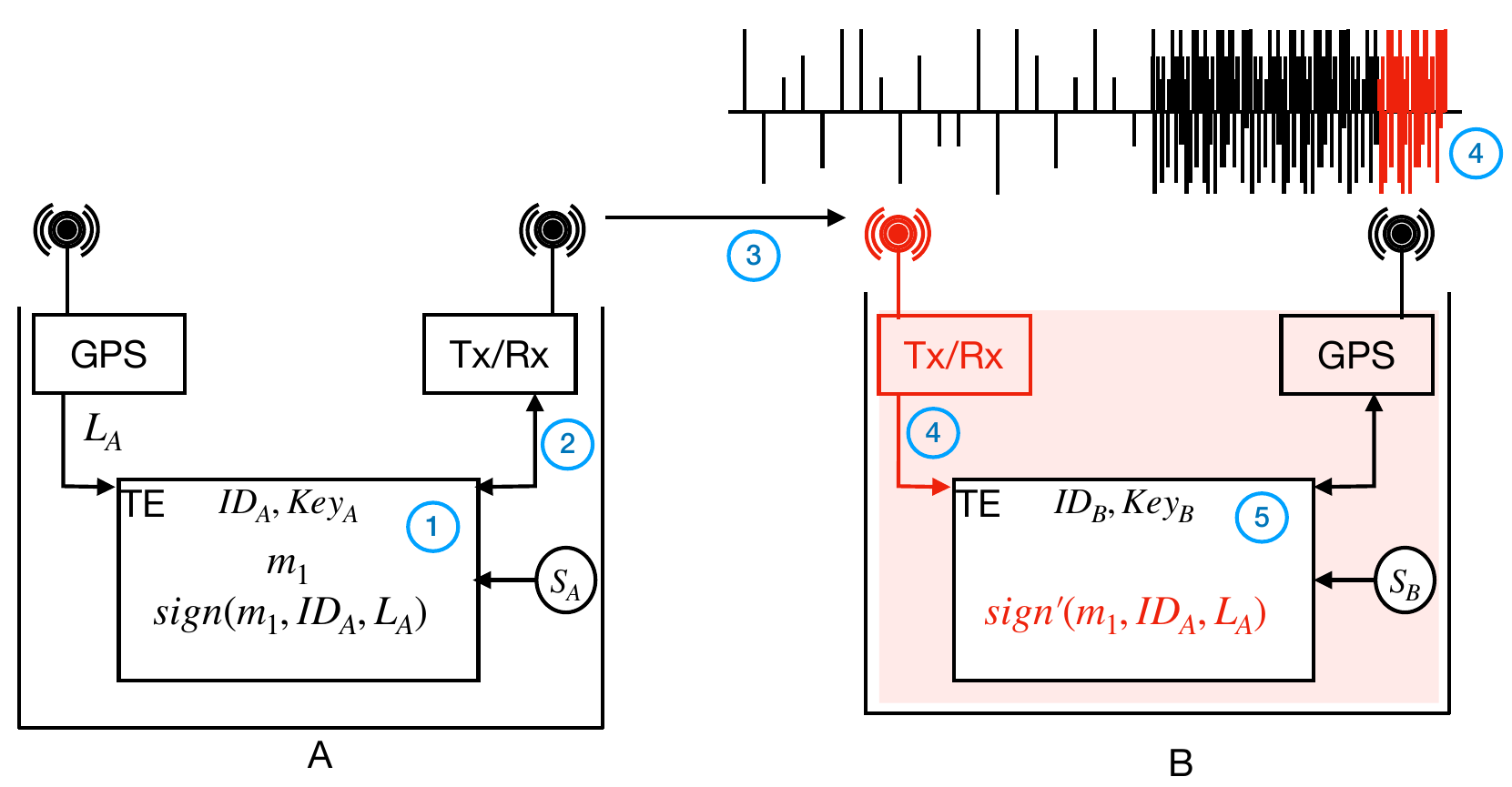}
  \caption{TE and Fraud relation.
  }
  \label{fig:fraud_analysis}
\end{figure}

Attack success depends on the number of bits an attacker want to predict and the symbol duration. As shown in Figure~\ref{fig:fraud_analysis}, MTAC and data modulated on the OFDM symbols are needed to produce the correct response. Suppose the receiver has received the whole message except for the last OFDM symbol. In order to advance the arrival time by the OFDM symbol duration $T_{Sym}$, the adversary needs to predict data bits modulated on this symbol. Since this symbol carries the digital signature of the freshly generated nonce, an attacker would need to guess the bits. For the OFDM symbols with more than 4000 subcarriers, and 16-QAM modulation, it would be around guessing 16000 bits. 
Therefore, an attacker would resort to ED/LC attack. It is shown that the multicarrier time-of-flight ranging is also vulnerable to ED/LC attack. However, the attacker would need $n_s/4+1$ subcarriers for early detection, where $n_s$ is the number of subcarriers of the OFDM symbols and symbols are 4-QAM or 16-QAM modulated~\cite{Patrick_ACSAC}. If we consider processing time zero (processing time will depend on the attacker's capabilities, but it would be non-zero), the maximum time advancement attacker can achieve is $3/4$ of the symbol duration. If we use subcarrier spacing of 480 kHz 
$(T_{sym}= 2.08 \mu s)$, then attacker can achieve maximum distance shortening of $\approx 200m$  
% \msnote{It is on one way... do we need to include both device being malicious.... We need to check how error scale.... }

% This system is unlike UWB systems, where one pulse carry only one bit of information, here we have a multicarroer system. 

% \begin{figure}[t]
%   \centering
%   \includegraphics[width=1\linewidth]{images/Fraud_signalLevel}
%   \caption{TEE and Fraud relation... 
%   }
%   \label{fig:Fraud_signalLevel}
% \end{figure}

The trusted environment does not have a trusted local clock. Therefore, an attacker can also manipulate the processing time. Instead of transmitting the response exactly after $T_p$ time since the arrival of the challenge, an attacker can send it after $T_p \pm \theta$ time duration. Given that the $T_p$ is in the order of nanoseconds to a few $\mu s$, the maximum error occurring due to this reason would still be limited to a few hundred $ns$. 

This analysis shows that there are many ways an attacker can manipulate arrival time or the processing time as malicious prover, therefore, we consider that distance reduction and enlargement attacks are possible up to a certain extent due to Distance Fraud. However, it is upper bounded by the duration of the symbol $T_{sym}$ and processing time $T_p$. Therefore, the attacker still needed to be co-located to perform these attacks and cannot perform them if it is remotely located.

\vspace{10pt}
\noindent\textbf{Terrorist Fraud.}
In a Terrorist Fraud attack, an attacker close to the Verifier tries to impersonate the prover. In our scenario, this would mean that a malicious entity in the communication region colludes with an entity outside of the communication region. The assumption in terrorist fraud is that prover outside communication regions share short-term keys but do not want to share their long-term secret. Since revealing their long-term secret can lead to impersonation by other malicious entities. 

Since our protocol's security depends on the entities' diversity that would meet the legitimate entity and help in realizing the GNSS spoofing detection, it is important to prevent this attack. As shown the Figure~\ref{fig:fraud_analysis}, our challenge and response are generated inside the trusted environment. While a long-term secret exists outside the trusted environment, the temporary $(ID, Key)$ is available only inside the trusted environment. This application is also responsible for generating nonces and digital signatures sent as part of the protocol. If the application closes, the $(ID,Key)$ pair is destroyed. The key required for digital signatures never leaves the trusted environment, so it cannot be shared with other colluders. We consider that Terrorist fraud is not possible without leaking long-term secrets. The sharing of the long-term secret can be further prevented by keeping it in temper-resistant storage. 

\vspace{10pt}
\noindent\textbf{Distance Hijacking.}
In Distance Hijacking attacks, a dishonest prover convinces the verifier that it is at a different distance than it actually is, by exploiting the presence of an honest prover. For example, one of the ways in which the dishonest prover can achieve this is by hijacking the distance measurement phase of a distance bounding protocol from an honest entity. In our scenario, that would mean that the attacker forces one of the non-malicious entities which are present near the target entity to perform the exchange of MTAC codes. Conceptually, Distance Hijacking can be placed between Distance Fraud and Terrorist Fraud. Unlike Terrorist Fraud, in which a dishonest prover colludes with another attacker, Distance Hijacking involves a dishonest prover interacting with other honest provers. In our protocol, we perform explicit linking to generate the response ($m_{i+1} = m_i \oplus sign(r_i)$), which ensures that the response from different provers is distinguishable by explicitly including identity information in the response, combined with integrity protection.

\vspace{10pt}
\noindent\textbf{Communication Jamming.} 
The strength of our protocol design is the density and dynamic nature of the traffic. The more entities encountered in the path increase the confidence in the GNSS location and detect GNSS spoofing. In order to prevent GNSS spoofing detection, an attacker can prevent communication with other entities by jamming signals sent by other entities. In the LER protocol, we consider the frequency at which we meet new entities. The information which is acquired before the $T_k$ becomes redundant. An attacker can prevent communication with other entities, but that would reduce the entity's confidence in its location and will enter in the recovery phase.

\subsection{Attacker Influence} 
%Error an attacker introduce should be undetectable, such that error region depend on the other nodes used for the ranging, and if they are 
%- Maximum bound being the communication error, which attacker can build over time, if it can control traffic.
%some other check on after how much time, the location information become redundant 

The location estimation using the location coordinates and the distance measurement from the neighboring entities requires solving a quadratic equation to find the point at which these circles intersect formed by each node. A single attacker can send malicious data, introduce delays, and do other forms of data manipulation, but it is a single node in a network, since our LER protocol prevents terrorist fraud by its design.  
 
\begin{figure}[t]
  \centering
  \includegraphics[width=1\linewidth]{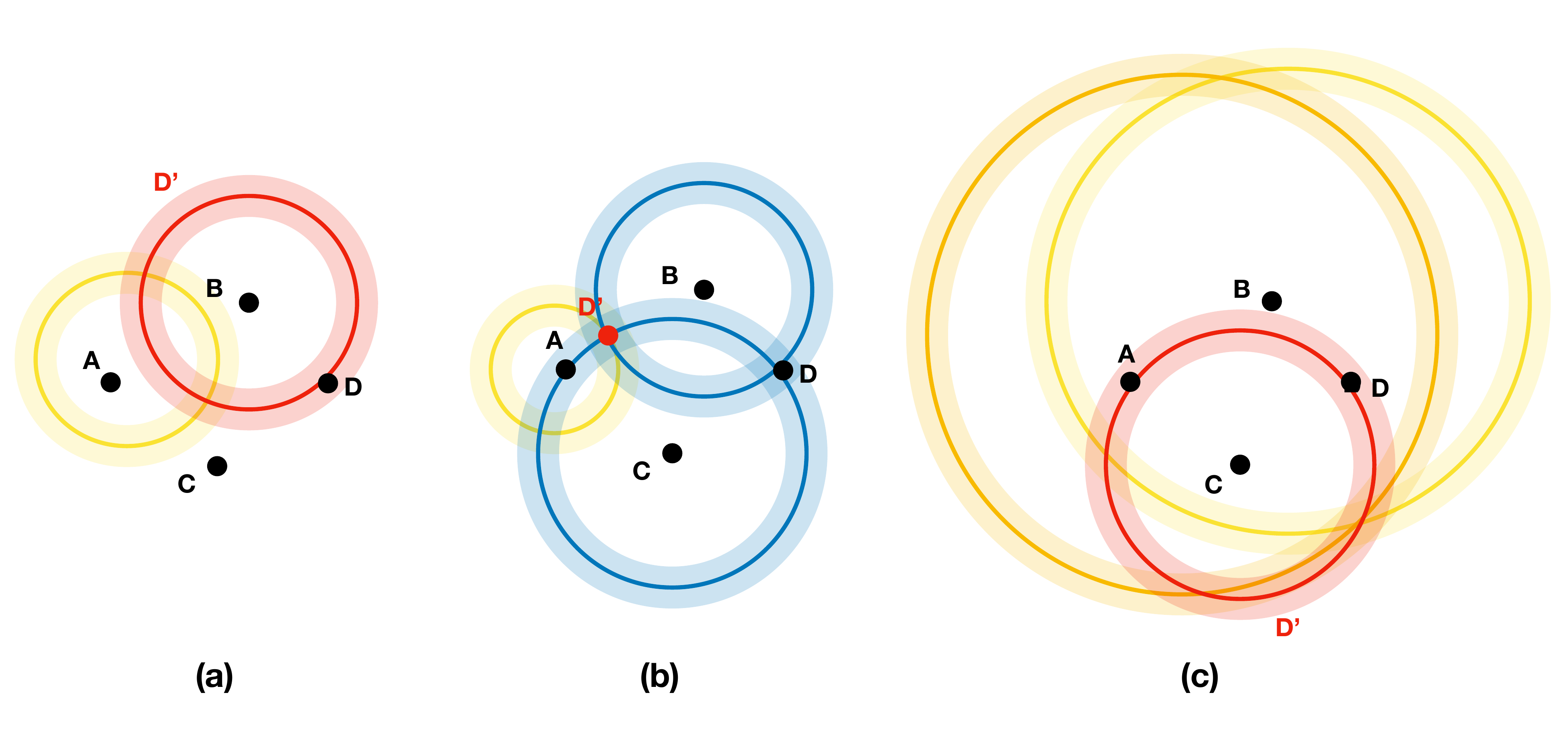}
  \caption{What happens when a subset of active nodes is malicious. %\snote{more concrete caption?} 
  }
  \label{fig:Malicious_error_region}
  \vspace{-20pt}
\end{figure}

Figure \ref{fig:Malicious_error_region}(a) and (b) explains the scenario when one entity is malicious. The non-malicious entities will share the correct information, and the circles of the respective non-malicious entities will intersect at most two points. In order to spoof the location, the malicious node will have only one choice other than the correct location to spoof. Therefore, even if a malicious entity is active, the scope of the attacker is limited to spoofing to a single incorrect location.

Figure \ref{fig:Malicious_error_region}(c) explains the scenario when two of three entities are malicious. The non-malicious entities will share the correct information. For successful spoofing to an incorrect location, the malicious nodes will have a choice of anywhere in the circumference of the circle of the non-malicious entity. The choice of attack space increases with the number of malicious entities. For $n$ active entities of which $k$ are malicious nodes, successful spoofing is possible only when $n < 2k+3$. The two circles can, at maximum, have two intersecting points. If two active nodes are non-malicious, the third malicious entity only has one choice for a successful attack. 

This shows the limitations on the attacker to spoof the location of any entity to any arbitrary location. Also, in order to spoof or fail the verification of any entity, attacker needs to know or preemptively guess the location of the other honest neighboring entities with whom the verification will be done to decide what wrong location it should broadcast and how much distance manipulation it should perform to satisfy the necessary conditions for accepting their wrong location to be a valid one. Also, even if the attacker can figure out the location and the distance manipulation necessary by guessing the neighboring entities(possible in case of less traffic situation) but due to the security guarantee of the LER protocol the attacker might not be able to perform the distance manipulation required. Therefore, in the dynamic setting of vehicular network with continuous secure location verification protocol in place, the chance of success of the attacker is limited. Also, since our protocol prevents terrorist fraud attack the attackers can not collude among each other to fakely increase their presence in the network. Therefore, the number of malicious nodes in the network that will enable failed verification of legitimate GNSS location will be very less.

%% file: evaluation.tex
\section{Evaluation of the Recovery Protocol}
\label{sec:recoveval}

% \subsection{Evaluation of the Recovery Protocol}
% \label{sec:recoveval}
In this section, we evaluate the performance of the recovery protocol described in Section~\ref{sec:recproc}.
The design of the recovery protocol ensures that the expected number of verification steps until GPS spoofing is detected does not exceed the value $E$ selected by the user.
In other words, the guarantee that the recovery protocol satisfies in the case that the GPS is spoofed is fully specified by the user.
Therefore, the natural way to evaluate the recovery protocol is to consider its performance in the case that the GPS is not spoofed, which is precisely characterized by the relative number of true negatives, i.e., the fraction of times the recovery protocol is not fooled into mistakenly going into recovery mode by malicious vehicles.

\begin{figure}[t]
  \centering
  \includegraphics[width=0.3\textwidth]{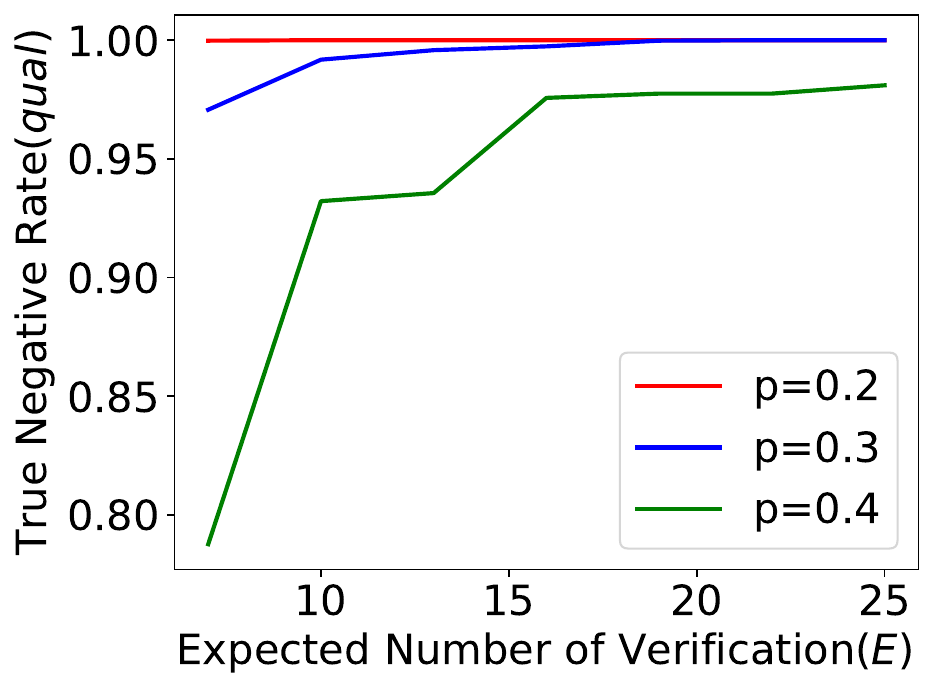}
  \caption{True negative rate achieved for different settings of $p$ and $E$. $p$ remains same for evaluation \& in the meta-protocol.}
  \label{fig:select_qual}
\vspace{-20pt}
\end{figure}

In Figure~\ref{fig:select_qual}, we see the true negative rate of the recovery protocol plotted against the expected number of verifications (specified by the user) until GNSS spoofing is detected.
The figure shows the respective curves for three different scenarios ($p = 0.2, 0.3, 0.4$) depending on the assumed probability $p$ of a vehicle being malicious (also specified by the user).
%In either case, the window size $T$ in the recovery protocol is set to $T = (E - 1)/0.3$, which suffices to obtain the displayed curves.
We see that even for the highly unrealistic scenario of each vehicle being malicious with probability $p = 0.4$, we reach a fairly large true negative rate by setting the expected number of verifications until GNSS spoofing is detected to a small $2$-digit number.
In the more realistic setting of $p = 0.2$, the true negative rate is very close to $1$ (meaning that the recovery protocol is almost never mistakenly going into recovery mode) already for very small values of $E$, %i.e., if we require GNSS spoofing to be detected after only a handful of verifications on average.
In conclusion, we get almost optimal results already for the scenario of $p = 0.2$; for smaller $p$ the true negative rate comes even closer to $1$.

\begin{figure}[t]
     \centering
      \includegraphics[width=0.3\textwidth]{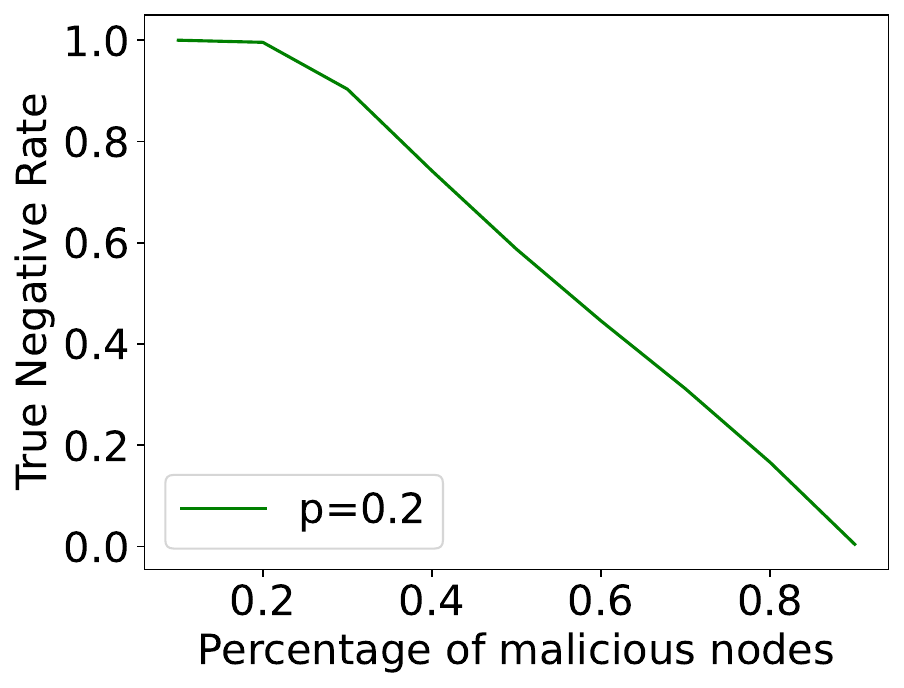}
      \caption{True negative rate for different settings of $p$ when evaluated for the meta-protocol parameter setting($p=0.2,E=7,T=20$)
      %\snote{something is missing here. Also ``perecentage'' below the picture should be ``percentage''}}
}
       \label{fig:trueneg_all}
       \vspace{-20pt}
    \end{figure}

In Figure \ref{fig:trueneg_all}, we exemplify the performance of the recovery protocol for the scenarios when we optimized our weight functions with respect to a given percentage $p = 0.2$ of malicious vehicles (with the expected value for detecting GPS spoofing set to $E = 7$).
More precisely, the plotted curve describes the fraction of true negatives as a function of the percentage of malicious vehicles.
In particular, we consider the recovery protocol obtained from setting $p$ to $0.2$ (in the optimization performed by the meta-protocol) also in scenarios where the percentage of malicious vehicles is not $0.2$.
Indeed, in the overwhelming majority of realistic scenarios, the percentage of malicious vehicles can be expected to be much less than $0.2$.
Our results show that even when the recovery protocol is fine-tuned to the value $p = 0.2$, its (already quite good) performance (in terms of true negatives) improves further when the actual percentage of malicious nodes is decreased (e.g., from $0.2$ to $0.1$), i.e., if we go from a rare worst-case scenario to a much more common scenario.  The Figure~\ref{fig:trueneg_all} showcases the true negative rates and it is close to $1$ when evaluated with $p \leq 0.2$.
For values of $p$ larger than $0.2$ the true negative rate decreases quickly; however, this is essentially irrelevant in practice as such large percentages of malicious nodes are virtually impossible in the overwhelming majority of use cases.

\begin{figure*}
    \centering
    \begin{subfigure}[b]{0.32\textwidth}
        \includegraphics[width=\linewidth]{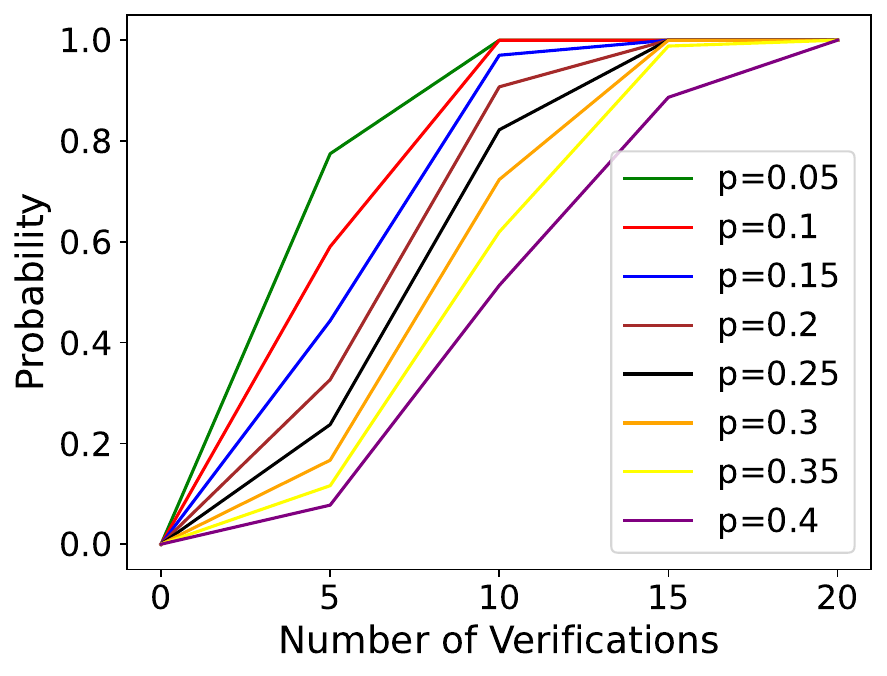}
        %\vspace{5pt}
        \caption{$p=0.2$, $E=7$ and $T=20$}
        \label{fig:detection_0.2}
    \end{subfigure}
    ~ %add desired spacing between images, e. g. ~, \quad, \qquad, \hfill etc. 
    \begin{subfigure}[b]{0.32\textwidth}
        \includegraphics[width=\linewidth]{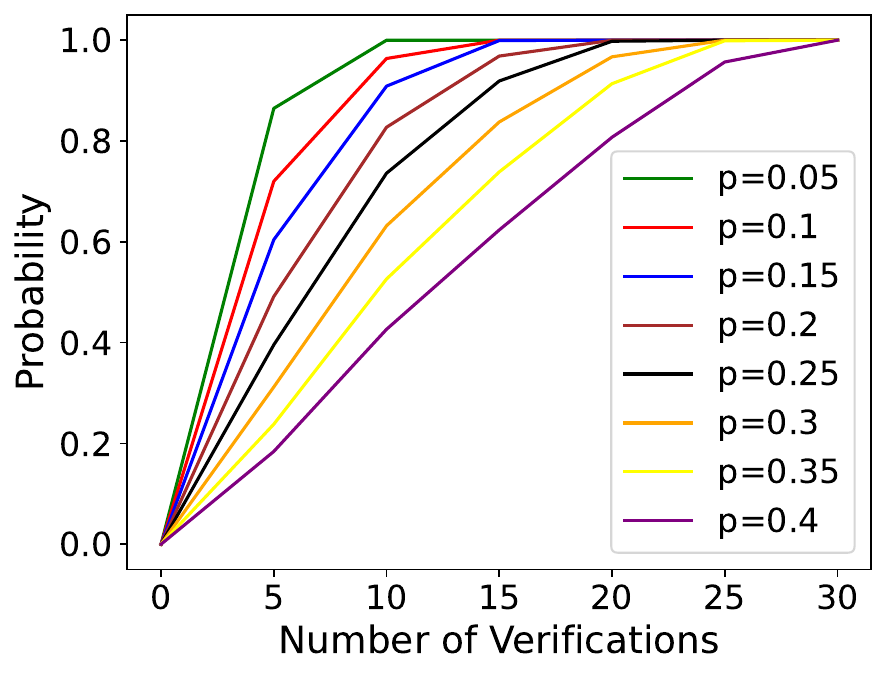}
        \caption{$p=0.3$, $E=10$ and $T=30$}
        \label{fig:detection_0.3}
    \end{subfigure}
    ~ %add desired spacing between images, e. g. ~, \quad, \qquad, \hfill etc. 
    \begin{subfigure}[b]{0.32\textwidth}
        \includegraphics[width=\linewidth]{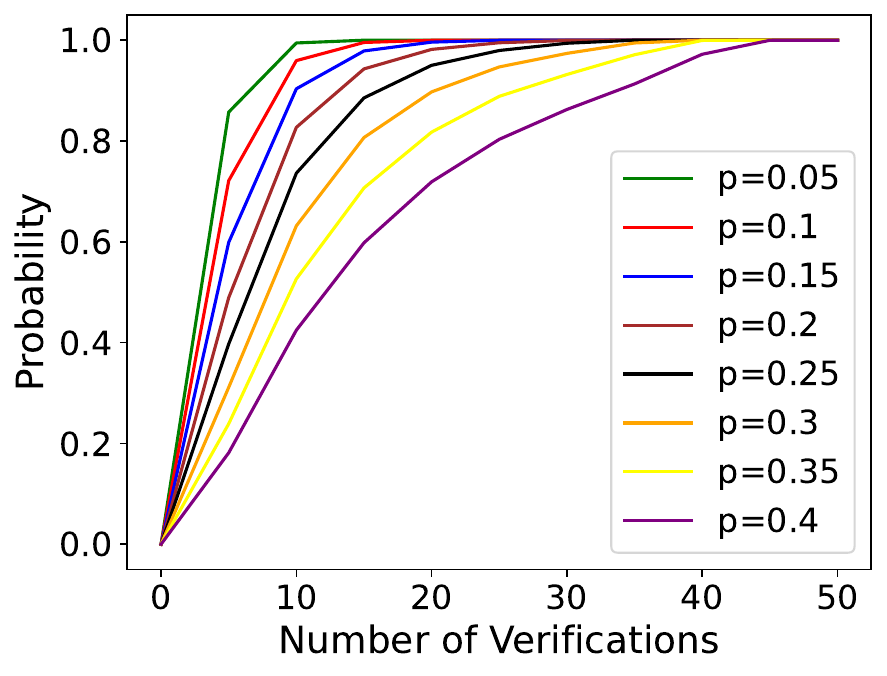}
        \caption{$p=0.4$, $E=16$ and $T=50$}
        \label{fig:detection_0.4}
    \end{subfigure}
    \caption{Probability of GPS spoofing detection with respect to the number of verification steps required for different $p$ settings. The $p$-value given in each figure caption indicates the assumed probability \emph{during the execution of the meta-protocol}. The $p$-values given in each picture indicate the assumed probability \emph{in the actual setting in which the recovery protocol is applied}.}\label{fig:detection_all}
    \vspace{-8pt}
\end{figure*}

While, as explained before, the performance of the recovery protocol is customizable in the case of a spoofed GPS, it is only customizable in terms of the expected value $E$, $p$ and $T$.
While an expected value is a perfect indicator for the average number of verification steps until GPS spoofing is detected, sometimes users prefer to have concrete (and small) bounds on the \emph{probability} that the number of steps until spoofing is detected exceeds some desired value.
In Figures~\ref{fig:detection_0.2},~\ref{fig:detection_0.3} and ~\ref{fig:detection_0.4}, we plot for each number of verification steps
%(after the point in time when the spoofed GPS starts providing false coordinates
the probability that after this many steps the spoofing has been detected.
We provide the evaluation results for the settings given by percentages $p = 0.05, 0.1, 0.15, 0.2, 0.25, 0.3, 0.35, 0.4$ of malicious vehicles. To evaluate the optimized weight and threshold pair received from the meta-protocol fine-tuned for different sets of parameter $p = 0.2, 0.3$, and $0.4$ and  $T=20, 30$ and $50$ respectively.
As we can see, even in (fairly unrealistic) scenarios characterized by $p$-values that exceed even $0.2$, the probability that we have detected GPS spoofing after a reasonably small number of verification steps, e.g., $20$ steps, is fairly close to $1$.
In the much more common scenario of $p$-values (far) below $0.2$, we detect the GPS spoofing with much smaller number of verifications.

% \begin{figure}[hbtp]
 %     \includegraphics[width=0.9\linewidth]{images/detection_0.3.pdf}
 %     \caption{Probability of GPS spoofing detection with respect to the number of verification steps required for different $p$ settings. The weight function and optimal threshold is obtained from the setting $p=0.3$, $E=10$ and $T=30$.}
  %     \label{fig:performance}
  %  \end{figure}

% \begin{figure}[hbtp]
 %     \includegraphics[width=0.9\linewidth]{images/detection_0.4.pdf}
  %    \caption{Probability of GPS spoofing detection with respect to the number of verification steps required for different $p$ settings. The weight function and optimal threshold is obtained from the setting $p=0.3$, $E=10$ and $T=30$.}
  %     \label{fig:performance}
  %  \end{figure}

%\subsection{Results}

%1. For any k number of malicious nodes in the verifier group, the protocol will always recover the correct location if the total number of verifiers $n$ $\geq$ $2k+4$.\\

%2. Limitation on the attacker choice.

%3. Probability of successful attack. 

 %  $\sum_{A=1}^{A=M}{\frac{\binom{M}{A}*\binom{N-M}{K-A}}{\binom{N}{K}}*\frac{1}{\binom{N-M}{K-A}}}$
%   $\sum_{A=1}^{A=M}{\frac{\binom{M}{A}}{\binom{N}{K}}}$

%4. Privacy

%5. Security \& Performance trade-off

\subsection{Interpreting the Results using Real-World Traffic Simulation}

%\snote{Add some introductory sentence to give a smooth introduction to this section.}

The proposed protocol is evaluated using Simulation of Urban MObility (SUMO), a traffic simulation package where we simulated the network of Monaco city\footnote{https://github.com/lcodeca/MoSTScenario} and Berlin\footnote{https://github.com/mosaic-addons/best-scenario} with the mentioned configuration setting. The simulation results in this section is to showcase how the the numbers provided by the recovery protocol can be interpreted in real-world traffic. 

\begin{figure}[t]
  \centering
  \includegraphics[width=0.75\linewidth]{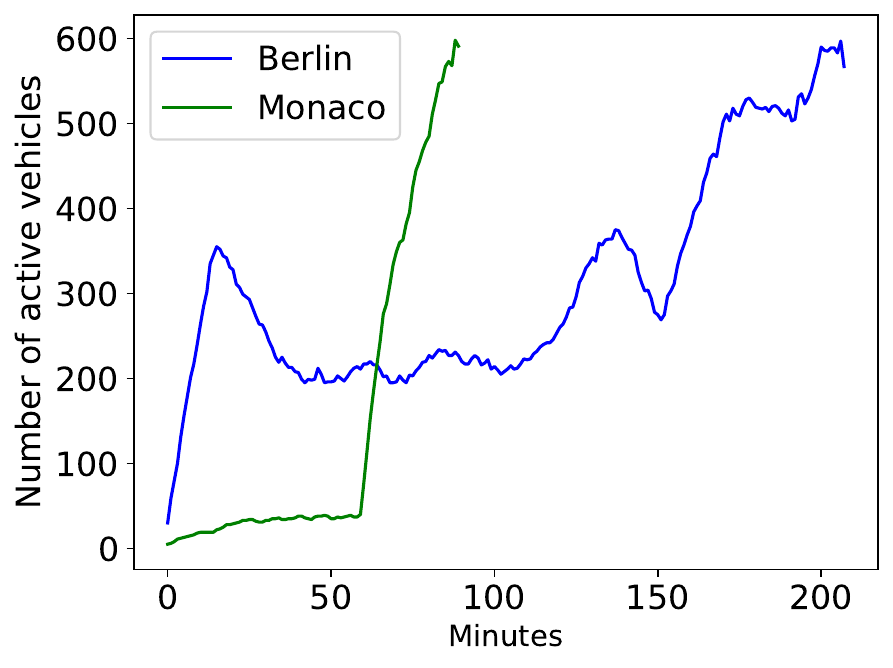}
  \vspace{-10pt}
  \caption{Number of active vehicles in the network in consecutive time periods of $1$ minute}
  \label{fig:sumo_active}
  \vspace{-10pt}
\end{figure}

The SuMo simulation for Monaco City is run for a total of $96$ minutes and $207$ minutes for Berlin. Figure \ref{fig:sumo_active} shows us how many number of active vehicles are present in the network at an interval of $1$ minute. It also explains how the traffic situation is dynamic in nature and what are the peak traffic time and what is the low or moderate traffic time. The Figure \ref{fig:sumo_active} for Monaco shows that initially the number of vehicles inserted in the simulation is less(like early morning/late night traffic in real scenario) and gradually it increases with passing time.
The Figure \ref{fig:sumo_active} for Berlin demonstrates a completely different traffic pattern from Monaco and thus our protocol is discussed with varying traffic conditions. The simulation data consists of only the neighboring vehicles and not any base stations or RSU information. 

%\begin{figure}[t]
 % \centering
 % \includegraphics[width=0.9\linewidth]{images/active_vehicles_monaco.pdf}
 % \caption{Number of active vehicles in the network in consecutive time periods of 1 minute in Monaco Traffic}
 % \label{fig:sumo_monaco}
%\end{figure}

\begin{figure}[t]
  \centering
  \includegraphics[width=0.75\linewidth]{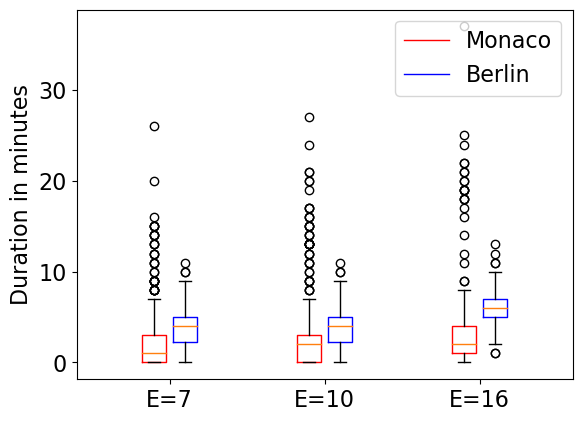}
  \caption{Duration of meeting the expected number of vehicles in different setting when communication range is $100 m$}
  \label{fig:sumo_E_100}
  \vspace{-10pt}
\end{figure}

In order to utilize the mathematical formulation for GPS detection and recovery, a certain number of vehicles are required. Figure \ref{fig:sumo_E_100} shows the number of unique neighbors the vehicles are encountering with respect to time. The higher the window size $T$, the more unique vehicles the entity needs to encounter GNSS spoofing, in turn, higher will be the duration.
%\begin{figure}[t]
 % \centering
 % \includegraphics[width=0.9\linewidth]{images/unique_neighbors.pdf}
 % \caption{Number of unique neighbors on the path of vehicles covering different route lengths %\msnote{Mention diversity of traffic point here..If traffic is scarce, then resources at the RSUs and BSs are available to better support entity for ranging purpose. Otherwise, dynamic and dense traffic is anyways better for the LER protocol...}}
 % }
 % \label{fig:unique}
%\end{figure}

Figure \ref{fig:sumo_E_100} demonstrates the time encountered by different entities to meet the required number of vehicles($E$) to detect GNSS spoofing. Reporting the result of the setting when the percentage of malicious nodes($p=0.2$) and the communication range is $100 m$, the average duration in Monaco is $2.62$ minutes and for Berlin is $4.05$ minutes. The simulation results are given in Appendix~\ref{fig:sumo_T_100} for the duration with respect to $T$.

%Since the number of unique vehicles is $20$ derived from the meta protocol \ref{fig:performance} \snote{please rephrase this sentence, it's hard to understand} to detect the GPS spoofing the average time it required to meet $20$ unique vehicles so that we can guarantee GPS spoofing detection with probability close to 1 is $4$ minutes. When the communication range is increased to $500m$ the detection time is within $2$ minutes.

%is $T_k=114.551$ seconds  which is roughly $2$ minutes when communication range is $100m$. When the communication range is increased to $500m$ $T_k$ is within 1 minutes. 

%$T_k=114.857142857$ seconds for detection \snote{What's $T_k$? And can this sentence be expanded and connected to the rest?}

%In an exemplary scenario in our simulation, we have fixed the range of communication to be 100m for a vehicle with id mentioned as bus\_1:Monaco.2. Within a given time frame of 60sec we found out how many vehicles are available within its close proximity and number of verification it can perform. Since, the vehicle is moving we have increased the range to see how much distance it needs to travel from its current location such that the verification is successful. We have plotted a graph to show how the number of unique neighbors increases with increase in the distance. 

From the results of the SuMO simulation we can derive the following insights:
\begin{itemize}[noitemsep]
    \item The real-world traffic is dynamic in nature and thus it will be difficult for the attackers to manipulate $20\%$ of the vehicles considering we limit the chance of attacker's success by our secure distance bounding protocol. We have considered, $p=0.2$ as a baseline scenario which gives us strong security guarantees and thus, the time estimated in GNSS spoofing detection can be much less in real-world scenario. The unique neighbors encountered does not include the base stations and RSUs, therefore percentage of malicious nodes can also be reduced.
    \item Based on the traffic scenario and the availability of resources(base-station,RSU), the user can choose its corresponding set of parameters mainly $p$. Depending on the $p$ and $T$, $E$ will vary. The more the percentage of malicious nodes assumed in the network the higher will be the window size($T$) and longer will be the duration to detect the spoofing attack.
\end{itemize}

%% file: discussion.tex
\vspace{-10pt}
\section{Discussion and Future work}
\label{sec:discussion}
% \textbf{Recovery from spoofing.} 

% \textbf{Density of vehicles.}

\noindent\textbf{Device Architecture to prevent Distance Fraud.}
Our design takes a step towards using TE to provide security against Terrorist Fraud. However, if the transceiver is malicious, or there does not exist a secure path between the transceiver and TE, then a malicious user can perform distance fraud. Future work should explore the architecture that can overcome this shortcoming of our design by using a secure architecture. If such an architecture can be designed, then the attack would be reduced to distance enlargement by distance fraud, where an attacker can shield the antenna and relay it later, further reducing the effectiveness of the attack. 

% \vspace{10pt}
% \noindent\textbf{Key Validity.} In our protocol, we consider that the $ID,key$ pair is valid for $T_k$ duration in order to reduce overhead and increase the privacy of the active nodes. However, it allows for the replay attacks within $T_k$ duration, affecting passive entities. Better key exchange algorithms should be explored for these use cases. 

\noindent\textbf{Deployment of LER Design.}
This protocol can use the transceivers that road users use for communicating with the RSUs and BSs. The protocols need to be run in a secure computing environment. For ride-hailing services, it can be part of the applications. It can be executed inside the ankle bracelet of the offender. For privately owned vehicles, it can be part of the usage-based insurance dongles. In many of these scenarios, the user may not even have root access to the device that performs computation, further reducing the chances of distance fraud.

\noindent\textbf{Implementation of TE in LER Design.} In this paper we have proposed the use of TE as a concept for secure distance bounding that provides the root-of-trust in the proposed protocol.
The implementation and the performance analysis is still an area to be explored in the context of positioning systems.

%% file: appendix.tex
\section{Appendix}

\RestyleAlgo{ruled}

\begin{algorithm}
\scriptsize
\caption{Weight Optimization}\label{alg:meta}

\textbf{Input:}{Weight function$(W(i))$)}\\

\textbf{Output:}{($W(i),t_{\opt},qual$)}

initialization;\\
 
 $N=$number of times the experiment is run\\
 $T$=window size;\\
 $E=$ Expected number of verification for GPS detection;\\

  $t=$Optimal Threshold($W(i)$)\\

  $max=2 \cdot t$ \\

  \For{$index\gets0$ \KwTo $T$}{
  $value1,value2=$Function1($W,0,max$)\\
  $lower\_bound_{index}=value1$\\
  $higher\_bound_{index}=value2$\\

  }

\end{algorithm}

\RestyleAlgo{ruled}

\begin{algorithm}
\scriptsize
\caption{Optimal Threshold($t_{\opt}$)}\label{alg:one}

\textbf{Input:}{$p=$percentage of malicious nodes in the network\\
Weight function$(W(i))$\\
}

\textbf{Output:}{Optimal Threshold($t_{\opt}$)}

%\KwResult{Write here the expected result}
 initialization;\\
 $learning\_rate=1$\\
 $flag1=0$;\\
 $flag2=0$;
 
 \While{True}{
   X=[];\\
  \If{$flag==1$ \&\& $flag2==1$}{$learning\_rate=learning\_rate/2$;}
  \For{$i\gets0$ \KwTo $N$}{
    r=[];\\
    $flag=0$; \\
    \For{$j\gets0$ \KwTo $T$}{
    r.append(flip($1-p$));\\
    $l=0$;\\
    $s=0$;\\
    \For{$k\gets len(r)-1$ \KwTo $0$}{
    $s=s+w[l]*r[k]$; \\
    $l=l+1$ \\
    \If{$s>t$ \&\& $flag==0$}{
    X.append(len(r)); \\
    $flag=1$;
    }
    }
    }
    }
    $Y=\sum_{i=1}^{N}(X)$;\\
    $L=Y/N$;\\
  \If{$L>E$}{$t=t-learning\_rate$;}
  \ElseIf{$L<E$}{
   $t=t+learning\_rate$;
   
   }
   \Else{$t_{\opt}=t$; \\
   \Return $(t_{\opt})$}
}
\end{algorithm}

\RestyleAlgo{ruled}

\begin{algorithm}
\scriptsize
\caption{Qual Calculation}\label{alg:qual}

\textbf{Input:}{Weight function$(W(i))$, Optimal threshold($t_{\opt}$)}\\

\textbf{Output:}{True Negative Rate($qual$)}

   $count=0$;\\
   \For{$i\gets0$ \KwTo $N$}{
   $s=0$;\\
   r=[];\\
   \For{$j\gets0$ \KwTo $T$}{
     r.append(flip($p$));\\}
    \For{$j\gets0$ \KwTo $T$}{
    $s=s+r[k]*w_0[k]$\\
    }
    \If{$s<t_{opt}$}{$count=count+1$};\\
   
   }
   $qual=count/N$;\\

   \Return($qual$);

\end{algorithm}

\RestyleAlgo{ruled}

\begin{algorithm}
\scriptsize
\caption{Function1}\label{alg:case1}

\textbf{Input:}{Weight function$(W(i))$, lower index($low$), higher index($high$})\\

\textbf{Output:}{($value1$,$value2$)}

  $mid=(low+high)/2$\\

  $t_{low}=$ Optimal Threshold($W_{index}[low]$)\\
  $qual_{low}=$ Optimal Threshold($W_{index}[low]$,$t_{low}$)\\

  $t_{mid}=$ Optimal Threshold($W_{index}[mid]$)\\
  $qual_{mid}=$ Optimal Threshold($W_{index}[mid]$,$t_{mid}$)\\

  $t_{high}=$ Optimal Threshold($W_{index}[high]$)\\
  $qual_{high}=$ Optimal Threshold($W_{index}[high]$,$t_{high}$)\\

  \If{$qual_{mid}<qual_{low}$}{
  $value1,value2=$ Function1($W(i)$,$low$,$mid$)
  }
  \ElseIf{$qual_{mid}<qual_{high}$}{
  $value1,value2=$ Function1($W(i)$,$mid$,$high$)
  }
  \Else{
  $value1,value2=$Function2($W(i)$,$low$,$mid$,$mid$,$high$)
  }

   \Return($value1$,$value2$);

\end{algorithm}

\RestyleAlgo{ruled}

\begin{algorithm}
\scriptsize
\caption{Function2}\label{alg:case2}

\textbf{Input:}{Weight function$(W(i))$, $low$, $mid1$, $mid2$, $high$}\\

\textbf{Output:}{($value1$,$value2$)}

  \If{$mid1-low<= 0.00001$}{$z=1$}
  \If{$high-mid2<=0.00001$ \&\& $mid1-low<=0.00001$}{

   \If{$qual_{low}<qual_{mid1}$}{$value1=mid1$}
   \Else{$value1=low$}

   \If{$qual_{high}<qual_{mid2}$}{$value2=mid2$}
   \Else{$value2=high$}

   \Return($value1$,$value2$)}

  \Else{
  
  \If{$z==0$}{

  $mid_{new}=(low+mid1)/2$\\

  $t_{mid1}=$ Optimal Threshold($W_{index}[mid1]$)\\
  $qual_{mid1}=$ Optimal Threshold($W_{index}[mid1]$,$t_{mid1}$)\\

  $t_{mid_{new}}=$ Optimal Threshold($W_{index}[mid_{new}]$)\\
  $qual_{mid_{new}}=$ Optimal Threshold($W_{index}[mid_{new}]$,$t_{mid_{new}}$)\\

   \If{$qual_{mid_{new}}<qual_{mid}$}{
   $value1$,$value2=$Function2($W(i),mid_{new},mid1,mid2,high$)
   \Return($value1$,$value2$);
   }
   \ElseIf{$qual_{mid_{new}}>qual_{mid}$}{
   $value1$,$value2=$Function2($W(i),low,mid_{new},mid_{new},mid1$)
   \Return($value1$,$value2$);
   }
   \Else{
   $value1$,$value2=$Function2($W(i),low,mid_{new},mid2,high$)
   \Return($value1$,$value2$);
   }
   }
  
   \Else{

   $mid_{new}=(mid2+high)/2$\\

  $t_{mid2}=$ Optimal Threshold($W_{index}[mid2]$)\\
  $qual_{mid2}=$ Optimal Threshold($W_{index}[mid2]$,$t_{mid2}$)\\

  $t_{mid_{new}}=$ Optimal Threshold($W_{index}[mid_{new}]$)\\
  $qual_{mid_{new}}=$ Optimal Threshold($W_{index}[mid_{new}]$,$t_{mid_{new}}$)\\

   \If{$qual_{mid_{new}}<qual_{mid2}$}{
   $value1$,$value2=$Function2($W(i),low,mid1,mid2,mid_{new}$)
   \Return($value1$,$value2$);
   }
   \ElseIf{$qual_{mid_{new}}>qual_{mid}$}{
   $value1$,$value2=$Function2($W(i),mid2,mid_{new},mid_{new},high$)
   \Return($value1$,$value2$);
   }
   \Else{
   $value1$,$value2=$Function2($W(i),low,mid1,mid_{new},high$)
   \Return($value1$,$value2$);
   }
   
   }
  }

\end{algorithm}

\begin{figure}[t]
  \centering
  \includegraphics[width=0.8\linewidth]{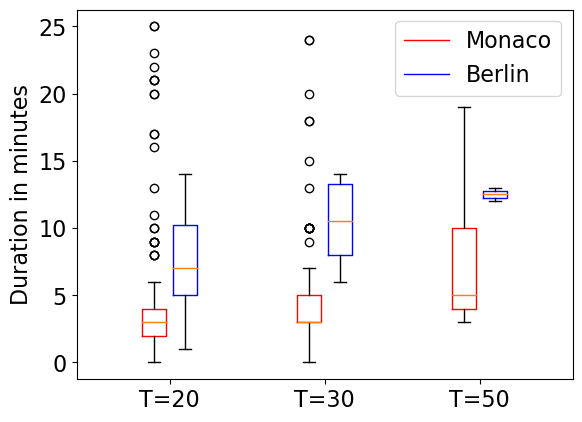}
  \caption{Duration of meeting $T$  number of vehicles in different setting when communication range is $100m$}
  \label{fig:sumo_T_100}
\end{figure}

% \begin{figure}[t]
%   \centering
%   \includegraphics[width=0.8\linewidth]{images/sumo_E_500.jpeg}
%   \caption{Duration of meeting the expected number of vehicles in different setting when communication range is $500m$}
%   \label{fig:sumo_E_500}
% \end{figure}

% \begin{figure}[t]
%   \centering
%   \includegraphics[width=0.8\linewidth]{images/sumo_T_500.jpeg}
%   \caption{Duration of meeting $T$ number of vehicles in different setting when communication range is $500m$}
%   \label{fig:sumo_T_500}
% \end{figure}

%% file: main.bbl
\begin{thebibliography}{10}

\bibitem{3GPP_38.855}
{5G - GPP 38.855 ;Technical Specification Group Radio Access Network; Study on NR positioning support}.
\newblock \url{https://www.3gpp.org/ftp/Specs/archive/38_series/38.855/}.
\newblock [Online; Accessed 1. November 2022].

\bibitem{testbed1}
{5GCAR}.
\newblock \url{https://www.ip45g.de/en/testbeds/5gcar/}.
\newblock [Online; Accessed 12. April 2023].

\bibitem{5G_positioning_Chalmers}
"beyond 5g positioning".
\newblock \url{https://research.chalmers.se/en/project/10832}.
\newblock [Online; Accessed October 10th 2023].

\bibitem{CarRentalSpoofing}
{Car Rental GPS Spoofing}.
\newblock \url{https://www.carrentalgateway.com/glossary/gps-spoofing/}.
\newblock [Online; Accessed 12. April 2023].

\bibitem{chimera}
{CHIMERA GPS}.
\newblock \url{https://www.gpsexpert.net/chimera-specification}.
\newblock [Online; Accessed 12. April 2023].

\bibitem{Parole}
{Crime Prevention System for Public Safety}.
\newblock \url{ https://www.unodc.org/documents/justice-and-prison-reform/ReducingReoffending/MS/Republic_of_Korea_Annex_-_.pdf}.
\newblock [Online; Accessed 13. April 2023].

\bibitem{tesla_gallilio}
{Galileo Open Service Navigation Message Authentication}.
\newblock \url{https://gssc.esa.int/navipedia/index.php/Galileo_Open_Service_Navigation_Message_Authentication/}.
\newblock [Online; Accessed 12. April 2023].

\bibitem{fleet_tracking_2}
{Geotab fleet tracking}.
\newblock \url{https://www.geotab.com}.
\newblock [Online; Accessed 13. April 2023].

\bibitem{GNSS_spoofing_road}
"gps spoofing attack sends 38 drivers the wrong way – and into possible danger".
\newblock \url{https://shorturl.at/lqt45}.
\newblock [Online; Accessed October 10th 2023].

\bibitem{GPS_spoofing_HackRF}
{How to Spoof GPS Location with HackRF}.
\newblock \url{https://drfone.wondershare.com/fake-location/gps-spoofing-with-hackrf-from-windows.html}.
\newblock [Online; Accessed 18. August 2023].

\bibitem{osnma}
{OSNMA}.
\newblock \url{https://shorturl.at/hmuIQ}.
\newblock [Online; Accessed 12. April 2023].

\bibitem{HybridPositioning_ESA}
{Proof of Concept of Hybrid 5G-NR/GNSS Positioning with AD-HOC Overlay}.
\newblock \url{https://navisp.esa.int/opportunity/details/72/show}.
\newblock [Online; Accessed 05. July 2021].

\bibitem{publicTransportGPS}
{Public Transportation: Bus \& Mass Transit GPS}.
\newblock \url{https://gpstrackit.com/solutions/public-transportation/}.
\newblock [Online; Accessed 12. April 2023].

\bibitem{3gpp_seal}
{Service Enabler Architecture Layer for Verticals (SEAL); Functional architecture and information flows (3GPP TS 23.434 version 16.4.0 Release 16)}.
\newblock \url{https://www.etsi.org/deliver/etsi_ts/123400_123499/123434/16.04.00_60/ts_123434v160400p.pdf}.
\newblock [Online; Accessed 18. April 2021].

\bibitem{gps_spoof_1}
{Spoofing GPS With an SDR}.
\newblock \url{https://kaitlyn.guru/projects/spoofing-gps-with-an-sdr/}.
\newblock [Online; Accessed 12. April 2023].

\bibitem{5AA_V2X_positioning}
{System Architecture and Solution Development; High-Accuracy Positioning for C-V2X}.
\newblock \url{https://5gaa.org/content/uploads/2021/02/5GAA_A-200118_TR_V2XHAP.pdf}.
\newblock [Online; Accessed 12. April 2023].

\bibitem{tesla_hack}
{Tesla Hack}.
\newblock \url{https://www.gpsworld.com/two-years-since-the-tesla-gps-hack}.
\newblock [Online; Accessed 13. April 2023].

\bibitem{tee}
{What is a Trusted Execution Environment (TEE)?}
\newblock \url{https://www.trustonic.com/technical-articles/what-is-a-trusted-execution-environment-tee/}.
\newblock [Online; Accessed 18. April 2021].

\bibitem{3GPP}
{3GPP}.
\newblock \url{https://www.3gpp.org/ftp/Specs/archive/38_series/38.211/}.
\newblock [Online; Accessed 1. November 2022].

\bibitem{Akos2012WhosAO}
Dennis~M. Akos.
\newblock Who's afraid of the spoofer? gps/gnss spoofing detection via automatic gain control (agc).
\newblock {\em Annual of Navigation}, 59:281--290, 2012.

\bibitem{DB_EPFL}
Ioana Boureanu, Aikaterini Mitrokotsa, and Serge Vaudenay.
\newblock {Towards Secure Distance Bounding}.
\newblock Cryptology ePrint Archive, Report 2015/208, 2015.
\newblock \url{https://eprint.iacr.org/2015/208}.

\bibitem{Brands1994}
Stefan Brands and David Chaum.
\newblock Distance-bounding protocols.
\newblock In {\em EUROCRYPT}, pages 344--359. Springer, 1994.

\bibitem{DB_Multi}
Agn{\`e}s Brelurut, David Gerault, and Pascal Lafourcade.
\newblock {Survey of Distance Bounding Protocols and Threats}.
\newblock In {\em {Foundations and Practice of Security (FPS)}}, pages 29 -- 49, 2015.

\bibitem{GDB_Capkun}
{\v{C}apkun, Srdjan and El Defrawy, Karim and Tsudik, Gene}.
\newblock {\em {Group Distance Bounding Protocols}}, pages 302--312.
\newblock Springer, 2011.

\bibitem{DistanceHijacking}
Cas Cremers, Kasper~B. Rasmussen, Benedikt Schmidt, and Srdjan Capkun.
\newblock Distance hijacking attacks on distance bounding protocols.
\newblock In {\em 2012 IEEE Symposium on Security and Privacy}, pages 113--127, 2012.

\bibitem{DolevYao}
D.~Dolev and A.~Yao.
\newblock {On the security of public key protocols}.
\newblock {\em IEEE Transactions on Information Theory}, 29(2):198--208, 1983.

\bibitem{DB_10}
Saar Drimer and Steven~J. Murdoch.
\newblock Keep your enemies close: Distance bounding against smartcard relay attacks.
\newblock In {\em 16th {USENIX} Security Symposium ({USENIX} Security 07)}, Boston, MA, August 2007. {USENIX} Association.

\bibitem{ESA_5G_GPS}
{Hybrid 5G and GPS}.
\newblock \url{ https://www.esa.int/Applications/Navigation/ESA_leads_drive_into_our_5G_positioning_future}.
\newblock [Online; Accessed June 16, 2020].

\bibitem{LRP_HRP_comparision}
Laura Flueratoru, Silvan Wehrli, Michele Magno, and Dragos Niculescu.
\newblock On the energy consumption and ranging accuracy of ultra-wideband physical interfaces.
\newblock In {\em GLOBECOM 2020 - 2020 IEEE Global Communications Conference}, pages 1--7, 2020.

\bibitem{Flury_reductionattack}
Manuel Flury, Marcin Poturalski, Panos Papadimitratos, Jean-Pierre Hubaux, and Jean-Yves Le~Boudec.
\newblock Effectiveness of distance-decreasing attacks against impulse radio ranging.
\newblock In {\em Proceedings of the Third ACM Conference on Wireless Network Security}, WiSec '10, 2010.

\bibitem{3gpp_rel16}
Mehdi Harounabadi, Dariush~Mohammad Soleymani, Shubhangi Bhadauria, Martin Leyh, and Elke Roth-Mandutz.
\newblock V2x in 3gpp standardization: Nr sidelink in release-16 and beyond.
\newblock {\em IEEE Communications Standards Magazine}, 5(1):12--21, 2021.

\bibitem{DB_Ariadne_forRouting}
Yih-chun Hu, Adrian Perrig, and David Johnson.
\newblock Ariadne: A secure on-demand routing protocol for ad hoc networks.
\newblock {\em Wireless Networks}, 11, 10 2002.

\bibitem{Humphreys_GPS_sec}
Todd~E. Humphreys.
\newblock Detection strategy for cryptographic gnss anti-spoofing.
\newblock {\em IEEE Transactions on Aerospace and Electronic Systems}, 49(2), 2013.

\bibitem{Kuhn_asymmetric_crypto}
Markus Kuhn.
\newblock An asymmetric security mechanism for navigation signals.
\newblock volume 3200, 07 2004.

\bibitem{MTAC}
P.~{Leu}, M.~{Singh}, M.~{Roeschlin}, K.~G. {Paterson}, and S.~{{\v C}apkun}.
\newblock Message time of arrival codes: A fundamental primitive for secure distance measurement.
\newblock In {\em 2020 IEEE Symposium on Security and Privacy (SP)}, pages 500--516, 2020.

\bibitem{Patrick_HRP}
Patrick Leu, Giovanni Camurati, Alexander Heinrich, Marc Roeschlin, Claudio Anliker, Matthias Hollick, Srdjan Capkun, and Jiska Classen.
\newblock Ghost peak: Practical distance reduction attacks against {HRP} {UWB} ranging.
\newblock In {\em 31st {USENIX} Security Symposium, {USENIX} Security 2022, Boston, MA, USA, August 10-12, 2022}, 2022.

\bibitem{Patrick_ACSAC}
Patrick Leu, Martin Kotuliak, Marc Roeschlin, and Srdjan Capkun.
\newblock Security of multicarrier time-of-flight ranging.
\newblock In {\em Annual Computer Security Applications Conference}, ACSAC '21, page 887–899, New York, NY, USA, 2021. Association for Computing Machinery.

\bibitem{maryam_GNSS}
Maryam Motallebighomi, Harshad Sathaye, Mridula Singh, and Aanjhan Ranganathan.
\newblock Cryptography is not enough: Relay attacks on authenticated gnss signals, 2022.

\bibitem{GPS_aanjhan}
Sashank Narain, Aanjhan Ranganathan, and Guevara Noubir.
\newblock Security of gps/ins based on-road location tracking systems.
\newblock In {\em 2019 IEEE Symposium on Security and Privacy (SP)}, pages 587--601, 2019.

\bibitem{fleet_tracking_1}
Yekutiel~A Novik.
\newblock {System and method for fleet tracking}.
\newblock 2000.

\bibitem{TESLA_perrig}
Adrian Perrig, Ran Canetti, J.~Doug Tygar, Dawn~Xiaodong Song, Leonid Reyzin, and Itsik Mantin.
\newblock The tesla broadcast authentication protocol.
\newblock 2002.

\bibitem{cicadaEPFL}
M~Poturalski, Mand~Flury, P~Papadimitratos, JP~Hubaux, and JY~Le~Boudec.
\newblock The cicada attack: degradation and denial of service in ir ranging.
\newblock In {\em Ultra-Wideband (ICUWB), 2010 IEEE International Conference on}, volume~2, pages 1--4. IEEE, 2010.

\bibitem{edlcEPFl}
Marcin Poturalski, Manuel Flury, Panos Papadimitratos, Jean-Pierre Hubaux, and Jean-Yves Le~Boudec.
\newblock Distance bounding with ieee 802.15.4a: Attacks and countermeasures.
\newblock {\em IEEE Transactions on Wireless Communications}, 10(4):1334--1344, 2011.

\bibitem{rabinowitz_capabilities_2000}
M.~Rabinowitz, B.~W. Parksinson, and J.~J. Spilker.
\newblock Some {Capabilities} of a {Joint} {GPS}-{LEO} {Navigation} {System}.
\newblock pages 255--265, September 2000.

\bibitem{Aanjhan_Spree}
Aanjhan Ranganathan, Hildur \'{O}lafsd\'{o}ttir, and Srdjan Capkun.
\newblock Spree: A spoofing resistant gps receiver.
\newblock In {\em Proceedings of the 22nd Annual International Conference on Mobile Computing and Networking}, MobiCom '16. ACM, 2016.

\bibitem{kasper_DB}
Kasper Rasmussen and Srdjan Capkun.
\newblock Realization of rf distance bounding.
\newblock pages 389--402, 09 2010.

\bibitem{GPS_harshad}
Harshad Sathaye, Gerald LaMountain, Pau Closas, and Aanjhan Ranganathan.
\newblock Semperfi: {A} spoofer eliminating {GPS} receiver for uavs.
\newblock {\em CoRR}, abs/2105.01860, 2021.

\bibitem{UWB-ED}
Mridula Singh, Patrick Leu, AbdelRahman Abdou, and Srdjan Capkun.
\newblock {UWB-ED}: Distance enlargement attack detection in {Ultra-Wideband}.
\newblock In {\em 28th USENIX Security Symposium (USENIX Security 19)}, pages 73--88, Santa Clara, CA, August 2019. USENIX Association.

\bibitem{vrange}
Mridula Singh, Marc Roeschlin, Aanjhan Ranganathan, and Srdjan Capkun.
\newblock V-range: Enabling secure ranging in 5g wireless networks.
\newblock In {\em NDSS 2022}, April 2022.

\bibitem{HRP_singh_wisec}
Mridula Singh, Marc Roeschlin, Ezzat Zalzala, Patrick Leu, and Srdjan \v{C}apkun.
\newblock Security analysis of ieee 802.15.4z/hrp uwb time-of-flight distance measurement.
\newblock In {\em Proceedings of the 14th ACM Conference on Security and Privacy in Wireless and Mobile Networks}. Association for Computing Machinery, 2021.

\bibitem{Nils_wisec2017}
Nils~Ole Tippenhauer, Heinrich Luecken, Marc Kuhn, and Srdjan Capkun.
\newblock Uwb rapid-bit-exchange system for distance bounding.
\newblock In {\em Proceedings of the 8th ACM Conference on Security \& Privacy in Wireless and Mobile Networks}, WiSec '15, pages 2:1--2:12. ACM, 2015.

\bibitem{Nils2011requirements}
Nils~Ole Tippenhauer, Christina P{\"o}pper, Kasper~Bonne Rasmussen, and Srdjan Capkun.
\newblock On the requirements for successful gps spoofing attacks.
\newblock In {\em Proceedings of the 18th ACM conference on Computer and communications security}, pages 75--86. ACM, 2011.

\bibitem{Valero2022}
Jos{\'e} Mar{\'i}a~Jorquera Valero, Pedro Miguel~S{\'a}nchez S{\'a}nchez, Alexios Lekidis, Pedro Martins, Pedro Diogo, Manuel~Gil P{\'e}rez, Alberto~Huertas Celdr{\'a}n, and Gregorio~Mart{\'i}nez P{\'e}rez.
\newblock {\em Trusted Execution Environment-Enabled Platform for 5G Security and Privacy Enhancement}, pages 203--223.
\newblock Springer International Publishing, Cham, 2022.

\bibitem{VerifiableMultilateration}
S.~\v{C}apkun and J.~Hubaux.
\newblock Secure positioning of wireless devices with application to sensor networks.
\newblock In {\em IEEE Computer and Communications Societies.}, volume~3, pages 1917--1928, 2005.

\end{thebibliography}


@inproceedings{TESLA_perrig,
	author = {Adrian Perrig and Ran Canetti and J. Doug Tygar and Dawn Xiaodong Song and Leonid Reyzin and Itsik Mantin},
	title = {The TESLA Broadcast Authentication Protocol},
	url = {https://api.semanticscholar.org/CorpusID:261251531},
	year = {2002},
	Bdsk-Url-1 = {https://api.semanticscholar.org/CorpusID:261251531}}


@misc{5G_positioning_Chalmers,
	howpublished = {\url{https://research.chalmers.se/en/project/10832}},
	lastvisited = {'10.10.2023},
	note = {[Online; Accessed October 10th 2023]},
	title = {"Beyond 5G Positioning"}}

@inproceedings{rabinowitz_capabilities_2000,
	author = {Rabinowitz, M. and Parksinson, B. W. and Spilker, J. J.},
	language = {en},
	month = sep,
	pages = {255--265},
	title = {Some {Capabilities} of a {Joint} {GPS}-{LEO} {Navigation} {System}},
	url = {http://www.ion.org/publications/abstract.cfm?jp=p&articleID=1411},
	urldate = {2023-09-29},
	year = {2000},
	Bdsk-Url-1 = {http://www.ion.org/publications/abstract.cfm?jp=p&articleID=1411}}


 @misc{gps_spoof_1,
	Howpublished = {\url{https://kaitlyn.guru/projects/spoofing-gps-with-an-sdr/}},
	Lastvisitsed = {'12.04.2023},
	Note = {[Online; Accessed 12. April 2023]},
	Title = {{Spoofing GPS With an SDR}}
	}


@misc{GNSS_spoofing_road,
	howpublished = {\url{https://shorturl.at/lqt45}},
	lastvisited = {'10.10.2023},
	note = {[Online; Accessed October 10th 2023]},
	title = {"GPS spoofing attack sends 38 drivers the wrong way – and into possible danger"}}

@inproceedings{Patrick_ACSAC, author = {Leu, Patrick and Kotuliak, Martin and Roeschlin, Marc and Capkun, Srdjan}, title = {Security of Multicarrier Time-of-Flight Ranging}, year = {2021}, isbn = {9781450385794}, publisher = {Association for Computing Machinery}, address = {New York, NY, USA}, url = {https://doi.org/10.1145/3485832.3485898}, doi = {10.1145/3485832.3485898}, booktitle = {Annual Computer Security Applications Conference}, pages = {887–899}, numpages = {13}, keywords = {Secure Ranging, OFDM, IEEE 802.11az}, location = {Virtual Event, USA}, series = {ACSAC '21} }


@inproceedings{GPS-free,

author={Capkun, S. and Hamdi, M. and Hubaux, J.-P.},

booktitle={Proceedings of the 34th Annual Hawaii International Conference on System Sciences}, 

title={GPS-free positioning in mobile ad-hoc networks}, 

year={2001},

volume={},

number={},

pages={10 pp.-},

doi={10.1109/HICSS.2001.927202}}

@inproceedings{secure_positioning_capkun,

author={Capkun, S. and Hubaux, J.-P.},

booktitle={Proceedings IEEE 24th Annual Joint Conference of the IEEE Computer and Communications Societies.}, 

title={Secure positioning of wireless devices with application to sensor networks}, 

year={2005},

doi={10.1109/INFCOM.2005.1498470}}



@Article{efficient_location_verification,
AUTHOR = {Kim, In-hwan and Kim, Bo-sung and Song, JooSeok},
TITLE = {An Efficient Location Verification Scheme for Static Wireless Sensor Networks},
JOURNAL = {Sensors},
VOLUME = {17},
YEAR = {2017},
NUMBER = {2},
ARTICLE-NUMBER = {225},
URL = {https://www.mdpi.com/1424-8220/17/2/225},
PubMedID = {28125007},
ISSN = {1424-8220},

DOI = {10.3390/s17020225}
}



@Article{collab_localization_verification,
AUTHOR = {Miao, Chunyu and Dai, Guoyong and Ying, Kezhen and Chen, Qingzhang},
TITLE = {Collaborative Localization and Location Verification in WSNs},
JOURNAL = {Sensors},
VOLUME = {15},
YEAR = {2015},
NUMBER = {5},
PAGES = {10631--10649},
URL = {https://www.mdpi.com/1424-8220/15/5/10631},
PubMedID = {25954948},
ISSN = {1424-8220},

DOI = {10.3390/s150510631}
}

@ARTICLE{collab_geo,

  author={Xu, Liyuan and Yao, Lin and He, Jie and Wang, Peng and Long, Keping and Wang, Qin},

  journal={IEEE Access}, 

  title={Collaborative Geolocation Based on Imprecise Initial Coordinates for Internet of Things}, 

  year={2018},

  volume={6},

  number={},

  pages={48850-48858},

  doi={10.1109/ACCESS.2018.2866957}}









  




@inproceedings{UWB-PR,
	author = {Mridula Singh and Patrick Leu and Srdjan \v{C}apkun},
	booktitle = {NDSS},
	title = {{UWB with Pulse Reordering: Securing Ranging against Relay and Physical Layer Attacks}},
	year = {2019}}

@inproceedings{vrange,
       booktitle = {NDSS 2022},
          author = {Mridula Singh and Marc Roeschlin and Aanjhan Ranganathan and Srdjan Capkun},
           month = {April},
           title = {V-Range: Enabling Secure Ranging in 5G Wireless Networks},
            year = {2022},
         journal = {NDSS},
             url = {https://publications.cispa.saarland/3568/}
}

misc{mtac,
      author = {Patrick Leu and Mridula Singh and Marc Roeschlin and Kenneth G.  Paterson and Srdjan Capkun},
      title = {Message Time of Arrival Codes: A Fundamental Primitive for Secure Distance Measurement},
      howpublished = {Cryptology ePrint Archive, Paper 2019/1350},
      year = {2019},
      note = {\url{https://eprint.iacr.org/2019/1350}},
      url = {https://eprint.iacr.org/2019/1350}
}


@inproceedings {UWB-ED,
author = {Mridula Singh and Patrick Leu and AbdelRahman Abdou and Srdjan Capkun},
title = {{UWB-ED}: Distance Enlargement Attack Detection in {Ultra-Wideband}},
booktitle = {28th USENIX Security Symposium (USENIX Security 19)},
year = {2019},
isbn = {978-1-939133-06-9},
address = {Santa Clara, CA},
pages = {73--88},
url = {https://www.usenix.org/conference/usenixsecurity19/presentation/singh},
publisher = {USENIX Association},
month = aug,
}


@INPROCEEDINGS{GPS_INS,
  author={Narain, Sashank and Ranganathan, Aanjhan and Noubir, Guevara},
  booktitle={2019 IEEE Symposium on Security and Privacy (SP)}, 
  title={Security of GPS/INS Based On-road Location Tracking Systems}, 
  year={2019},
  volume={},
  number={},
  pages={587-601},
  doi={10.1109/SP.2019.00068}}


@inproceedings {drift_with_devil,
author = {Junjie Shen and Jun Yeon Won and Zeyuan Chen and Qi Alfred Chen},
title = {Drift with Devil: Security of {Multi-Sensor} Fusion based Localization in {High-Level} Autonomous Driving under {GPS} Spoofing},
booktitle = {29th USENIX Security Symposium (USENIX Security 20)},
year = {2020},
isbn = {978-1-939133-17-5},
pages = {931--948},
url = {https://www.usenix.org/conference/usenixsecurity20/presentation/shen},
publisher = {USENIX Association},
month = aug,
}



@INPROCEEDINGS{Overview5GPositioning,
	author={Keating, Ryan and Säily, Mikko and Hulkkonen, Jari and Karjalainen, Juha},
	booktitle={2019 16th International Symposium on Wireless Communication Systems (ISWCS)}, 
	title={Overview of Positioning in 5G New Radio}, 
	year={2019},
	volume={},
	number={},
	pages={320-324},
	doi={10.1109/ISWCS.2019.8877160}}

@ARTICLE{Release18_expectation,
  author={Lin, Xingqin},
  journal={IEEE Communications Standards Magazine}, 
  title={An Overview of 5G Advanced Evolution in 3GPP Release 18}, 
  year={2022},
  volume={6},
  number={3},
  pages={77-83},
  doi={10.1109/MCOMSTD.0001.2200001}}


 @inproceedings{GPS_spoofing, author = {Zeng, Kexiong Curtis and Shu, Yuanchao and Liu, Shinan and Dou, Yanzhi and Yang, Yaling}, title = {A Practical GPS Location Spoofing Attack in Road Navigation Scenario}, year = {2017}, isbn = {9781450349079}, publisher = {Association for Computing Machinery}, address = {New York, NY, USA}, url = {https://doi.org/10.1145/3032970.3032983}, doi = {10.1145/3032970.3032983}, abstract = {High value of GPS location information and easy availability of portable GPS signal spoofing devices incentivize attackers to launch GPS spoofing attacks against location-based applications. In this paper, we propose an attack model in road navigation scenario, and develop a complete framework to analyze, simulate and evaluate the spoofing attacks under practical constraints. To launch an attack, the framework first constructs a road network, and then searches for an attack route that smoothly diverts a victim without his awareness. In extensive data-driven simulations in College Point, New York City, we managed to navigate a victim to locations 1km away from his original destination.}, booktitle = {Proceedings of the 18th International Workshop on Mobile Computing Systems and Applications}, pages = {85–90}, numpages = {6}, keywords = {Route planning, Navigation, GPS spoofing}, location = {Sonoma, CA, USA}, series = {HotMobile '17} }


 %% This BibTeX bibliography file was created using BibDesk.
%% http://bibdesk.sourceforge.net/

%% Created for Mridula  at 2021-06-04 00:22:34 +0200 


%% Saved with string encoding Unicode (UTF-8) 

@misc{5AA_V2X_positioning,
	Howpublished = {\url{https://5gaa.org/content/uploads/2021/02/5GAA_A-200118_TR_V2XHAP.pdf}},
	Lastvisited = {'12.04.2023},
	Note = {[Online; Accessed 12. April 2023]},
	Title = {{System Architecture and Solution Development; High-Accuracy Positioning for C-V2X}}
	}


 @misc{CarRentalSpoofing,
	Howpublished = {\url{https://www.carrentalgateway.com/glossary/gps-spoofing/}},
	Lastvisited = {'12.04.2023},
	Note = {[Online; Accessed 12. April 2023]},
	Title = {{Car Rental GPS Spoofing}}
	}
 @misc{osnma,
	Howpublished = {\url{https://shorturl.at/hmuIQ}},
	Lastvisited = {'12.04.2023},
	Note = {[Online; Accessed 12. April 2023]},
	Title = {{OSNMA}}
	}


 @misc{chimera,
	Howpublished = {\url{https://www.gpsexpert.net/chimera-specification}},
	Lastvisited = {'12.04.2023},
	Note = {[Online; Accessed 12. April 2023]},
	Title = {{CHIMERA GPS}}
	}

 @misc{gps_spoof_1,
	Howpublished = {\url{https://kaitlyn.guru/projects/spoofing-gps-with-an-sdr/}},
	Lastvisited = {'12.04.2023},
	Note = {[Online; Accessed 12. April 2023]},
	Title = {{Spoofing GPS With an SDR}}
	}






 @misc{publicTransportGPS,
	Howpublished = {\url{https://gpstrackit.com/solutions/public-transportation/}},
	Lastvisited = {'12.04.2023},
	Note = {[Online; Accessed 12. April 2023]},
	Title = {{Public Transportation: Bus \& Mass Transit GPS}}
	}


 @misc{illegalGPSinterferenec,
	Howpublished = {\url{https://www.cbsnews.com/newyork/news/n-j-man-in-a-jam-after-illegal-gps-device-interferes-with-newark-liberty-operations/}},
	Lastvisited = {'12.04.2023},
	Note = {[Online; Accessed 12. April 2023]},
	Title = {{Illegal GPS Device Interferes }}
	}

 @misc{GPS_hacking_road_navigation,
	Howpublished = {\url{https://www.bleepingcomputer.com/news/security/researchers-mount-successful-gps-spoofing-attack-against-road-navigation-systems/}},
	Lastvisited = {'12.04.2023},
	Note = {[Online; Accessed 12. April 2023]},
	Title = {{Successful GPS Spoofing Attack Against Road Navigation Systems}}
	}

 @misc{testbed1,
	Howpublished = {\url{https://www.ip45g.de/en/testbeds/5gcar/}},
	Lastvisited = {'12.04.2023},
	Note = {[Online; Accessed 12. April 2023]},
	Title = {{5GCAR}}
	}
 

@ARTICLE{3gpp_rel16,
  author={Harounabadi, Mehdi and Soleymani, Dariush Mohammad and Bhadauria, Shubhangi and Leyh, Martin and Roth-Mandutz, Elke},
  journal={IEEE Communications Standards Magazine}, 
  title={V2X in 3GPP Standardization: NR Sidelink in Release-16 and Beyond}, 
  year={2021},
  volume={5},
  number={1},
  pages={12-21},
  doi={10.1109/MCOMSTD.001.2000070}}


 @misc{tesla_gallilio,
	Howpublished = {\url{https://gssc.esa.int/navipedia/index.php/Galileo_Open_Service_Navigation_Message_Authentication/}},
	Lastvisited = {'12.04.2023},
	Note = {[Online; Accessed 12. April 2023]},
	Title = {{Galileo Open Service Navigation Message Authentication}}
	}

@ARTICLE{Humphreys_GPS_sec,
  author={Humphreys, Todd E.},
  journal={IEEE Transactions on Aerospace and Electronic Systems}, 
  title={Detection Strategy for Cryptographic GNSS Anti-Spoofing}, 
  year={2013},
  volume={49},
  number={2},
  doi={10.1109/TAES.2013.6494400}}


@article{Akos2012WhosAO,
  title={Who's Afraid of the Spoofer? GPS/GNSS Spoofing Detection via Automatic Gain Control (AGC)},
  author={Dennis M. Akos},
  journal={Annual of Navigation},
  year={2012},
  volume={59},
  pages={281-290}
}

@inproceedings{Kuhn_asymmetric_crypto,
author = {Kuhn, Markus},
year = {2004},
month = {07},
pages = {},
title = {An Asymmetric Security Mechanism for Navigation Signals},
volume = {3200},
isbn = {978-3-540-24207-9},
doi = {10.1007/978-3-540-30114-1_17}
}



@data{kim_cooperatviePos,
doi = {10.21227/hyfa-fa96},
url = {https://dx.doi.org/10.21227/hyfa-fa96},
author = {Kim, Hyowon and Karl Granstrom, Karl and Gao, Lin and Battistelli, Giorgio and Kim, Sunwoo and Wymmersch, Henk},
publisher = {IEEE Dataport},
title = {5G mmWave Cooperative Positioning and Mapping using Multi-Model PHD Filter and Map Fusion},
year = {2019} }


@misc{maryam_GNSS,
  doi = {10.48550/ARXIV.2204.11641},
  url = {https://arxiv.org/abs/2204.11641},  
  author = {Motallebighomi, Maryam and Sathaye, Harshad and Singh, Mridula and Ranganathan, Aanjhan},  
  keywords = {Cryptography and Security (cs.CR), FOS: Computer and information sciences, FOS: Computer and information sciences}, 
  title = {Cryptography Is Not Enough: Relay Attacks on Authenticated GNSS Signals},
  publisher = {arXiv},
   year = {2022},
  copyright = {arXiv.org perpetual, non-exclusive license}
}


@misc{NeedPrecisePositioningAutonomousDriving,
	Howpublished = {\url{https://www.wardsauto.com/industry-voices/six-ways-autonomous-driving-relying-precise-positioning}},
	Lastvisited = {'01.07.2021},
	Note = {[Online; Accessed 01. July 2021]},
	Title = {{Six Ways Autonomous Driving is Relying on Precise Positioning}}
	}


@misc{NFC_door_lock,
	Howpublished = {\url{https://www.getkisi.com/academy/lessons/how-to-use-nfc-door-locks}},
	Lastvisited = {'01.07.2021},
	Note = {[Online; Accessed 01. July 2021]},
	Title = {{How to Use NFC Door Locks}}
	}


@misc{apple_unlock,
	Howpublished = {\url{https://support.apple.com/en-gb/guide/security/secc7d85209d/web}},
	Lastvisited = {'01.07.2021},
	Note = {[Online; Accessed 01. July 2021]},
	Title = {{A Mac can be unlocked by an Apple Watch}}
	}


@misc{smart_unlock,
	Howpublished = {\url{https://www.blemobileapps.com/blog/smart-locks-entry-key-less-world/}},
	Lastvisited = {'01.07.2021},
	Note = {[Online; Accessed 01. July 2021]},
	Title = {{Smart Locks: Your Entry to the Keyless World}}
	}


@misc{nearlock,
	Howpublished = {\url{https://nearlock.me}},
	Lastvisited = {'01.07.2021},
	Note = {[Online; Accessed 01. July 2021]},
	Title = {{Near Lock}}
	}





@misc{Contactless_token_future,
	Howpublished = {\url{https://www.csiweb.com/what-to-know/content-hub/blog/contactless-payments-the-future-of-digital-payment-technologies/}},
	Lastvisited = {'01.07.2021},
	Note = {[Online; Accessed 01. July 2021]},
	Title = {{Contactless Payments: The Future Of Digital Payment Technologies}}
	}


@inproceedings{HRP_singh_wisec, author = {Singh, Mridula and Roeschlin, Marc and Zalzala, Ezzat and Leu, Patrick and \v{C}apkun, Srdjan}, title = {Security Analysis of IEEE 802.15.4z/HRP UWB Time-of-Flight Distance Measurement}, year = {2021}, isbn = {9781450383493}, publisher = {Association for Computing Machinery}, url = {https://doi.org/10.1145/3448300.3467831}, doi = {10.1145/3448300.3467831}, booktitle = {Proceedings of the 14th ACM Conference on Security and Privacy in Wireless and Mobile Networks} }

@inproceedings{Patrick_HRP,
	author = {Patrick Leu and Giovanni Camurati and Alexander Heinrich and Marc Roeschlin and Claudio Anliker and Matthias Hollick and Srdjan Capkun and Jiska Classen},
	bibsource = {dblp computer science bibliography, https://dblp.org},
	biburl = {https://dblp.org/rec/conf/uss/LeuCHRAHCC22.bib},
	booktitle = {31st {USENIX} Security Symposium, {USENIX} Security 2022, Boston, MA, USA, August 10-12, 2022},
	title = {Ghost Peak: Practical Distance Reduction Attacks Against {HRP} {UWB} Ranging},
	url = {https://www.usenix.org/conference/usenixsecurity22/presentation/leu},
	year = {2022}}


@ARTICLE{DB_KeyAgreement,
  author={Čapkun, Srdjan and Čagalj, Mario and Karame, Ghassan and Tippenhauer, Nils Ole},
  journal={IEEE Transactions on Mobile Computing}, 
  title={Integrity Regions: Authentication through Presence in Wireless Networks}, 
  year={2010},
  volume={9},
  number={11},
  pages={1608-1621},
  doi={10.1109/TMC.2010.127}}


@inproceedings{DB_AccessControl, author = {Rasmussen, Kasper Bonne and Castelluccia, Claude and Heydt-Benjamin, Thomas S. and Capkun, Srdjan}, title = {Proximity-Based Access Control for Implantable Medical Devices}, year = {2009}, isbn = {9781605588940}, publisher = {Association for Computing Machinery}, address = {New York, NY, USA}, url = {https://doi.org/10.1145/1653662.1653712}, doi = {10.1145/1653662.1653712}, abstract = {We propose a proximity-based access control scheme for implantable medical devices (IMDs). Our scheme is based on ultrasonic distance-bounding and enables an implanted medical device to grant access to its resources only to those devices that are in its close proximity. We demonstrate the feasibility of our approach through tests in an emulated patient environment. We show that, although implanted, IMDs can successfully verify the proximity of other devices with high accuracy. We propose a set of protocols that support our scheme, analyze their security in detail and discuss possible extensions. We make new observations about the security of implementations of ultrasonic distance-bounding protocols. Finally, we discuss the integration of our scheme with existing IMD devices and with their existing security measures.}, booktitle = {Proceedings of the 16th ACM Conference on Computer and Communications Security}, pages = {410–419}, numpages = {10}, keywords = {ultrasonic communication, access control, distance bounding, medical devices, secure pairing}, location = {Chicago, Illinois, USA}, series = {CCS '09} }


@article{DB_Ariadne_forRouting,
author = {Hu, Yih-chun and Perrig, Adrian and Johnson, David},
year = {2002},
month = {10},
pages = {},
title = {Ariadne: A Secure On-Demand Routing Protocol for Ad Hoc Networks},
volume = {11},
journal = {Wireless Networks},
doi = {10.1145/570645.570648}
}


@INPROCEEDINGS{DB_forLocationVerification,
  author={Singelee, D. and Preneel, B.},
  booktitle={IEEE International Conference on Mobile Adhoc and Sensor Systems Conference, 2005.}, 
  title={Location verification using secure distance bounding protocols}, 
  year={2005},
  volume={},
  number={},
  pages={7 pp.-840},
  doi={10.1109/MAHSS.2005.1542879}}


@inproceedings {DB_10,
author = {Saar Drimer and Steven J. Murdoch},
title = {Keep Your Enemies Close: Distance Bounding Against Smartcard Relay Attacks},
booktitle = {16th {USENIX} Security Symposium ({USENIX} Security 07)},
year = {2007},
address = {Boston, MA },
url = {https://www.usenix.org/conference/16th-usenix-security-symposium/keep-your-enemies-close-distance-bounding-against},
publisher = {{USENIX} Association},
month = aug,
}


@article{5G_Vehicular_Networks_Positioning_mmWaves,
	Author = {H. {Wymeersch} and G. {Seco-Granados} and G. {Destino} and D. {Dardari} and F. {Tufvesson}},
	Doi = {10.1109/MWC.2017.1600374},
	Issn = {1536-1284},
	Journal = {IEEE Wireless Communications},
	Keywords = {5G mobile communication;vehicular ad hoc networks;wireless channels;wireless LAN;5G mmWave positioning;vehicular networks;high data-rate services;IEEE 802.11p;specific signal characteristics;5G communication turn;on-vehicle positioning;mapping systems;cellular positioning;vehicular positioning;inter-vehicle communication standards;5G mobile communication;Bandwidth;Global navigation satellite system;Device-to-device communication;Sensors;Directive antennas;Signal to noise ratio;Intelligent vehicles;Vehicular ad hoc networks},
	Month = {Dec},
	Number = {6},
	Pages = {80-86},
	Title = {{5G} mmWave Positioning for Vehicular Networks},
	Volume = {24},
	Year = {2017},
	Bdsk-Url-1 = {https://doi.org/10.1109/MWC.2017.1600374}}


@misc{5G-NR-Ericsson,
	Archiveprefix = {arXiv},
	Author = {Xingqin Lin and Jingya Li and Robert Baldemair and Thomas Cheng and Stefan Parkvall and Daniel Larsson and Havish Koorapaty and Mattias Frenne and Sorour Falahati and Asbj{\"o}rn Gr{\"o}vlen and Karl Werner},
	Eprint = {1806.06898},
	Primaryclass = {cs.NI},
	Title = {{5G New Radio: Unveiling the Essentials of the Next Generation Wireless Access Technology}},
	Year = {2018}}



@INPROCEEDINGS{DistanceHijacking,
  author={Cremers, Cas and Rasmussen, Kasper B. and Schmidt, Benedikt and Capkun, Srdjan},
  booktitle={2012 IEEE Symposium on Security and Privacy}, 
  title={Distance Hijacking Attacks on Distance Bounding Protocols}, 
  year={2012},
  volume={},
  number={},
  pages={113-127},
  doi={10.1109/SP.2012.17}}


@inproceedings{Ultrasound_Relay, author = {Sedighpour, Sahar and \v{C}apkun, Srdjan and Ganeriwal, Saurabh and Srivastava, Mani}, title = {Distance Enlargement and Reduction Attacks on Ultrasound Ranging}, year = {2005}, isbn = {159593054X}, publisher = {Association for Computing Machinery}, address = {New York, NY, USA}, url = {https://doi.org/10.1145/1098918.1098977}, doi = {10.1145/1098918.1098977}, booktitle = {Proceedings of the 3rd International Conference on Embedded Networked Sensor Systems}, pages = {312}, numpages = {1}, keywords = {attacks on ranging, security, wormholes, ultrasonic ranging}, location = {San Diego, California, USA}, series = {SenSys '05} }



@inproceedings{kasper_DB,
author = {Rasmussen, Kasper and Capkun, Srdjan},
year = {2010},
month = {09},
pages = {389-402},
title = {Realization of RF Distance Bounding},
journal = {Proceedings of the USENIX Security Symposium}
}


@InProceedings{Frauds_Terrorist,
author="Bussard, Laurent
and Bagga, Walid",
editor="Sasaki, Ryoichi
and Qing, Sihan
and Okamoto, Eiji
and Yoshiura, Hiroshi",
title="Distance-Bounding Proof of Knowledge to Avoid Real-Time Attacks",
booktitle="Security and Privacy in the Age of Ubiquitous Computing",
year="2005",
publisher="Springer US",
address="Boston, MA",
pages="223--238",
abstract="Traditional authentication is based on proving the knowledge of a private key corresponding to a given public key. In some situations, especially in the context of pervasive computing, it is additionally required to verify the physical proximity of the authenticated party in order to avoid a set of real-time attacks. Brands and Chaum proposed distance-bounding protocols as a way to compute a practical upper bound on the distance between a prover and a verifier during an authentication process. Their protocol prevents frauds where an intruder sits between a legitimate prover and a verifier and succeeds to perform the distance-bounding process. However, frauds where a malicious prover and an intruder collaborate to cheat a verifier have been left as an open issue. In this paper, we provide a solution preventing both types of attacks.",
isbn="978-0-387-25660-3"
}




@inproceedings{Location_privacy2, author = {Gruteser, Marco and Grunwald, Dirk}, title = {Enhancing Location Privacy in Wireless LAN through Disposable Interface Identifiers: A Quantitative Analysis}, year = {2003}, isbn = {1581137680}, publisher = {Association for Computing Machinery}, address = {New York, NY, USA}, url = {https://doi.org/10.1145/941326.941334}, doi = {10.1145/941326.941334}, abstract = {The recent proliferation of wireless local area networks (WLAN) has introduced new location privacy risks. An adversary controlling several access points could triangulate a client's position. In addition, interface identifiers uniquely identify each client, allowing tracking of location over time. We enhance location privacy through frequent disposal of a client's interface identifier. The described system curbs the adversary's ability to continuously track a client's position. Design challenges include selecting new interface identifiers, detecting address collisions at the MAC layer, and timing identifier switches to balance network disruptions against privacy protection. Using a modified authentication protocol, network operators can still control access to their network. An analysis of a public WLAN usage trace shows that disposing addresses before reassociation already yields significant privacy improvements.}, booktitle = {Proceedings of the 1st ACM International Workshop on Wireless Mobile Applications and Services on WLAN Hotspots}, pages = {46–55}, numpages = {10}, keywords = {location privacy, wireless LAN, interface identifiers}, location = {San Diego, CA, USA}, series = {WMASH '03} }


@inproceedings{LocationLeak_Yongdae,
  title={Location leaks over the GSM air interface},
  author={D. Kune and John K{\"o}lndorfer and N. Hopper and Yongdae Kim},
  booktitle={NDSS},
  year={2012}
}


@misc{DP3T,
      title={Decentralized Privacy-Preserving Proximity Tracing}, 
      author={Carmela Troncoso and Mathias Payer and Jean-Pierre Hubaux and Marcel Salathé and James Larus and Edouard Bugnion and Wouter Lueks and Theresa Stadler and Apostolos Pyrgelis and Daniele Antonioli and Ludovic Barman and Sylvain Chatel and Kenneth Paterson and Srdjan Čapkun and David Basin and Jan Beutel and Dennis Jackson and Marc Roeschlin and Patrick Leu and Bart Preneel and Nigel Smart and Aysajan Abidin and Seda Gürses and Michael Veale and Cas Cremers and Michael Backes and Nils Ole Tippenhauer and Reuben Binns and Ciro Cattuto and Alain Barrat and Dario Fiore and Manuel Barbosa and Rui Oliveira and José Pereira},
      year={2020},
      eprint={2005.12273},
      archivePrefix={arXiv},
      primaryClass={cs.CR}
}


@Inbook{Relay_toDB,
author="Avoine, Gildas
and Boureanu, Ioana
and G{\'e}rault, David
and Hancke, Gerhard P.
and Lafourcade, Pascal
and Onete, Cristina",
editor="Avoine, Gildas
and Hernandez-Castro, Julio",
title="From Relay Attacks to Distance-Bounding Protocols",
bookTitle="Security of Ubiquitous Computing Systems: Selected Topics",
year="2021",
publisher="Springer International Publishing",
address="Cham",
pages="113--130",
abstract="We present the concept of relay attacks, and discuss distance-bounding schemes as the main countermeasure. We give details on relaying mechanisms, we review canonical distance-bounding protocols, as well as their threat-model (i.e., covering attacks beyond relaying) stemming from the authentication dimension in distance bounding. Advanced aspects of distance-bounding security are also covered. We conclude by presenting what we consider to be the most important challenges in distance bounding.",
isbn="978-3-030-10591-4",
doi="10.1007/978-3-030-10591-4_7",
url="https://doi.org/10.1007/978-3-030-10591-4_7"
}






@inproceedings{HybridPositioning_5G_2,
  TITLE = {{5G Positioning And Hybridization With GNSS Observations}},
  AUTHOR = {Crapart, Romain and Maymo-Camps , Roc and Vautherin, Benoit and Saloranta, Jani},
  URL = {https://hal-enac.archives-ouvertes.fr/hal-01942264},
  BOOKTITLE = {{ITSNT 2018,  International Technical Symposium on Navigation and Timing}},
  ADDRESS = {Toulouse, France},
  YEAR = {2018},
  MONTH = Oct,
  DOI = {10.31701/itsnt2018.10},
  PDF = {https://hal-enac.archives-ouvertes.fr/hal-01942264/file/ITSNT-PAPER-5G_Final.pdf},
  HAL_ID = {hal-01942264},
  HAL_VERSION = {v1},
}

@INPROCEEDINGS{HybridPositioning_5G_1,
  author={Destino, Giuseppe and Saloranta, Jani and Seco-Granados, Gonzalo and Wymeersch, Henk},
  booktitle={2018 52nd Asilomar Conference on Signals, Systems, and Computers}, 
  title={Performance Analysis of Hybrid 5G-GNSS Localization}, 
  year={2018},
  volume={},
  number={},
  pages={8-12},
  doi={10.1109/ACSSC.2018.8645207}}



@inproceedings{fleet_tracking_1,
  TITLE = {{System and method for fleet tracking}},
  AUTHOR = {Yekutiel A Novik},
  YEAR = {2000},
}

@misc{fleet_tracking_2,
	Howpublished = {\url{https://www.geotab.com}},
	Lastvisited = {'13.04.2023},
	Note = {[Online; Accessed 13. April 2023]},
	Title = {{Geotab fleet tracking}}}


@misc{gizmodo,
	Howpublished = {\url{https://gizmodo.com/jamming-gps-signals-is-illegal-dangerous-cheap-and-e-1796778955}},
	Lastvisited = {'13.04.2023},
	Note = {[Online; Accessed 13. April 2023]},
	Title = {{Geotab fleet tracking}}}


@misc{tesla_hack,
	Howpublished = {\url{https://www.gpsworld.com/two-years-since-the-tesla-gps-hack}},
	Lastvisited = {'13.04.2023},
	Note = {[Online; Accessed 13. April 2023]},
	Title = {{Tesla Hack}}}

/


@INPROCEEDINGS{Channel_impact_PRS_LTE,
  author={Medbo, Jonas and Siomina, Iana and Kangas, Ari and Furuskog, Johan},
  booktitle={2009 IEEE 20th International Symposium on Personal, Indoor and Mobile Radio Communications}, 
  title={Propagation channel impact on LTE positioning accuracy: A study based on real measurements of observed time difference of arrival}, 
  year={2009},
  volume={},
  number={},
  pages={2213-2217},
  doi={10.1109/PIMRC.2009.5450144}}






@misc{5G_positioning_blog,
	Howpublished = {\url{https://www.pointr.tech/blog/5g-indoor-positioning}},
	Lastvisited = {'01.06.2021},
	Note = {[Online; Accessed 05. June 2021]},
	Title = {{5G for Positioning}}}

@misc{5G_positioning_blog_Nokia_bell_labs,
	Howpublished = {\url{https://www.bell-labs.com/institute/blog/5g-will-open-new-possibilities-positioning/#gref}},
	Lastvisited = {'01.06.2021},
	Note = {[Online; Accessed 05. June 2021]},
	Title = {{5G will open new possibilities in positioning}}
	}


@misc{Lidar_spoofingArticle,
	Howpublished = {\url{https://gcn.com/articles/2020/03/06/lidar-spoofs-autonomous-vehicle-hack.aspx}},
	Lastvisited = {'01.06.2021},
	Note = {[Online; Accessed 05. June 2021]},
	Title = {{Autonomous vehicles can be fooled to ‘see’ nonexistent obstacles}}}

@inproceedings{LIDAR_attack,
	Address = {New York, NY, USA},
	Author = {Cao, Yulong and Xiao, Chaowei and Cyr, Benjamin and Zhou, Yimeng and Park, Won and Rampazzi, Sara and Chen, Qi Alfred and Fu, Kevin and Mao, Z. Morley},
	Booktitle = {Proceedings of the 2019 ACM SIGSAC Conference on Computer and Communications Security},
	Doi = {10.1145/3319535.3339815},
	Isbn = {9781450367479},
	Keywords = {adversarial machine learning, autonomous driving, sensor attack},
	Location = {London, United Kingdom},
	Numpages = {15},
	Pages = {2267--2281},
	Publisher = {Association for Computing Machinery},
	Series = {CCS '19},
	Title = {Adversarial Sensor Attack on LiDAR-Based Perception in Autonomous Driving},
	Url = {https://doi.org/10.1145/3319535.3339815},
	Year = {2019},
	Bdsk-Url-1 = {https://doi.org/10.1145/3319535.3339815}}


@misc{LIDAR_attack_2,
      title={Towards Robust LiDAR-based Perception in Autonomous Driving: General Black-box Adversarial Sensor Attack and Countermeasures}, 
      author={Jiachen Sun and Yulong Cao and Qi Alfred Chen and Z. Morley Mao},
      year={2020},
      eprint={2006.16974},
      archivePrefix={arXiv},
      primaryClass={cs.CR}
}


@misc{Positioningin5G_Erricson,
      title={Positioning in 5G networks}, 
      author={Satyam Dwivedi and Ritesh Shreevastav and Florent Munier and Johannes Nygren and Iana Siomina and Yazid Lyazidi and Deep Shrestha and Gustav Lindmark and Per Ernström and Erik Stare and Sara M. Razavi and Siva Muruganathan and Gino Masini and Åke Busin and Fredrik Gunnarsson},
      year={2021},
      eprint={2102.03361},
      archivePrefix={arXiv},
      primaryClass={cs.NI}
}





@inproceedings{5G_localizationWaveform_dlr_2,
author = {Staudinger, Emanuel and Walter, Michael and Dammann, Armin},
year = {2016},
month = {09},
pages = {xx - xx},
title = {The 5G Localization Waveform Ranging Accuracy over Time-Dispersive Channels -- An Evaluation},
doi = {10.33012/2016.14867}
}


@inproceedings{PRS_atReceiver,
author = {Ferre, Ruben and Seco-Granados, Gonzalo and Lohan, Elena Simona},
year = {2019},
month = {10},
pages = {},
title = {Positioning Reference Signal design for positioning via 5G},
doi = {10.5281/zenodo.3537686}
}


@inproceedings{PRS_atReceiver_2,
author = {Peral-Rosado, Jose and López-Salcedo, José A. and Seco-Granados, Gonzalo and Zanier, Francesca and Massimo, Crisci},
year = {2012},
month = {09},
pages = {139-146},
title = {Evaluation of the LTE positioning capabilities under typical multipath channels},
isbn = {978-1-4673-2676-6},
doi = {10.1109/ASMS-SPSC.2012.6333065}
}


@inproceedings{PRS_atReceiver_3,
  title={Software-Defined Radio LTE Positioning Receiver Towards Future Hybrid Localization Systems},
  author={J. A. D. Peral-Rosado and J. L{\'o}pez-Salcedo and G. Seco-Granados and F. Zanier and P. Crosta and R. Ioannides and M. Crisci},
  year={2013}
}


@inproceedings{5G_localizationWaveform_dlr,
author = {Raulefs, Ronald and Dammann, Armin and Jost, Thomas and Walter, Michael and Zhang, Siwei},
year = {2016},
month = {01},
pages = {},
title = {The 5G Localization Waveform}
}

@misc{fcc_locationAccuracyRequirements,
	Howpublished = {\url{https://www.fcc.gov/public-safety-and-homeland-security/policy-and-licensing-division/911-services/general/location-accuracy-indoor-benchmarks}},
	Lastvisited = {'01.11.2022},
	Note = {[Online; Accessed 01. November 2022]},
	Title = {{Indoor Location Accuracy Timeline and Live Call Data Reporting Template}}}


@misc{Parole,
	Howpublished = {\url{ https://www.unodc.org/documents/justice-and-prison-reform/ReducingReoffending/MS/Republic_of_Korea_Annex_-_.pdf}},
	Lastvisited = {'13.04.2023},
	Note = {[Online; Accessed 13. April 2023]},
	Title = {{Crime Prevention System for Public Safety}}}
 



 


@article{MiladPaper,
  author    = {Milad Rezaee and
               Dave Singel{\'{e}}e and
               Bart Preneel},
  title     = {A Novel Demodulation Scheme for Secure and Reliable {UWB} Distance
               Bounding},
  journal   = {CoRR},
  volume    = {abs/2010.10387},
  year      = {2020},
  url       = {https://arxiv.org/abs/2010.10387},
  archivePrefix = {arXiv},
  eprint    = {2010.10387},
  timestamp = {Mon, 26 Oct 2020 15:39:44 +0100},
  biburl    = {https://dblp.org/rec/journals/corr/abs-2010-10387.bib},
  bibsource = {dblp computer science bibliography, https://dblp.org}
}



@INPROCEEDINGS{UWB_HRP_TuGRAZ,
  author={Stocker, Michael and Großwindhager, Bernhard and Boano, Carlo Alberto and Römer, Kay},
  booktitle={2020 IEEE 17th International Conference on Mobile Ad Hoc and Sensor Systems (MASS)}, 
  title={Towards Secure and Scalable UWB-based Positioning Systems}, 
  year={2020},
  volume={},
  number={},
  pages={247-255},
  doi={10.1109/MASS50613.2020.00039}}



@article{Channel_effectUWBRanging,
	affiliation = {Centre for Wireless Communications, University of Oulu; Biomimetics and Intelligent Systems Group, University of Oulu; Division of Physics, Electrical and Computer Engineering, Yokohama National University; University of Oulu Research Institute Japan CWC-Nippon Co. Ltd.},
	author = {Mikhaylov, Konstantin and Petäjäjärvi, Juha and Hämäläinen, Matti and Tikanmäki, Antti and Kohno, Ryuji},
	copyright = {Springer Science+Business Media New York},
	doi = {10.1007/s10776-017-0340-9},
	journal = {International Journal of Wireless Information Networks},
	keywords = {Ultra wideband; IEEE 802.15.4; Localization; Asynchronous; Ranging; Indoor; Body area network; Accuracy; Experiment; Measurements; Performance; DW1000},
	language = {English},
	month = {2},
	number = {2},
	pages = {124-139},
	title = {Impact of IEEE 802.15.4 Communication Settings on Performance in Asynchronous Two Way UWB Ranging},
	volume = {24},
	year = {2017},
}



@misc{FIRA_HRP,
	Howpublished = {\url{https://www.firaconsortium.org/sites/default/files/2020-04/fira-introduction-impulse-radio-uwb-wp-en.pdf}},
	Lastvisited = {'05.07.2021},
	Note = {[Online; Accessed 05. July 2021]},
	Title = {{Introduction to Impulse Radio UWB Seamless Access Systems}}}



@misc{HybridPositioning_ESA,
	Howpublished = {\url{https://navisp.esa.int/opportunity/details/72/show}},
	Lastvisited = {'05.07.2021},
	Note = {[Online; Accessed 05. July 2021]},
	Title = {{Proof of Concept of Hybrid 5G-NR/GNSS Positioning with AD-HOC Overlay}}}



@inproceedings{SpectrumFlexible_WisecPaper, author = {Vo-Huu, Tien Dang and Vo-Huu, Triet Dang and Noubir, Guevara}, title = {Spectrum-Flexible Secure Broadcast Ranging}, year = {2021}, isbn = {9781450383493}, publisher = {Association for Computing Machinery}, address = {New York, NY, USA}, url = {https://doi.org/10.1145/3448300.3467819}, doi = {10.1145/3448300.3467819},  booktitle = {Proceedings of the 14th ACM Conference on Security and Privacy in Wireless and Mobile Networks}, pages = {300–310}, numpages = {11}, keywords = {protocols, privacy, wireless ranging, software defined radio}, location = {Abu Dhabi, United Arab Emirates}, series = {WiSec '21} }


@INPROCEEDINGS{RelayPreventionCombineSensors,
  author={Wang, Juan and Lounis, Karim and Zulkernine, Mohammad},
  booktitle={2019 IEEE 43rd Annual Computer Software and Applications Conference (COMPSAC)}, 
  title={CSKES: A Context-Based Secure Keyless Entry System}, 
  year={2019},
  volume={1},
  number={},
  pages={817-822},
  doi={10.1109/COMPSAC.2019.00120}}

@article{RelayAttackPreventionUsing2ndFactor,
author = {Choi, Wonsuk and Seo, Minhye and Lee, Dong},
year = {2018},
month = {01},
pages = {1-13},
title = {Sound-Proximity: 2-Factor Authentication against Relay Attack on Passive Keyless Entry and Start System},
volume = {2018},
journal = {Journal of Advanced Transportation},
doi = {10.1155/2018/1935974}
}


@INPROCEEDINGS{LRP_HRP_comparision,
  author={Flueratoru, Laura and Wehrli, Silvan and Magno, Michele and Niculescu, Dragos},
  booktitle={GLOBECOM 2020 - 2020 IEEE Global Communications Conference}, 
  title={On the Energy Consumption and Ranging Accuracy of Ultra-Wideband Physical Interfaces}, 
  year={2020},
  volume={},
  number={},
  pages={1-7},
  doi={10.1109/GLOBECOM42002.2020.9347984}}


@ARTICLE{UWB_PKES_ETSI,
  author={},
  journal={IEEE Std 802.15.4a-2007 (Amendment to IEEE Std 802.15.4-2006)}, 
  title={System Reference document (SRdoc); Short Range Devices (SRD) using Ultra Wide Band (UWB); Technical characteristics and spectrum requirements for UWB based vehicular access systems for operation in the 3,4 GHz to 4,8 GHz and 6 GHz to 8,5 GHz frequency ranges}, 
  year={2016},
  volume={},
  number={},
  pages={}}

@misc{3dB_article_standard,
	Howpublished = {\url{https://www.3db-access.com/article/15}},
	Lastvisited = {'29.11.19},
	Note = {[Online; Accessed 29. November 2019]},
	Title = {{UWB Secure Ranging-3dB}}}

	@misc{3dbVW,
	Howpublished = {\url{https://www.3db-access.com/article/18}},
	Lastvisited = {'25.03.2021},
	Note = {[Online; Accessed 25. March 2021]},
	Title = {{LRP deployment in automotive.}}}


	@misc{3db,
	Howpublished = {\url{https://www.3db-access.com/Product.3.html}},
	Lastvisited = {'23.10.17},
	Note = {[Online; Accessed 23. October 2017]},
	Title = {{3db Access AG - 3DB6830} (``Proximity based access control")}}


@misc{TaskGroup802.15.4az,
	Howpublished = {\url{http://www.ieee802.org/15/pub/TG4z.html}},
	Lastvisited = {'17.06.2020},
	Note = {[Online; Accessed 17. June 2020]},
	Title = {{802.15.4z Task Group}}}


@misc{UWB_IEEE802.15.4z,
	Date-Modified = {2021-01-18 21:17:30 +0100},
	Howpublished = {\url{https://standards.ieee.org/develop/project/802.15.4z.html}},
	Lastvisited = {'7.10.18},
	Note = {[Online; Accessed 7. August 2018]},
	Title = {802.15.4z - Standard for Low-Rate Wireless Networks Amendment: Enhanced High Rate Pulse (HRP) and Low Rate Pulse (LRP) Ultra Wide-Band (UWB) Physical Layers (PHYs) and Associated Ranging Techniques}}


@ARTICLE{UWB_IEEE802.15.4a,
  author={},
  journal={IEEE Std 802.15.4a-2007 (Amendment to IEEE Std 802.15.4-2006)}, 
  title={IEEE Standard for Information technology-- Local and metropolitan area networks-- Specific requirements-- Part 15.4: Wireless Medium Access Control (MAC) and Physical Layer (PHY) Specifications for Low-Rate Wireless Personal Area Networks (WPANs): Amendment 1: Add Alternate PHYs}, 
  year={2007},
  volume={},
  number={},
  pages={1-210},
  doi={10.1109/IEEESTD.2007.4299496}}

@ARTICLE{UWB_IEEE802.15.4f,
  author={},
  journal={IEEE Std 802.15.4f-2012 (Amendment to IEEE Std 802.15.4-2011)}, 
  title={IEEE Standard for Local and metropolitan area networks-- Part 15.4: Low-Rate Wireless Personal Area Networks (LR-WPANs) Amendment 2: Active Radio Frequency Identification (RFID) System Physical Layer (PHY)}, 
  year={2012},
  volume={},
  number={},
  pages={1-72},
  doi={10.1109/IEEESTD.2012.6188488}}





@inproceedings{Brands1994,
	Author = {Brands, Stefan and Chaum, David},
	Booktitle = {EUROCRYPT},
	Numpages = {16},
	Pages = {344--359},
	Publisher = {Springer},
	Title = {Distance-bounding Protocols},
	Year = {1994}}

	@inproceedings{Nils_wisec2017,
	Acmid = {2766504},
	Author = {Tippenhauer, Nils Ole and Luecken, Heinrich and Kuhn, Marc and Capkun, Srdjan},
	Booktitle = {Proceedings of the 8th ACM Conference on Security \& Privacy in Wireless and Mobile Networks},
	Doi = {10.1145/2766498.2766504},
	Pages = {2:1--2:12},
	Publisher = {ACM},
	Series = {WiSec '15},
	Title = {UWB Rapid-bit-exchange System for Distance Bounding},
	Url = {http://doi.acm.org/10.1145/2766498.2766504},
	Year = {2015},
	Bdsk-Url-1 = {http://doi.acm.org/10.1145/2766498.2766504},
	Bdsk-Url-2 = {https://doi.org/10.1145/2766498.2766504}}

	@inproceedings{rasmussenUsenix10,
	Author = {Rasmussen,, Kasper Bonne and Capkun,, Srdjan},
	Booktitle = {Proceedings of the USENIX Security Symposium},
	Title = {Realization of RF Distance Bounding},
	Year = {2010}}

	@inproceedings{kim2008swiss,
	Author = {Kim, Chong Hee and Avoine, Gildas and Koeune, Fran{\c{c}}ois and Standaert, Fran{\c{c}}ois-Xavier and Pereira, Olivier},
	Booktitle = {ICISC},
	Organization = {Springer},
	Pages = {98--115},
	Title = {The Swiss-Knife RFID Distance Bounding Protocol.},
	Volume = {5461},
	Year = {2008}}


	@inproceedings{DB_Multi,
	Author = {Brelurut, Agn{\`e}s and Gerault, David and Lafourcade, Pascal},
	Booktitle = {{Foundations and Practice of Security (FPS)}},
	Doi = {10.1007/978-3-319-30303-1\_3},
	Pages = {29 - 49},
	Title = {{Survey of Distance Bounding Protocols and Threats}},
	Url = {https://hal.archives-ouvertes.fr/hal-01588557},
	Year = {2015},
	Bdsk-Url-1 = {https://hal.archives-ouvertes.fr/hal-01588557},
	Bdsk-Url-2 = {https://doi.org/10.1007/978-3-319-30303-1%5C_3}}


	@inproceedings{Hancke_Kuhn_DB,
	Acmid = {1128472},
	Author = {Hancke, Gerhard P. and Kuhn, Markus G.},
	Booktitle = {Proceedings of the First International Conference on Security and Privacy for Emerging Areas in Communications Networks},
	Doi = {10.1109/SECURECOMM.2005.56},
	Isbn = {0-7695-2369-2},
	Pages = {67--73},
	Publisher = {IEEE Computer Society},
	Series = {SECURECOMM '05},
	Title = {An RFID Distance Bounding Protocol},
	Url = {http://dx.doi.org/10.1109/SECURECOMM.2005.56},
	Year = {2005},
	Bdsk-Url-1 = {http://dx.doi.org/10.1109/SECURECOMM.2005.56}}


@inproceedings{singelee2005location,
	Author = {Singelee, Dave and Preneel, Bart},
	Booktitle = {Mobile Adhoc and Sensor Systems Conference, 2005. IEEE International Conference on},
	Organization = {IEEE},
	Pages = {7--pp},
	Title = {Location verification using secure distance bounding protocols},
	Year = {2005}}

	@misc{DB_EPFL,
	Author = {Ioana Boureanu and Aikaterini Mitrokotsa and Serge Vaudenay},
	Howpublished = {Cryptology ePrint Archive, Report 2015/208},
	Note = {\url{https://eprint.iacr.org/2015/208}},
	Title = {{Towards Secure Distance Bounding}},
	Year = {2015}}

@inproceedings{SoNearYetSoFar,
	Acmid = {2174091},
	Author = {Clulow, Jolyon and Hancke, Gerhard P. and Kuhn, Markus G. and Moore, Tyler},
	Booktitle = {Proceedings of the Third European Conference on Security and Privacy in Ad-Hoc and Sensor Networks},
	Doi = {10.1007/11964254_9},
	Isbn = {3-540-69172-3, 978-3-540-69172-3},
	Location = {Hamburg, Germany},
	Pages = {83--97},
	Publisher = {Springer},
	Series = {ESAS'06},
	Title = {So Near and Yet So Far: Distance-bounding Attacks in Wireless Networks},
	Url = {http://dx.doi.org/10.1007/11964254_9},
	Year = {2006},
	Bdsk-Url-1 = {http://dx.doi.org/10.1007/11964254_9}}


	@article{avoine2018security,
	Author = {Avoine, Gildas and Bing{\"o}l, Muhammed Ali and Boureanu, Ioana and {\v{C}}apkun, Srdjan and Hancke, Gerhard and Karda{\c{s}}, S{\"u}leyman and Kim, Chong Hee and Lauradoux, C{\'e}dric and Martin, Benjamin and Munilla, Jorge and others},
	Journal = {ACM Computing Surveys (CSUR)},
	Number = {5},
	Pages = {1--33},
	Publisher = {ACM New York, NY, USA},
	Title = {Security of distance-bounding: A survey},
	Volume = {51},
	Year = {2018}}

@inbook{GDB_Capkun,
	Author = {{\v{C}apkun, Srdjan and El Defrawy, Karim and Tsudik, Gene}},
	Booktitle = {Springer TRUST 2011},
	Pages = {302--312},
	Publisher = {Springer},
	Title = {{Group Distance Bounding Protocols}},
	Year = {2011}}


	@article{edlcEPFl,
	Author = {Poturalski, Marcin and Flury, Manuel and Papadimitratos, Panos and Hubaux, Jean-Pierre and Le Boudec, Jean-Yves},
	Journal = {IEEE Transactions on Wireless Communications},
	Number = {4},
	Pages = {1334--1344},
	Publisher = {IEEE},
	Title = {Distance bounding with IEEE 802.15.4a: Attacks and countermeasures},
	Volume = {10},
	Year = {2011}}

	@inproceedings{Flury_reductionattack,
	Acmid = {1741887},
	Author = {Flury, Manuel and Poturalski, Marcin and Papadimitratos, Panos and Hubaux, Jean-Pierre and Le Boudec, Jean-Yves},
	Booktitle = {Proceedings of the Third ACM Conference on Wireless Network Security},
	Series = {WiSec '10},
	Title = {Effectiveness of Distance-decreasing Attacks Against Impulse Radio Ranging},
	Year = {2010}}


	@inproceedings{cicadaEPFL,
	Author = {Poturalski, Mand Flury, M and Papadimitratos, P and Hubaux, JP and Le Boudec, JY},
	Booktitle = {Ultra-Wideband (ICUWB), 2010 IEEE International Conference on},
	Organization = {IEEE},
	Pages = {1--4},
	Title = {The cicada attack: degradation and denial of service in IR ranging},
	Volume = {2},
	Year = {2010}}


	@article{cicadaEPFL_2,
	Author = {Poturalski, Marcin and Flury, Manuel and Papadimitratos, Panos and Hubaux, Jean-Pierre and Le Boudec, Jean-Yves},
	Journal = {IEEE transactions on wireless communications},
	Number = {3},
	Pages = {1087--1099},
	Publisher = {IEEE},
	Title = {On secure and precise IR-UWB ranging},
	Volume = {11},
	Year = {2012}}



@misc{802.11az,
	Howpublished = {\url{http://www.ieee802.org/11/Reports/tgaz_update.htm}},
	Lastvisited = {'24.09.19},
	Note = {[Online; Accessed 24. September 2019]},
	Title = {{802.11az}}}


	@misc{CP_replay_ppt,
	Howpublished = {\url{https://mentor.ieee.org/802.11/dcn/17/11-17-1122-00-00az-cp-replay-threat-model-for-11az.pptx}},
	Lastvisited = {'24.09.19},
	Note = {[Online; Accessed 24. September 2019]},
	Title = {{Cyclic Prefix Replay Attack}}}

@misc{CP_replay_ppt_2,
	Howpublished = {\url{https://mentor.ieee.org/802.11/dcn/17/11-17-1372-01-00az-cp-replay-attack-protection.pptx}},
	Lastvisited = {'24.09.19},
	Note = {[Online; Accessed 24. September 2019]},
	Title = {{Cyclic Prefix Replay Attack Protection}}}

	@misc{ieee2016,
	Author = {IEEE Standards Association and others},
	Title = {IEEE Std 802.11-2016, IEEE Standard for Local and Metropolitan Area Networks---Part 11: Wireless LAN Medium Access Control (MAC) and Physical Layer (PHY) Specifications, 2016}}

@article{passportRelay,
  title={A Note on the Relay Attacks on e-passports: The Case of Czech e-passports},
  author={M. Hlav{\'a}c and Tom{\'a}s Rosa},
  journal={IACR Cryptol. ePrint Arch.},
  year={2007},
  volume={2007},
  pages={244}
}

@misc{dirichlet,
	Author = {MathWorks},
	Howpublished = {\url{https://ch.mathworks.com/help/signal/ref/diric.html}},
	Key = {mathworks_dirichlet},
	Lastvisited = {'20.06.19},
	Note = {[Online; Accessed 20. June 2019]},
	Publisher = {MathWorks},
	Title = {{Dirichlet or periodic sinc function}}}

@misc{snapdragon,
	Author = {Qualcomm},
	Howpublished = {\url{https://www.qualcomm.com/products/snapdragon-processors-800}},
	Lastvisited = {'30.10.20},
	Note = {[Online; Accessed 30. October 2020]},
	Publisher = {Qualcomm},
	Title = {{Snapdragon 800 Processor}}}

@misc{5G_clock_link,
	Howpublished = {\url{https://www.edn.com/electronics-products/electronic-product-reviews/other/4461327/Deploying-precision-timing-for-5G}},
	Lastvisited = {'18.06.19},
	Note = {[Online; Accessed 18. June 2019]},
	Title = {{Deploying precision timing for 5G}}}


@misc{WifiChannelModel,
	Howpublished = {\url{https://www.mathworks.com/help/wlan/gs/wlan-channel-models.html}},
	Lastvisited = {'21.06.2020},
	Note = {[Online; Accessed 21. June 2020]},
	Title = {{WLAN Channel Models}}}




@book{molisch2012wireless,
	Author = {Molisch, Andreas F},
	Publisher = {John Wiley \& Sons},
	Title = {Wireless communications},
	Volume = {34},
	Year = {2012}}

@article{5G_clock_req,
	Author = {J. {Lin}},
	Doi = {10.1109/MVT.2018.2813339},
	Issn = {1556-6072},
	Journal = {IEEE Vehicular Technology Magazine},
	Keywords = {4G mobile communication;5G mobile communication;cellular radio;radio spectrum management;telecommunication network reliability;cellular networks;radio communications;fifth-generation wireless communication;5G;high spectral efficiency;fourth-generation mobile communications;4G;Synchronization;5G mobile communication;Frequency synchronization;3G mobile communication},
	Month = {Sep.},
	Number = {3},
	Pages = {91-99},
	Title = {Synchronization Requirements for 5G: An Overview of Standards and Specifications for Cellular Networks},
	Volume = {13},
	Year = {2018},
	Bdsk-Url-1 = {https://doi.org/10.1109/MVT.2018.2813339}}

@article{Maziar2016,
	Author = {Maziar, Nekovee and Yue, Wang and Milos, Tesanovic and Shangbin, Wu and Yinan, Qi and Mohammed, Al-Imari},
	Day = {01},
	Doi = {10.1007/BF03391545},
	Issn = {2509-3312},
	Journal = {Journal of Communications and Information Networks},
	Month = {Jun},
	Number = {1},
	Pages = {44--60},
	Title = {Overview of 5G modulation and waveforms candidates},
	Volume = {1},
	Year = {2016},
	Bdsk-Url-1 = {https://doi.org/10.1007/BF03391545}}

@article{parkvall2017nr,
	Author = {Parkvall, Stefan and Dahlman, Erik and Furuskar, Anders and Frenne, Mattias},
	Journal = {IEEE Communications Standards Magazine},
	Number = {4},
	Pages = {24--30},
	Publisher = {IEEE},
	Title = {NR: The new 5G radio access technology},
	Volume = {1},
	Year = {2017}}



@article{5G_technical_report,
	Title = {{5G}; Study on Scenarios and Requirements for Next Generation Access Technologies (3GPP TR 38.913 version 14.2.0 Release 14)}}

@article{6923528,
	Author = {P. Banelli and S. Buzzi and G. Colavolpe and A. Modenini and F. Rusek and A. Ugolini},
	Doi = {10.1109/MSP.2014.2337391},
	Issn = {1053-5888},
	Journal = {IEEE Signal Processing Magazine},
	Keywords = {cellular radio;frequency division multiple access;Long Term Evolution;MIMO communication;OFDM modulation;wireless channels;5G network;OFDM;spectral efficiency improvement;fifth-generation cellular communication;mobile user;Long Term Evolution system;LTE system;innovative technique;physical layer;orthogonal frequency division multiplexing modulation format;orthogonal frequency division multiple access;OFDMA;cellular environment;5G ingredient;massive multiple-input multiple-output system;MIMO system;real channel measurement;OFDM;Time-frequency analysis;Bandwidth allocation;Frequency modulation;Quadrature amplitude modulation;5G mobile communication;Next generation networking;Mobile communication;Wireless cellular networks;Cellular networks},
	Month = {Nov},
	Number = {6},
	Pages = {80-93},
	Title = {Modulation Formats and Waveforms for 5G Networks: Who Will Be the Heir of OFDM?: An overview of alternative modulation schemes for improved spectral efficiency},
	Volume = {31},
	Year = {2014},
	Bdsk-Url-1 = {https://doi.org/10.1109/MSP.2014.2337391}}

@article{SC_OFDM_5G,
	Author = {S. Buzzi and C. D'Andrea and T. Foggi and A. Ugolini and G. Colavolpe},
	Doi = {10.1109/TCOMM.2017.2771334},
	Issn = {0090-6778},
	Journal = {IEEE Transactions on Communications},
	Keywords = {5G mobile communication;array signal processing;equalisers;fast Fourier transforms;frequency-domain analysis;MIMO communication;modems;OFDM modulation;power amplifiers;precoding;single-carrier modulation;millimeter-wave wireless MIMO;achievable spectral efficiency;energy efficiency;millimeter wave frequencies;linear equalization;frequency-domain equalization;MIMO channels;mmWave frequencies;post-coding beamformers;time-domain equalization;advanced equalization schemes;wireless multiple-input-multiple-output link;typical 5G scenario;single-carrier modem schemes;fast Fourier transform;hybrid pre-coding;transmitter power amplifiers;nonlinear distortion;OFDM;Frequency modulation;Radio frequency;Transceivers;Array signal processing;MIMO;mmWave;5G;MIMO;single-carrier modulation;spectral efficiency;energy efficiency;MIMO-OFDM;time-domain equalization;frequency-domain equalization;hybrid decoding},
	Month = {March},
	Number = {3},
	Pages = {1335-1348},
	Title = {Single-Carrier Modulation Versus OFDM for Millimeter-Wave Wireless MIMO},
	Volume = {66},
	Year = {2018},
	Bdsk-Url-1 = {https://doi.org/10.1109/TCOMM.2017.2771334}}

@inproceedings{Null_CP1,
	Author = {M. Cudak and A. Ghosh and T. Kovarik and R. Ratasuk and T. A. Thomas and F. W. Vook and P. Moorut},
	Booktitle = {2013 IEEE 77th Vehicular Technology Conference (VTC Spring)},
	Doi = {10.1109/VTCSpring.2013.6692638},
	Issn = {1550-2252},
	Keywords = {4G mobile communication;cellular radio;local area networks;losses;microwave links;millimetre wave antenna arrays;millimetre wave propagation;MMIC power amplifiers;radio spectrum management;mm-wave-based beyond-4G small cell technology;untapped spectrum resource availability;cellular systems;millimeter wave transmission;shadowing loss;link budgets;low power amplifier;path loss;mm-wave circuits;antenna arrays fabrication;antenna elements;backhaul;IC technology;local area network;Antenna arrays;Radiofrequency integrated circuits;Receivers;Wireless communication;Bandwidth;Transmitting antennas;Propagation losses},
	Month = {June},
	Pages = {1-5},
	Title = {Moving Towards Mmwave-Based Beyond-4G (B-4G) Technology},
	Year = {2013},
	Bdsk-Url-1 = {https://doi.org/10.1109/VTCSpring.2013.6692638}}

@article{mmWave_1,
	Author = {A. Ghosh and T. A. Thomas and M. C. Cudak and R. Ratasuk and P. Moorut and F. W. Vook and T. S. Rappaport and G. R. MacCartney and S. Sun and S. Nie},
	Doi = {10.1109/JSAC.2014.2328111},
	Issn = {0733-8716},
	Journal = {IEEE Journal on Selected Areas in Communications},
	Keywords = {channel capacity;millimetre wave propagation;mobile radio;multifrequency antennas;radio spectrum management;telecommunication traffic;wireless channels;wireless LAN;local area system;high data rate approach;future wireless network;wireless data traffic;system complexity;fundamental limit;hardware implementation;wireless data usage;unused nontraditional spectrum;bandwidth availability;millimeter wave;air interface;spectral efficiency;mmWave systems;higher system capacity;5G wireless system;ultradense networks;enhanced local area;eLA technology;spectrum consideration;spectrum propagation;channel modeling;multiantenna design;network architecture solution;Loss measurement;Antenna measurements;Bandwidth;OFDM;Data models;Wireless communication;Urban areas;mmWave;5G;enhanced local area;spectrum;propagation;channel modeling;air-interface design;modulation},
	Month = {June},
	Number = {6},
	Pages = {1152-1163},
	Title = {Millimeter-Wave Enhanced Local Area Systems: A High-Data-Rate Approach for Future Wireless Networks},
	Volume = {32},
	Year = {2014},
	Bdsk-Url-1 = {https://doi.org/10.1109/JSAC.2014.2328111}}

@inproceedings{5g_Bounds_waveform,
	Author = {Staudinger, Emanuel and Walter, Michael and Dammann, Armin},
	Doi = {10.33012/2016.14867},
	Month = {09},
	Title = {The 5G Localization Waveform Ranging Accuracy over Time-Dispersive Channels -- An Evaluation},
	Year = {2016},
	Bdsk-Url-1 = {https://doi.org/10.33012/2016.14867}}

@article{What_Will_5G_Be,
	Author = {J. G. {Andrews} and S. {Buzzi} and W. {Choi} and S. V. {Hanly} and A. {Lozano} and A. C. K. {Soong} and J. C. {Zhang}},
	Doi = {10.1109/JSAC.2014.2328098},
	Issn = {1558-0008},
	Journal = {IEEE Journal on Selected Areas in Communications},
	Keywords = {4G mobile communication;cellular radio;Long Term Evolution;mobile antennas;wireless LAN;5G mobile communication;4G mobile communication;cellular technology;extreme base station;device density;antennas;5G air interface;LTE;Wi-Fi;universal high-rate coverage;seamless user experience;spectrum regulation;cost efficiency;energy efficiency;core network;Special issues and sections;Macrocell networks;Wireless communication;IEEE 802.11 Standards;Energy efficiency;Bandwidth;Millimeter wave technology;MIMO;Mobile communication;Cellular systems;energy efficiency;HetNets;massive MIMO;millimeter wave;small cells},
	Month = {June},
	Number = {6},
	Pages = {1065-1082},
	Title = {What Will 5G Be?},
	Volume = {32},
	Year = {2014},
	Bdsk-Url-1 = {https://doi.org/10.1109/JSAC.2014.2328098}}

@article{channel_model,
	Author = {3GPP, R1-161636, NTT DOCOMO, ATT,CMCC, Ericsson, Huawei, HiSilicon, Intel, KT Corporation, Nokia Networks, Qualcomm,Samsung},
	Month = {Mar},
	Title = {Updated White Paper on Channel Modelling for Bands up to 100 GHz},
	Year = {2016}}

@article{PTRS_samsung,
	Archiveprefix = {arXiv},
	Author = {Yinan Qi and Mythri Hunukumbure and Hyungju Nam and Hyunil Yoo and SaiDhiraj Amuru},
	Bibsource = {dblp computer science bibliography, https://dblp.org},
	Biburl = {https://dblp.org/rec/bib/journals/corr/abs-1807-07336},
	Eprint = {1807.07336},
	Journal = {CoRR},
	Timestamp = {Mon, 13 Aug 2018 16:47:42 +0200},
	Title = {On the Phase Tracking Reference Signal {(PT-RS)} Design for 5G New Radio {(NR)}},
	Url = {http://arxiv.org/abs/1807.07336},
	Volume = {abs/1807.07336},
	Year = {2018},
	Bdsk-Url-1 = {http://arxiv.org/abs/1807.07336}}

@article{DolevYao,
	Author = {D. Dolev and A. Yao},
	Journal = {IEEE Transactions on Information Theory},
	Number = {2},
	Pages = {198-208},
	Title = {{On the security of public key protocols}},
	Volume = {29},
	Year = {1983}}

@misc{5G_V2X,
	Howpublished = {\url{https://www.5gamericas.org/wp-content/uploads/2019/07/2018_5G_Americas_White_Paper_Cellular_V2X_Communications_Towards_5G__Final_for_Distribution.pdf}},
	Key = {5G_V2X},
	Lastvisited = {16.06.2020},
	Note = {[Online; Accessed 16. June 2020]},
	Title = {{5G Americas Whitepaper Cellular V2X Communications towards 5G}}}

@misc{5G_slots,
	Author = {Ericsson},
	Howpublished = {\url{https://www.ericsson.com/assets/local/publications/ericsson-technology-review/docs/2017/designing-for-the-future---the-5g-nr-physical-layer.pdf}},
	Key = {5G_slots},
	Lastvisited = {16.06.2020},
	Note = {[Online; Accessed 16. June 2020]},
	Title = {{5G New Radio: Designing for the future}}}

@misc{ESA_5G_GPS,
	Howpublished = {\url{ https://www.esa.int/Applications/Navigation/ESA_leads_drive_into_our_5G_positioning_future}},
	Key = {ESA_5G_GPS},
	Lastvisited = {16.06.2020},
	Note = {[Online; Accessed June 16, 2020]},
	Title = {{Hybrid 5G and GPS}}}

@misc{3GPP,
	Howpublished = {\url{https://www.3gpp.org/ftp/Specs/archive/38_series/38.211/}},
	Key = {3GPP},
	Lastvisited = {17.06.2020},
	Note = {[Online; Accessed 1. November 2022]},
	Title = {{3GPP}}}

@misc{Release13_LTE_LPP,
	Howpublished = {\url{https://www.etsi.org/deliver/etsi_ts/136300_136399/136355/13.00.00_60}},
	Key = {Release13_LTE_LPP},
	Lastvisited = {12.01.2021},
	Note = {[Online; Accessed 12. January 2021]},
	Title = {{LTE Positioning Protocol (LPP)}}}


@misc{LTE_Nutshell,
	Howpublished = {\url{https://home.zhaw.ch/kunr/NTM1/literatur/LTE%20in%20a%20Nutshell%20-%20Physical%20Layer.pdf}},
	Key = {Release13_LTE_LPP},
	Lastvisited = {12.01.2021},
	Note = {[Online; Accessed 12. January 2021]},
	Title = {{LTE in a Nutshell}}}




@inproceedings{RADAR_RSSI,
	Author = {P. Bahl and V. N. Padmanabhan},
	Booktitle = {IEEE INFOCOM},
	Pages = {775-784},
	Title = {{RADAR: an in-building RF-based user location and tracking system}},
	Volume = {2},
	Year = {2000}}

@inproceedings{VerifiableMultilateration,
	Author = {S. \v{C}apkun and J. Hubaux},
	Booktitle = {IEEE Computer and Communications Societies.},
	Pages = {1917-1928},
	Title = {Secure positioning of wireless devices with application to sensor networks},
	Volume = {3},
	Year = {2005}}

@misc{5G_slotsOLD2,
	Title = {{Whitepaper on New Localization Methods for 5G Wireless Systems and the Internet-of-Things, COST Action CA15104, IRACON ; Pedersen, Troels; Fleury, Bernard Henri}}}



@article{mm_wave_lidar_cost,
	Author = {F. {Guidi} and A. {Guerra} and D. {Dardari}},
	Doi = {10.1109/TMC.2015.2467373},
	Issn = {1536-1233},
	Journal = {IEEE Transactions on Mobile Computing},
	Keywords = {5G mobile communication;Bayes methods;indoor radio;millimetre wave antenna arrays;mobile antennas;quantisation (signal);radar antennas;personal mobile radars;millimeter-wave massive arrays;indoor mapping;millimeter-wave technology;5G user mobile devices;frequency selectivity;phase quantization effects;environmental information;grid-based Bayesian mapping approach;state-space model;massive antenna array characteristics;antenna elements;Radar antennas;Radar;Distance measurement;Manganese;Phased arrays;Bandwidth;Personal radar;indoor mapping;millimeter-wave;massive antenna arrays;Personal radar;indoor mapping;millimeter-wave;massive antenna arrays},
	Month = {June},
	Number = {6},
	Pages = {1471-1484},
	Title = {Personal Mobile Radars with Millimeter-Wave Massive Arrays for Indoor Mapping},
	Volume = {15},
	Year = {2016},
	Bdsk-Url-1 = {https://doi.org/10.1109/TMC.2015.2467373}}

@inproceedings{5G_AoA_TDoA,
	Author = {Anastasios Kakkavas and Mario Hern{\'{a}}n Casta{\~{n}}eda Garc{\'{\i}}a and Richard A. Stirling{-}Gallacher and Josef A. Nossek},
	Bibsource = {dblp computer science bibliography, https://dblp.org},
	Biburl = {https://dblp.org/rec/bib/conf/globecom/KakkavasGSN18},
	Booktitle = {{IEEE} Global Communications Conference, {GLOBECOM} 2018, Abu Dhabi, United Arab Emirates, December 9-13, 2018},
	Doi = {10.1109/GLOCOM.2018.8647812},
	Pages = {206--212},
	Timestamp = {Tue, 26 Feb 2019 16:24:13 +0100},
	Title = {Multi-Array 5G {V2V} Relative Positioning: Performance Bounds},
	Year = {2018},
	Bdsk-Url-1 = {https://doi.org/10.1109/GLOCOM.2018.8647812}}

@article{5G_multipath,
	Author = {K. {Witrisal} and P. {Meissner} and E. {Leitinger} and Y. {Shen} and C. {Gustafson} and F. {Tufvesson} and K. {Haneda} and D. {Dardari} and A. F. {Molisch} and A. {Conti} and M. Z. {Win}},
	Doi = {10.1109/MSP.2015.2504328},
	Issn = {1053-5888},
	Journal = {IEEE Signal Processing Magazine},
	Keywords = {5G mobile communication;assisted living;handicapped aids;multipath channels;radionavigation;sensor placement;assisted living technology;multipath channels;AL technology;Internet of Things;high-accuracy localization systems;location information;caretaking cost reduction;total cost of ownership;sensor systems;radio-based indoor localization system;reflected multipath components;MPCs;millimeter-wave technology;mm-wave technology;5G communications systems;centimeter-accuracy indoor localization;aging society;propagation environment;Interference;Signal to noise ratio;Robustness;Delays;Covariance matrices;Mathematical model;Assisted living;Assistive technology},
	Month = {March},
	Number = {2},
	Pages = {59-70},
	Title = {High-Accuracy Localization for Assisted Living: 5G systems will turn multipath channels from foe to friend},
	Volume = {33},
	Year = {2016},
	Bdsk-Url-1 = {https://doi.org/10.1109/MSP.2015.2504328}}

@article{mm_wave_waveform,
	Author = {X. {Cui} and T. A. {Gulliver} and H. {Song} and J. {Li}},
	Doi = {10.1109/ACCESS.2016.2604360},
	Issn = {2169-3536},
	Journal = {IEEE Access},
	Keywords = {5G mobile communication;millimetre wave devices;neural nets;radio equipment;real-time systems;millimeter wave device to device communications;mobile wireless communication networks;real-time positioning applications;5G cellular networks;threshold selection algorithm;mmWave waveforms;artificial neural network;Distance measurement;5G mobile communication;Real-time systems;Artificial neural networks;Millimeter wave communication;Wireless communication;Millimeter wave devices;Heuristic algorithms;Position measurement;millimeter wave technology;artificial neural networks;ranging;device to device;5G},
	Pages = {5520-5530},
	Title = {Real-Time Positioning Based on Millimeter Wave Device to Device Communications},
	Volume = {4},
	Year = {2016},
	Bdsk-Url-1 = {https://doi.org/10.1109/ACCESS.2016.2604360}}




@inproceedings{mm_wave_multipath,
	Author = {Z. {Abu-Shaban} and X. {Zhou} and T. {Abhayapala} and G. {Seco-Granados} and H. {Wymeersch}},
	Booktitle = {2018 IEEE Wireless Communications and Networking Conference (WCNC)},
	Doi = {10.1109/WCNC.2018.8376990},
	Issn = {1558-2612},
	Keywords = {5G mobile communication;millimetre wave communication;multipath channels;next generation networks;orientation estimation;mobile communications;location-aware communication systems;millimeter-wave technology;single-anchor localization limits;mmWave multipath channels;uplink localization;orientation angle;sub-meter position error;sub-degree orientation error;5G mmWave systems;Uplink;Downlink;Antenna arrays;Three-dimensional displays;5G mobile communication;Two dimensional displays;Conferences},
	Month = {April},
	Pages = {1-6},
	Title = {Performance of location and orientation estimation in 5G mmWave systems: Uplink vs downlink},
	Year = {2018},
	Bdsk-Url-1 = {https://doi.org/10.1109/WCNC.2018.8376990}}

@inproceedings{ofdm_comm_pos,
	Author = {R. {Montalban} and J. A. {L{\'o}pez-Salcedo} and G. {Seco-Granados} and A. L. {Swindlehurst}},
	Booktitle = {2013 IEEE 14th Workshop on Signal Processing Advances in Wireless Communications (SPAWC)},
	Doi = {10.1109/SPAWC.2013.6612139},
	Issn = {1948-3244},
	Keywords = {channel estimation;data communication;OFDM modulation;power allocation approaches;combined positioning-communications OFDM systems;data power allocations;multicarrier OFDM signals;high-data-rate communications systems;capacity-maximizing pilot;equi-powered pilot structures;equi-spaced structures;channel estimation;time-delay estimation;data power distributions;equi-powered pilot structures;equi-spaced pilot structures;OFDM;Estimation;Channel estimation;Accuracy;Vectors;Power distribution;Signal processing},
	Month = {June},
	Pages = {694-698},
	Title = {Power allocation approaches for combined positioning and communications OFDM systems},
	Year = {2013},
	Bdsk-Url-1 = {https://doi.org/10.1109/SPAWC.2013.6612139}}

@inproceedings{offset_60GHz,
	Author = {L. {Koschel} and A. {Kortke}},
	Booktitle = {2012 IEEE 23rd International Symposium on Personal, Indoor and Mobile Radio Communications - (PIMRC)},
	Doi = {10.1109/PIMRC.2012.6362736},
	Issn = {2166-9589},
	Keywords = {field programmable gate arrays;MIMO communication;numerical analysis;OFDM modulation;phase noise;synchronisation;frequency synchronization;phase offset tracking;real-time CS-OFDM MIMO system;orthogonal frequency-division-multiplexing;carrier frequency impairment;phase noise;PN;frequency synchronization scheme;phase offset tracking scheme;non-line-of-sight;wireless communication system;NLOS;carrier frequency offset estimation;CFO estimation;numerical investigation;residual frequency offset;RFO;common phase error;CPE;Altera Startix III FPGA;frequency 60 GHz;OFDM;Signal to noise ratio;Estimation;Antennas;Bit error rate;Oscillators;Receivers;OFDM;60GHz;CFO;RFO;CPE;phase noise;FPGA;Walsh-Hadamard;Alamouti},
	Month = {Sep.},
	Pages = {2281-2286},
	Title = {Frequency synchronization and phase offset tracking in a real-time 60-GHz CS-OFDM MIMO system},
	Year = {2012},
	Bdsk-Url-1 = {https://doi.org/10.1109/PIMRC.2012.6362736}}

@inproceedings{GPS_Spoofing1,
	Author = {Todd E. Humphreys},
	Booktitle = {Institute of Navigation GNSS (ION GNSS)},
	Title = {Assessing the Spoofing Threat: Development of a Portable GPS Civilian Spoofer},
	Year = {2008}}


	@inproceedings{Aanjhan_Spree,
 author = {Ranganathan, Aanjhan and \'{O}lafsd\'{o}ttir, Hildur and Capkun, Srdjan},
 title = {SPREE: A Spoofing Resistant GPS Receiver},
 booktitle = {Proceedings of the 22nd Annual International Conference on Mobile Computing and Networking},
 series = {MobiCom '16},
 year = {2016},
 publisher = {ACM},
} 


@inproceedings{Nils2011requirements,
	author = {Tippenhauer, Nils Ole and P{\"o}pper, Christina and Rasmussen, Kasper Bonne and Capkun, Srdjan},
	booktitle = {Proceedings of the 18th ACM conference on Computer and communications security},
	organization = {ACM},
	pages = {75--86},
	title = {On the requirements for successful GPS spoofing attacks},
	year = 2011}


	@INPROCEEDINGS{GPS_aanjhan,
  author={Narain, Sashank and Ranganathan, Aanjhan and Noubir, Guevara},
  booktitle={2019 IEEE Symposium on Security and Privacy (SP)}, 
  title={Security of GPS/INS Based On-road Location Tracking Systems}, 
  year={2019},
  volume={},
  number={},
  pages={587-601},
  doi={10.1109/SP.2019.00068}}


	@article{GPS_harshad,
  author    = {Harshad Sathaye and
               Gerald LaMountain and
               Pau Closas and
               Aanjhan Ranganathan},
  title     = {SemperFi: {A} Spoofer Eliminating {GPS} Receiver for UAVs},
  journal   = {CoRR},
  volume    = {abs/2105.01860},
  year      = {2021},
  url       = {https://arxiv.org/abs/2105.01860},
  archivePrefix = {arXiv},
  eprint    = {2105.01860},
  timestamp = {Wed, 12 May 2021 15:54:31 +0200},
  biburl    = {https://dblp.org/rec/journals/corr/abs-2105-01860.bib},
  bibsource = {dblp computer science bibliography, https://dblp.org}
}

@inproceedings{Spoofing2,
	Author = {K. Bauer and D. McCoy and E. Anderson and M. Breitenbach and G. Grudic and D. Grunwald and D. Sicker},
	Booktitle = {IEEE GLOBECOM},
	Pages = {1-6},
	Title = {{The Directional Attack on Wireless Localization -or- How to Spoof Your Location with a Tin Can}},
	Year = {2009}}

@inproceedings{ICodes,
	Author = {M. Cagalj and S. \v{C}apkun and R. Rengaswamy and I. Tsigkogiannis and M. Srivastava and J. Hubaux},
	Booktitle = {IEEE Symposium on Security and Privacy (S\&P)},
	Pages = {15 pp.-294},
	Title = {{Integrity (I) codes: message integrity protection and authentication over insecure channels}},
	Year = {2006}}

@inproceedings{ICodes_Gollakota,
	Author = {Gollakota, Shyamnath and Ahmed, Nabeel and Zeldovich, Nickolai and Katabi, Dina},
	Booktitle = {USENIX Security Symposium},
	Title = {Secure In-band Wireless Pairing},
	Year = {2011}}


@inproceedings{SecNav_Kasper,
	Author = {Kasper Bonne Rasmussen and Srdjan Capkun and Mario Cagalj},
	Booktitle = {MobiCom},
	Title = {SecNav: secure broadcast localization and time synchronization in wireless networks},
	Year = {2007}}



@article{OFDM_CSI_ToF,
	Author = {N. {Tadayon} and M. T. {Rahman} and S. {Han} and S. {Valaee} and W. {Yu}},
	Doi = {10.1109/TWC.2019.2914194},
	Journal = {IEEE Transactions on Wireless Communications},
	Keywords = {channel estimation;Global Positioning System;MIMO communication;OFDM modulation;signal classification;signal resolution;wireless channels;wireless LAN;CSI;contamination;channel state information;multiple-input-multiple-output orthogonal frequency-division multiplexing;MIMO-OFDM wireless local area network receivers;decimeter ranging;ToF estimation;WLAN receiver;MUSIC superresolution algorithm;time-of-flight estimation;range estimation;ToF-based;size 0.7 m;size 0.8 m;size 0.9 m;distance 5.0 m;distance 15.0 m;distance 10.0 m;Receivers;Estimation;Distance measurement;Channel estimation;Transmitters;Wireless fidelity;Radio frequency;Indoor positioning;MIMO;OFDM;CSI;calibration},
	Month = {July},
	Number = {7},
	Pages = {3453-3468},
	Title = {Decimeter Ranging With Channel State Information},
	Volume = {18},
	Year = {2019},
	Bdsk-Url-1 = {https://doi.org/10.1109/TWC.2019.2914194}}

@misc{mercedes_theft,
	Howpublished = {\url{http://www.bbc.com/news/uk-england-birmingham-42132689}},
	Lastvisited = {'15.06.2020},
	Note = {[Online; Accessed 15. June 2020]},
	Title = {"Mercedes 'relay' box thieves caught on CCTV in Solihull."}}

@inproceedings{Boris_cars,
	Author = {Francillon,, Aur\'elien and Danev, Boris and Capkun,, Srdjan},
	Booktitle = {Network and Distributed System Security Symposium (NDSS)},
	Title = {Relay Attacks on Passive Keyless Entry and Start Systems in Modern Cars},
	Year = {2011}}

@misc{HANCKE-NFC-RELAY,
	Author = {Lishoy Francis and Gerhard Hancke and Keith Mayes and Konstantinos Markantonakis},
	Title = {Practical Relay Attack on Contactless Transactions by Using NFC Mobile Phones},
	Year = {2012}}

@misc{relaySetup,
	Howpublished = {\url{https://www.thesun.co.uk/motors/7804489/keyless-car-100-ebay-gadgets-relay-attacks/}},
	Lastvisited = {'24.09.19},
	Note = {[Online; Accessed 1. Feb 2021]},
	Title = {{Relay Setup}}}



@misc{mini_circuits,
	Howpublished = {\url{https://www.minicircuits.com}},
	Lastvisited = {'29.11.19},
	Note = {[Online; Accessed 29. November 2019]},
	Title = {{Mini Circuits}}}


@misc{VectorSignalGenerator,
	Howpublished = {\url{https://www.rohde-schwarz.com/us/manual/r-s-smu200a-vector-signal-generator-operating-manual-manuals-gb1_78701-28893.html}},
	Lastvisited = {'29.11.19},
	Note = {[Online; Accessed 29. November 2019]},
	Title = {{Vector Signal Generator}}}

@misc{USRP_X310,
	Howpublished = {\url{https://www.ettus.com/all-products/x310-kit}},
	Lastvisited = {'29.11.19},
	Note = {[Online; Accessed 29. November 2019]},
	Title = {{USRP X310}}}

@misc{USRP_B210,
	Howpublished = {\url{https://www.ettus.com/all-products/usrp-b200mini-i-2/}},
	Lastvisited = {'10.11.2020},
	Note = {[Online; Accessed 10. November 2020]},
	Title = {{USRP B210}}}


@misc{NXP_article,
	Howpublished = {\url{https://www.rfidjournal.com/articles/view?18936}},
	Lastvisited = {'29.11.19},
	Note = {[Online; Accessed 29. November 2019]},
	Title = {{NXP Offers UWB Fine-Ranging Chipset for Mobile Devices}}}

@inproceedings{Overshadow_Usenix,
	Address = {Santa Clara, CA},
	Author = {Hojoon Yang and Sangwook Bae and Mincheol Son and Hongil Kim and Song Min Kim and Yongdae Kim},
	Booktitle = {28th {USENIX} Security Symposium ({USENIX} Security 19)},
	Isbn = {978-1-939133-06-9},
	Month = aug,
	Pages = {55--72},
	Publisher = {{USENIX} Association},
	Title = {Hiding in Plain Signal: Physical Signal Overshadowing Attack on {LTE}},
	Url = {https://www.usenix.org/conference/usenixsecurity19/presentation/yang-hojoon},
	Year = {2019},
	Bdsk-Url-1 = {https://www.usenix.org/conference/usenixsecurity19/presentation/yang-hojoon}}



@misc{OpenAir5G,
	Howpublished = {\url{https://www.openairinterface.org}},
	Lastvisited = {'16.10.19},
	Note = {[Online; Accessed 16. October 2019]},
	Title = {{Open Air Interface}}}

@inproceedings{MTAC,
	Author = {P. {Leu} and M. {Singh} and M. {Roeschlin} and K. G. {Paterson} and S. {{\v C}apkun}},
	Booktitle = {2020 IEEE Symposium on Security and Privacy (SP)},
	Doi = {10.1109/SP40000.2020.00010},
	Pages = {500-516},
	Title = {Message Time of Arrival Codes: A Fundamental Primitive for Secure Distance Measurement},
	Year = {2020},
	Bdsk-Url-1 = {https://doi.org/10.1109/SP40000.2020.00010}}



@article{mm-wave_delayspread,
	Author = {V. {Raghavan} and A. {Partyka} and L. {Akhoondzadeh-Asl} and M. A. {Tassoudji} and O. H. {Koymen} and J. {Sanelli}},
	Doi = {10.1109/TAP.2017.2758198},
	Issn = {1558-2221},
	Journal = {IEEE Transactions on Antennas and Propagation},
	Keywords = {electromagnetic wave reflection;indoor radio;microwave antenna arrays;microwave measurement;millimetre wave antenna arrays;millimetre wave measurement;millimetre wave propagation;signal detection;UHF antennas;UHF measurement;wireless channels;PHY layer design;millimeter wave frequency regime;next-generation wireless systems;mmW channel properties;mmW systems;physical layer design;simultaneous channel sounding measurements;indoor office;shopping mall;outdoor environments;path loss;outdoor-to-indoor coverage;material measurements;system design;millimeter wave channel measurements;transmit-receive location pairs;large-scale properties;delay spread;mmW reflection;mmW penetration;signal reception;frequency 2.9 GHz;frequency 29.0 GHz;frequency 61.0 GHz;Antenna measurements;Frequency measurement;Buildings;Loss measurement;Receivers;Transmitters;Delays;Beamforming;channel modeling;delay spread;millimeter wave (mmW) systems;path loss;penetration;reflection;system design},
	Month = {Dec},
	Number = {12},
	Pages = {6521-6533},
	Title = {Millimeter Wave Channel Measurements and Implications for PHY Layer Design},
	Volume = {65},
	Year = {2017},
	Bdsk-Url-1 = {https://doi.org/10.1109/TAP.2017.2758198}}

@inproceedings{XILINX_5G,
	Author = {Sassan Ahmadi},
	Title = {Toward 5 G Xilinx Solutions and Enablers for Next-Generation Wireless Systems},
	Year = {2016}}

@article{CFO_survey,
	Archiveprefix = {arXiv},
	Author = {Ali A. Nasir and Salman Durrani and Hani Mehrpouyan and Steven D. Blostein and Rodney A. Kennedy},
	Bibsource = {dblp computer science bibliography, https://dblp.org},
	Biburl = {https://dblp.org/rec/bib/journals/corr/NasirDMBK15},
	Eprint = {1507.02032},
	Journal = {CoRR},
	Timestamp = {Mon, 13 Aug 2018 16:46:30 +0200},
	Title = {Timing and Carrier Synchronization in Wireless Communication Systems: {A} Survey and Classification of Research in the Last Five Years},
	Url = {http://arxiv.org/abs/1507.02032},
	Volume = {abs/1507.02032},
	Year = {2015},
	Bdsk-Url-1 = {http://arxiv.org/abs/1507.02032}}

@unknown{v2X_security_survey,
	Author = {Ghosal, Amrita and Conti, Mauro},
	Doi = {10.13140/RG.2.2.10823.55200},
	Month = {03},
	Title = {Security Issues and Challenges in V2X: A Survey},
	Year = {2019},
	Bdsk-Url-1 = {https://doi.org/10.13140/RG.2.2.10823.55200}}

@misc{3GPP_38.855,
	Howpublished = {\url{https://www.3gpp.org/ftp/Specs/archive/38_series/38.855/}},
	Lastvisited = {'17.06.2020},
	Note = {[Online; Accessed 1. November 2022]},
	Title = {{5G - GPP 38.855 ;Technical Specification Group Radio Access Network; Study on NR positioning support}}}





@misc{mm-wavesetup,
	Howpublished = {\url{https://www.highfrequencyelectronics.com/index.php?option=com_content&view=article&id=1994:affordable-solutions-for-testing-28-ghz-5g-devices-with-your-6-ghz-lab-instrumentation&catid=167&Itemid=189}},
	Lastvisited = {'17.06.2020},
	Note = {[Online; Accessed 1. November 2022]},
	Title = {{mm-wave Setup}}}

@inproceedings{TDOA_80211ac,
	Author = {A. {Gaber} and A. {Omar}},
	Booktitle = {2012 Ubiquitous Positioning, Indoor Navigation, and Location Based Service (UPINLBS)},
	Pages = {1-8},
	Title = {A study of TDOA estimation using Matrix Pencil algorithms and IEEE 802.11ac},
	Year = {2012}}

@misc{TS36104,
	Howpublished = {\url{https://www.3gpp.org/ftp//Specs/archive/36_series/36.104/}},
	Lastvisited = {'12.12.20},
	Note = {[Online; Accessed 1. November 2022]},
	Title = {{TS-36.104}}}

@misc{MATLAB_PRS,
	Howpublished = {\url{https://www.mathworks.com/help/lte/ug/time-difference-of-arrival-positioning-using-prs.html}},
	Lastvisited = {'12.12.20},
	Note = {[Online; Accessed 12. December 2020]},
	Title = {{MATLAB PRS}}}


@misc{MATLAB_lte_toolbox,
	Howpublished = {\url{https://www.mathworks.com/products/lte.html}},
	Lastvisited = {'12.12.20},
	Note = {[Online; Accessed 12. December 2020]},
	Title = {{MATLAB LTE Toolbox}}}



@misc{srsLTE,
	Howpublished = {\url{https://github.com/srsLTE/srsLTE}},
	Lastvisited = {'12.12.20},
	Note = {[Online; Accessed 12. December 2020]},
	Title = {{srsLTE}}}

@inproceedings{GPS_attack,
	Address = {Baltimore, MD},
	Author = {Kexiong (Curtis) Zeng and Shinan Liu and Yuanchao Shu and Dong Wang and Haoyu Li and Yanzhi Dou and Gang Wang and Yaling Yang},
	Booktitle = {27th {USENIX} Security Symposium ({USENIX} Security 18)},
	Isbn = {978-1-939133-04-5},
	Month = aug,
	Pages = {1527--1544},
	Publisher = {{USENIX} Association},
	Title = {All Your {GPS} Are Belong To Us: Towards Stealthy Manipulation of Road Navigation Systems},
	Url = {https://www.usenix.org/conference/usenixsecurity18/presentation/zeng},
	Year = {2018},
	Bdsk-Url-1 = {https://www.usenix.org/conference/usenixsecurity18/presentation/zeng}}



@inproceedings{yu2012performance,
	Author = {Yu, Lei and Laaraiedh, Mohamed and Avrillon, St{\'e}phane and Uguen, Bernard and Keignart, Julien and Stephan, Julien},
	Booktitle = {European Wireless 2012; 18th European Wireless Conference 2012},
	Organization = {VDE},
	Pages = {1--7},
	Title = {Performance evaluation of threshold-based TOA estimation techniques using IR-UWB indoor measurements},
	Year = {2012}}

@article{channelmodel,
	Author = {Molisch, Andreas F and Balakrishnan, Kannan and Chong, Chia-Chin and Emami, Shahriar and Fort, Andrew and Karedal, Johan and Kunisch, Juergen and Schantz, Hans and Schuster, Ulrich and Siwiak, Kai},
	Journal = {IEEE P802},
	Number = {04},
	Pages = {0662},
	Publisher = {Citeseer},
	Title = {IEEE 802.15. 4a channel model-final report},
	Volume = {15},
	Year = {2004}}

@misc{AppleU1,
	Lastvisited = {22.03.2021},
	Note = {[Online; Accessed 24. March 2021]},
	Title = {{Apple U1 UWBChip}, howpublished="\url{https://support.apple.com/guide/security/ultra-wideband-security-sec1e6108efd/web"}}}



@misc{NXPTrimension,
	Howpublished = {\url{https://www.nxp.com/docs/en/fact-sheet/UWB-IOT-FS.pdf}},
	Lastvisited = {'25.03.2021},
	Note = {[Online; Accessed 25. March 2021]},
	Title = {{NXP Trimension}}}





@misc{MicrochipLRP,
	Howpublished = {\url{https://www.microchip.com/wwwproducts/en/ATA8352}},
	Lastvisited = {'25.03.2021},
	Note = {[Online; Accessed 25. March 2021]},
	Title = {{Microchip ATA8532}}}

@misc{SamsungArticle,
	Howpublished = {\url{https://news.samsung.com/global/samsung-expects-uwb-to-be-one-of-the-next-big-wireless-technologies/}},
	Lastvisited = {'22.03.2021},
	Note = {[Online; Accessed 24. March 2021]},
	Title = {{SamsungUWB}}}

@misc{volkswagenArticle,
	Howpublished = {\url{https://www.volkswagen-newsroom.com/en/stories/realtime-safety-with-uwb-5438}},
	Lastvisited = {'24.03.2021},
	Note = {[Online; Accessed 20. March 2021]},
	Title = {{Volkswagen UWB PKES}}}

@misc{UWBSocialDistance,
	Howpublished = {\url{https://www.uwb-social-distancing.com/}},
	Lastvisited = {'24.03.2021},
	Note = {[Online; Accessed 22. March 2021]},
	Title = {{UWB Social Distancing}}}


@misc{MeeblueSocialDistance,
	Howpublished = {\url{https://www.meeblue.com/blogs/UWB_For_Social_Alert/}},
	Lastvisited = {'24.03.2021},
	Note = {[online; Accessed 20. March 2021]},
	Title = {{UWB Social Distancing Meeblue}}}


@inproceedings{RSSI,
	Author = {Bahl, Paramvir and Padmanabhan, Venkata N},
	Booktitle = {Proceedings IEEE INFOCOM 2000. Conference on computer communications. Nineteenth annual joint conference of the IEEE computer and communications societies (Cat. No. 00CH37064)},
	Organization = {Ieee},
	Pages = {775--784},
	Title = {RADAR: An in-building RF-based user location and tracking system},
	Volume = {2},
	Year = {2000}}

@article{multicarrier,
	Author = {Huo, Kai and Deng, Bin and Liu, Yongxiang and Jiang, Weidong and Mao, Junjie},
	Journal = {Journal of Systems Engineering and Electronics},
	Number = {3},
	Pages = {421--427},
	Publisher = {BIAI},
	Title = {High resolution range profile analysis based on multicarrier phase-coded waveforms of OFDM radar},
	Volume = {22},
	Year = {2011}}

@article{wifitof,
	Author = {Golden, Stuart A and Bateman, Steve S},
	Journal = {IEEE Transactions on Mobile Computing},
	Number = {10},
	Pages = {1185--1198},
	Publisher = {IEEE},
	Title = {Sensor measurements for Wi-Fi location with emphasis on time-of-arrival ranging},
	Volume = {6},
	Year = {2007}}

@article{frequencymodulatedcontiniouswaveranging,
	Author = {Hymans, AJ and Lait, J},
	Journal = {Proceedings of the IEE-Part B: electronic and communication engineering},
	Number = {34},
	Pages = {365--372},
	Publisher = {IET},
	Title = {Analysis of a frequency-modulated continuous-wave ranging system},
	Volume = {107},
	Year = {1960}}

@article{compagno2016modeling,
	Author = {Compagno, Alberto and Conti, Mauro and D'Amico, Antonio Alberto and Dini, Gianluca and Perazzo, Pericle and Taponecco, Lorenzo},
	Journal = {IEEE Transactions on Information Forensics and Security},
	Number = {7},
	Pages = {1565--1577},
	Publisher = {IEEE},
	Title = {Modeling enlargement attacks against UWB distance bounding protocols},
	Volume = {11},
	Year = {2016}}

@article{ContactTracing1,
	Author = {Ferretti, Luca and Wymant, Chris and Kendall, Michelle and Zhao, Lele and Nurtay, Anel and Abeler-D{\"o}rner, Lucie and Parker, Michael and Bonsall, David and Fraser, Christophe},
	Doi = {10.1126/science.abb6936},
	Issn = {0036-8075},
	Journal = {Science (New York, N.Y.)},
	Month = {May},
	Number = {6491},
	Title = {Quantifying SARS-CoV-2 transmission suggests epidemic control with digital contact tracing},
	Volume = {368},
	Year = {2020},
	Bdsk-Url-1 = {https://doi.org/10.1126/science.abb6936}}

@misc{Contactles_payments_trendUS,
	Howpublished = {\url{https://www.forbes.com/sites/jordanmckee/2020/10/12/covid-19-is-changing-consumer-behavior-at-the-point-of-sale/?sh=7dc44adf375d}},
	Lastvisited = {'18.04.2021},
	Note = {[Online; Accessed 18. April 2021]},
	Title = {{Contactless Payments Trend US}}}

@misc{GPS_spoofing_HackRF,
	Howpublished = {\url{https://drfone.wondershare.com/fake-location/gps-spoofing-with-hackrf-from-windows.html}},
	Lastvisited = {'18.08.2023},
	Note = {[Online; Accessed 18. August 2023]},
	Title = {{How to Spoof GPS Location with HackRF}}}




@misc{tee,
	Howpublished = {\url{https://www.trustonic.com/technical-articles/what-is-a-trusted-execution-environment-tee/}},
	Lastvisited = {'18.04.2021},
	Note = {[Online; Accessed 18. April 2021]},
	Title = {{What is a Trusted Execution Environment (TEE)?}}}

@misc{3gpp_seal,
	Howpublished = {\url{https://www.etsi.org/deliver/etsi_ts/123400_123499/123434/16.04.00_60/ts_123434v160400p.pdf}},
	Lastvisited = {'18.04.2021},
	Note = {[Online; Accessed 18. April 2021]},
	Title = {{Service Enabler Architecture Layer for Verticals (SEAL); Functional architecture and information flows (3GPP TS 23.434 version 16.4.0 Release 16)}}}
 



@misc{Contactles_payments_behaviour,
	Howpublished = {\url{https://about.americanexpress.com/all-news/news-details/2020/COVID-19-is-Shifting-Consumer-Purchasing-Behavior-and-Driving-U.S.-Interest-in-Contactless-Payments-According-to-2020-American-Express-Digital-Payments-Survey/default.aspx}},
	Lastvisited = {'18.04.2021},
	Note = {[Online; Accessed 18. April 2021]},
	Title = {{Contactless Payments Trend}}}

@misc{Contactles_payments_Limit_increase,
	Howpublished = {\url{https://www.nfcw.com/2020/03/26/366173/table-contactless-payment-transaction-limit-increases-around-the-world/}},
	Lastvisited = {'18.04.2021},
	Note = {[Online; Accessed 18. April 2021]},
	Title = {{Contactless Payments Trend}}}

@misc{CovidAppDatabase,
	Howpublished = {\url{https://www.technologyreview.com/2020/05/07/1000961/launching-mittr-covid-tracing-tracker/}},
	Lastvisited = {'18.04.2021},
	Note = {[Online; Accessed 18. April 2021]},
	Title = {{COVID-19 Tracking Apps}}}


@misc{CovidAppDesign,
	Howpublished = {\url{https://news.mit.edu/2020/bluetooth-covid-19-contact-tracing-0409}},
	Lastvisited = {'18.04.2021},
	Note = {[Online; Accessed 18. April 2021]},
	Title = {{COVID-19 Apps Design}}}



@misc{CovidAppWiki,
	Howpublished = {\url{https://en.wikipedia.org/wiki/COVID-19_apps/}},
	Lastvisited = {'18.04.2021},
	Note = {[Online; Accessed 18. April 2021]},
	Title = {{COVID-19 Apps Wikipedia}}}

@misc{AV_perception,
	Howpublished = {\url{https://www.foley.com/en/insights/publications/2020/08/covid-19-adoption-autonomous-vehicle-technology}},
	Lastvisited = {'18.04.2021},
	Note = {[Online; Accessed 18. April 2021]},
	Title = {{The Impact of COVID-19 on Adoption of Autonomous Vehicle Technology}}}

@misc{Contactles_payments_growth,
	Howpublished = {\url{https://www.nfcw.com/2020/05/06/366460/contactless-payments-growth-rate-doubles-in-nordic-countries/}},
	Lastvisited = {'18.04.2021},
	Note = {[Online; Accessed 18. April 2021]},
	Title = {{Contactless Payments Growth}}}

@inproceedings{guo2005sequence,
	Author = {Guo, Fanglu and Chiueh, Tzi-cker},
	Booktitle = {International Workshop on Recent Advances in Intrusion Detection},
	Organization = {Springer},
	Pages = {309--329},
	Title = {Sequence number-based MAC address spoof detection},
	Year = {2005}}

@inproceedings{faria2006detecting,
	Author = {Faria, Daniel B and Cheriton, David R},
	Booktitle = {Proceedings of the 5th ACM workshop on Wireless security},
	Pages = {43--52},
	Title = {Detecting identity-based attacks in wireless networks using signalprints},
	Year = {2006}}

@manual{ieee80211e,
	Author = {IEEE Std 802.11e},
	Title = {Amendment 8: Medium Access Control ({MAC}) Quality of Service Enhancements},
	Year = 2005}

@inproceedings{tippenhauer2009attacks,
	Author = {Tippenhauer, Nils Ole and Rasmussen, Kasper Bonne and P{\"o}pper, Christina and {\v{C}}apkun, Srdjan},
	Booktitle = {Proceedings of the 7th international conference on Mobile systems, applications, and services},
	Pages = {29--40},
	Title = {Attacks on public WLAN-based positioning systems},
	Year = {2009}}

@inproceedings{rasmussen2008location,
	Author = {Rasmussen, Kasper Bonne and {\v{C}}apkun, Srdjan},
	Booktitle = {Proceedings of the 15th ACM conference on Computer and communications security},
	Pages = {149--160},
	Title = {Location privacy of distance bounding protocols},
	Year = {2008}}

@inproceedings{wibowo2009time,
	Author = {Wibowo, Sigit Basuki and Klepal, Martin and Pesch, Dirk},
	Booktitle = {Proceedings of the International Conference on Positioning and Context-Awareness (PoCA'09)},
	Title = {Time of flight ranging using off-the-self ieee802. 11 wifi tags},
	Year = {2009}}

@article{lanzisera2011radio,
	Author = {Lanzisera, Steven and Zats, David and Pister, Kristofer SJ},
	Journal = {IEEE Sensors Journal},
	Number = {3},
	Pages = {837--845},
	Publisher = {IEEE},
	Title = {Radio frequency time-of-flight distance measurement for low-cost wireless sensor localization},
	Volume = {11},
	Year = {2011}}

@inproceedings{ranganathan2012design,
	Author = {Ranganathan, Aanjhan and Tippenhauer, Nils Ole and {\v{S}}kori{\'c}, Boris and Singel{\'e}e, Dave and {\v{C}}apkun, Srdjan},
	Booktitle = {European Symposium on Research in Computer Security},
	Organization = {Springer},
	Pages = {415--432},
	Title = {Design and implementation of a terrorist fraud resilient distance bounding system},
	Year = {2012}}

@article{tippenhauer2012uwb,
	Author = {Tippenhauer, Nils Ole and Capkun, Srdjan},
	Journal = {Technical report/ETH Z{\"u}rich, Department of Computer Science},
	Publisher = {ETH Zurich},
	Title = {UWB-based secure ranging and localization},
	Volume = {586},
	Year = {2012}}

@inproceedings{marcaletti2014filtering,
	Author = {Marcaletti, Andreas and Rea, Maurizio and Giustiniano, Domenico and Lenders, Vincent and Fakhreddine, Aymen},
	Booktitle = {Proceedings of the 10th ACM International on Conference on emerging Networking Experiments and Technologies},
	Organization = {ACM},
	Pages = {13--20},
	Title = {Filtering noisy 802.11 time-of-flight ranging measurements},
	Year = {2014}}

@inproceedings{vanhoef2014advanced,
	Author = {Vanhoef, Mathy and Piessens, Frank},
	Booktitle = {Proceedings of the 30th Annual Computer Security Applications Conference},
	Pages = {256--265},
	Title = {Advanced Wi-Fi attacks using commodity hardware},
	Year = {2014}}

@article{sharp2014indoor,
	Author = {Sharp, Ian and Yu, Kegen},
	Journal = {IEEE Transactions on Instrumentation and Measurement},
	Number = {9},
	Pages = {2129--2144},
	Publisher = {IEEE},
	Title = {Indoor TOA error measurement, modeling, and analysis},
	Volume = {63},
	Year = {2014}}

@article{yang2015wifi,
	Author = {Yang, Chouchang and Shao, Huai-Rong},
	Journal = {IEEE Communications Magazine},
	Number = {3},
	Pages = {150--157},
	Publisher = {IEEE},
	Title = {WiFi-based indoor positioning},
	Volume = {53},
	Year = {2015}}

@article{makki2015indoor,
	Author = {Makki, Ahmed and Siddig, Abubakr and Saad, Mohamed and Cavallaro, Joseph R and Bleakley, Chris J},
	Journal = {IEEE Transactions on Instrumentation and Measurement},
	Number = {3},
	Pages = {614--623},
	Publisher = {IEEE},
	Title = {Indoor localization using 802.11 time differences of arrival},
	Volume = {65},
	Year = {2015}}

@article{makki2015survey,
	Author = {Makki, Ahmed and Siddig, Abubakr and Saad, Mohamed and Bleakley, Chris},
	Journal = {Computer Networks},
	Pages = {218--233},
	Publisher = {Elsevier},
	Title = {Survey of WiFi positioning using time-based techniques},
	Volume = {88},
	Year = {2015}}

@inproceedings{kotaru2015spotfi,
	Author = {Kotaru, Manikanta and Joshi, Kiran and Bharadia, Dinesh and Katti, Sachin},
	Booktitle = {ACM SIGCOMM computer communication review},
	Number = {4},
	Organization = {ACM},
	Pages = {269--282},
	Title = {Spotfi: Decimeter level localization using wifi},
	Volume = {45},
	Year = {2015}}



@inproceedings{vanhoef2016mac,
	Author = {Vanhoef, Mathy and Matte, C{\'e}lestin and Cunche, Mathieu and Cardoso, Leonardo S and Piessens, Frank},
	Booktitle = {Proceedings of the 11th ACM on Asia Conference on Computer and Communications Security},
	Pages = {413--424},
	Title = {Why MAC address randomization is not enough: An analysis of Wi-Fi network discovery mechanisms},
	Year = {2016}}

@inproceedings{banin2016wifi,
	Author = {Banin, Leor and Schatzberg, Uri and Amizur, Yuval},
	Booktitle = {2016 International Conference on Indoor Positioning and Indoor Navigation (IPIN)},
	Title = {WiFi FTM and map information fusion for accurate positioning},
	Year = {2016}}

@article{robyns2017noncooperative,
	Author = {Robyns, Pieter and Bonn{\'e}, Bram and Quax, Peter and Lamotte, Wim},
	Journal = {Security and Communication Networks},
	Publisher = {Hindawi},
	Title = {Noncooperative 802.11 mac layer fingerprinting and tracking of mobile devices},
	Volume = {2017},
	Year = {2017}}

@inproceedings{ibrahim2018verification,
	Author = {Ibrahim, Mohamed and Liu, Hansi and Jawahar, Minitha and Nguyen, Viet and Gruteser, Marco and Howard, Richard and Yu, Bo and Bai, Fan},
	Booktitle = {Proceedings of the 24th Annual International Conference on Mobile Computing and Networking},
	Organization = {ACM},
	Pages = {417--427},
	Title = {Verification: Accuracy evaluation of WiFi fine time measurements on an open platform},
	Year = {2018}}

@inproceedings{steinmetzer2018beam,
	Author = {Steinmetzer, Daniel and Yuan, Yimin and Hollick, Matthias},
	Booktitle = {Proceedings of the 11th ACM Conference on Security \& Privacy in Wireless and Mobile Networks},
	Pages = {12--22},
	Title = {Beam-stealing: intercepting the sector sweep to launch man-in-the-middle attacks on wireless IEEE 802.11 ad networks},
	Year = {2018}}

@article{lee2018easy,
	Author = {Lee, Byung Moo and Patil, Mayuresh and Hunt, Preston and Khan, Imran},
	Journal = {IEEE Access},
	Pages = {8763--8772},
	Publisher = {IEEE},
	Title = {An Easy Network Onboarding Scheme for Internet of Things Networks},
	Volume = {7},
	Year = {2018}}

@inproceedings{henry2019fingerprinting,
	Author = {Henry, J{\'e}r{\^o}me and Montavont, Nicolas},
	Booktitle = {Proceedings of the 17th ACM International Symposium on Mobility Management and Wireless Access},
	Pages = {49--56},
	Title = {Fingerprinting using Fine Timing Measurement},
	Year = {2019}}

@article{guo2019indoor,
	Author = {Guo, Guangyi and Chen, Ruizhi and Ye, Feng and Peng, Xuesheng and Liu, Zuoya and Pan, Yuanjin},
	Journal = {IEEE Access},
	Pages = {176767--176781},
	Publisher = {IEEE},
	Title = {Indoor Smartphone Localization: A Hybrid WiFi RTT-RSS Ranging Approach},
	Volume = {7},
	Year = {2019}}

@article{yu2019robust,
	Author = {Yu, Yue and Chen, Ruizhi and Chen, Liang and Guo, Guangyi and Ye, Feng and Liu, Zuoya},
	Journal = {Remote Sensing},
	Number = {5},
	Pages = {504},
	Publisher = {Multidisciplinary Digital Publishing Institute},
	Title = {A robust dead reckoning algorithm based on Wi-Fi FTM and multiple sensors},
	Volume = {11},
	Year = {2019}}

@inproceedings{rea2019smartphone,
	Author = {Rea, Maurizio and Abrudan, Traian Emanuel and Giustiniano, Domenico and Claussen, Holger and Kolmonen, Veli-Matti},
	Booktitle = {Proceedings of the 15th International Conference on Emerging Networking Experiments And Technologies},
	Pages = {200--206},
	Title = {Smartphone positioning with radio measurements from a single wifi access point},
	Year = {2019}}

@article{horn2020doubling,
	Author = {Horn, Berthold KP},
	Journal = {Sensors},
	Number = {5},
	Pages = {1489},
	Publisher = {Multidisciplinary Digital Publishing Institute},
	Title = {Doubling the Accuracy of Indoor Positioning: Frequency Diversity},
	Volume = {20},
	Year = {2020}}

@article{ogawa2020measurement,
	Author = {Ogawa, Masakatsu and Choi, Hoyeon},
	Journal = {IEICE Communications Express},
	Publisher = {The Institute of Electronics, Information and Communication Engineers},
	Title = {Measurement accuracy of Wi-Fi FTM on actual devices},
	Year = {2020}}

@inproceedings{irshad2020rethinking,
	Author = {Irshad, Shazal and Rozner, Eric and Bhartia, Apurv and Chen, Bo},
	Booktitle = {Proceedings of the 21st International Workshop on Mobile Computing Systems and Applications},
	Pages = {92--97},
	Title = {Rethinking Wireless Network Management Through Sensor-driven Contextual Analysis},
	Year = {2020}}

@misc{url-ftm-mit,
	Author = {Horn, Berthold KP},
	Howpublished = {\url{http://people.csail.mit.edu/bkph/ftmrtt_aps}},
	Title = {Indoor positioning using time of flight with respect to Wi-Fi access points},
	Year = {2020 (Accessed 22 February 2020)}}

@misc{url-80211az,
	Author = {IEEE},
	Howpublished = {\url{http://www.ieee802.org/11/Reports/tgaz_update.htm}},
	Title = {IEEE P802.11 - NEXT GENERATION POSITIONING STUDY GROUP},
	Year = {2020 (Accessed 29 March 2020)}}

@misc{url-alliance-product-finder,
	Author = {Wi-Fi Alliance},
	Howpublished = {\url{https://www.wi-fi.org/product-finder}},
	Title = {Product Finder | Wi-Fi Alliance},
	Year = {2020 (Accessed 1 November 2022)}}

@misc{url-android-privacy-mac,
	Author = {Android},
	Howpublished = {\url{https://source.android.com/devices/tech/connect/wifi-mac-randomization}},
	Title = {Privacy: MAC Randomization | Android Open Source Project},
	Year = {2020 (Accessed 3 April 2020)}}

@misc{wifirttscan,
	Author = {Android},
	Howpublished = {\url{https://github.com/android/connectivity-samples}},
	Title = {Connectivity Samples Repository},
	Year = {2020 (Accessed 17 June 2020)}}

@misc{android9-rtt,
	Author = {Android},
	Howpublished = {\url{https://developer.android.com/guide/topics/connectivity/wifi-rtt}},
	Title = {Wi-Fi location: ranging with RTT | Android Developers},
	Year = {2020 (Accessed 18 June 2020)}}

@misc{Apple_U1,
	Author = {Apple},
	Howpublished = {\url{https://www.techinsights.com/blog/apple-u1-tmka75-ultra-wideband-uwb-chip-analysis}},
	Title = {U1 TMKA75 Ultra Wideband (UWB) Chip Analysis},
	Year = {2020 (Accessed 17 June 2020)}}

@inproceedings{MathurMobocom08RSSI,
	Author = {Mathur, Suhas and Trappe, Wade and Mandayam, Narayan and Ye, Chunxuan and Reznik, Alex},
	Booktitle = {Proceedings of the 14th ACM International Conference on Mobile Computing and Networking},
	Isbn = {9781605580968},
	Pages = {128--139},
	Publisher = {Association for Computing Machinery},
	Series = {MobiCom '08},
	Title = {Radio-Telepathy: Extracting a Secret Key from an Unauthenticated Wireless Channel},
	Year = {2008}}

	@INPROCEEDINGS{RFID_proximity_attack,
  author={Hancke, G.P.},
  booktitle={2006 IEEE Symposium on Security and Privacy (S P'06)}, 
  title={Practical attacks on proximity identification systems}, 
  year={2006},
  volume={},
  number={},
  pages={6 pp.-333},
  doi={10.1109/SP.2006.30}}


@inproceedings{RFID_proximity_attack2,
	Author = {Hancke, Gerhard P.},
	Booktitle = {Proceedings of the 2006 IEEE Symposium on Security and Privacy},
	Pages = {328--333},
	Publisher = {IEEE Computer Society},
	Title = {Practical Attacks on Proximity Identification Systems (Short Paper)},
	Year = {2006}}

@inproceedings{ibrahim2020wi,
	Author = {Ibrahim, Mohamed and Rostami, Ali and Yu, Bo and Liu, Hansi and Jawahar, Minitha and Nguyen, Viet and Gruteser, Marco and Bai, Fan and Howard, Richard},
	Booktitle = {Proceedings of the 18th International Conference on Mobile Systems, Applications, and Services},
	Pages = {312--324},
	Title = {Wi-Go: accurate and scalable vehicle positioning using WiFi fine timing measurement},
	Year = {2020}}

@article{si2020wi,
	Author = {Si, Minghao and Wang, Yunjia and Xu, Shenglei and Sun, Meng and Cao, Hongji},
	Journal = {Applied Sciences},
	Number = {3},
	Pages = {956},
	Publisher = {Multidisciplinary Digital Publishing Institute},
	Title = {A Wi-Fi FTM-Based Indoor Positioning Method with LOS/NLOS Identification},
	Volume = {10},
	Year = {2020}}

@misc{gutierrez2017wireless,
	Author = {Gutierrez, Luis Javier and Wang, Qi and Erceg, Vinko and Ramakrishnan, Hariramanathan},
	Month = mar # {~7},
	Note = {US Patent 9,591,493},
	Publisher = {Google Patents},
	Title = {Wireless communication fine timing measurement PHY parameter control and negotiation},
	Year = {2017}}

@misc{chhabra2018secure,
	Author = {Chhabra, Kapil and Kasten, Welly},
	Month = jul # {~24},
	Note = {US Patent 10,033,760},
	Publisher = {Google Patents},
	Title = {Secure wireless ranging},
	Year = {2018}}

@misc{aldana2018secure,
	Author = {Aldana, Carlos Horacio and Raissinia, Alireza and Vamaraju, Santosh and Anand, Kumar},
	Month = aug # {~28},
	Note = {US Patent 10,064,057},
	Publisher = {Google Patents},
	Title = {Secure fine timing measurement exchange},
	Year = {2018}}

@misc{li2019authentication,
	Author = {Li, Qinghua and Jiang, Feng and Xiaogang, Chen and Segev, Jonathan and Stacey, Robert J},
	Month = mar # {~28},
	Note = {US Patent App. 16/202,559},
	Publisher = {Google Patents},
	Title = {Authentication of ranging device},
	Year = {2019}}

@misc{sun2019fine,
	Author = {Sun, Sheng and Au, Kwok Shum},
	Month = mar # {~7},
	Note = {US Patent App. 16/120,541},
	Publisher = {Google Patents},
	Title = {Fine timing measurement security with distance bounding protocol},
	Year = {2019}}

@misc{seok2020method,
	Author = {Seok, Yongho and Wang, Chao-Chun and Yee, James},
	Month = jun # {~23},
	Note = {US Patent 10,694,407},
	Publisher = {Google Patents},
	Title = {Method and devices for secure measurement exchange},
	Year = {2020}}

@misc{sridhar2019secure,
	Author = {Sridhar, Subash Marri and Aldana, Carlos Horacio},
	Month = aug # {~27},
	Note = {US Patent 10,397,779},
	Publisher = {Google Patents},
	Title = {Secure fine timing measurement protocol},
	Year = {2019}}

@article{warner2003gps,
	Author = {Warner, Jon S and Johnston, Roger G},
	Journal = {Homeland Security Journal},
	Number = {2},
	Pages = {19--27},
	Title = {GPS spoofing countermeasures},
	Volume = {25},
	Year = {2003}}

@article{jathe2019indoor,
	Author = {Jathe, Nicolas and L{\"u}tjen, Michael and Freitag, Michael},
	Journal = {IFAC-PapersOnLine},
	Number = {13},
	Pages = {857--862},
	Publisher = {Elsevier},
	Title = {Indoor Positioning in Car Parks by using Wi-Fi Round-Trip-Time to support Finished Vehicle Logistics on Port Terminals},
	Volume = {52},
	Year = {2019}}

@article{xu2019locating,
	Author = {Xu, Shihao and Chen, Ruizhi and Yu, Yue and Guo, Guangyi and Huang, Lixiong},
	Journal = {IEEE Access},
	Pages = {95140--95153},
	Publisher = {IEEE},
	Title = {Locating smartphones indoors using built-in sensors and Wi-Fi ranging with an enhanced particle filter},
	Volume = {7},
	Year = {2019}}

@article{bullmann2020comparison,
	Author = {Bullmann, Markus and Fetzer, Toni and Ebner, Frank and Ebner, Markus and Deinzer, Frank and Grzegorzek, Marcin},
	Journal = {Sensors},
	Number = {16},
	Pages = {4515},
	Publisher = {Multidisciplinary Digital Publishing Institute},
	Title = {Comparison of 2.4 GHz WiFi FTM-and RSSI-Based Indoor Positioning Methods in Realistic Scenarios},
	Volume = {20},
	Year = {2020}}

@article{martin2020passive,
	Author = {Martin-Escalona, Israel and Zola, Enrica},
	Journal = {Electronics},
	Number = {8},
	Pages = {1193},
	Publisher = {Multidisciplinary Digital Publishing Institute},
	Title = {Passive Round-Trip-Time Positioning in Dense IEEE 802.11 Networks},
	Volume = {9},
	Year = {2020}}

@inproceedings{roland_googlewallet,
	Author = {M. {Roland} and J. {Langer} and J. {Scharinger}},
	Booktitle = {2013 5th International Workshop on Near Field Communication (NFC)},
	Title = {Applying relay attacks to Google Wallet},
	Year = {2013}}

@misc{url-wifi-aware,
	Author = {Wi-Fi Alliance},
	Howpublished = {\url{https://www.wi-fi.org/discover-wi-fi/wi-fi-aware}},
	Title = {Wi-Fi Aware | Wi-Fi Alliance},
	Year = {2020 (Accessed 3 December 2020)}}


@inproceedings{ltrack,
	author = {Kotuliak, Martin and Erni, Simon and Leu, Patrick and R{\"o}schlin, Marc and Capkun, Srdjan},
	booktitle = {Proceedings of the 31st USENIX Security Symposium},
	copyright = {In Copyright - Non-Commercial Use Permitted},
	doi = {10.3929/ethz-b-000517466},
	editor = {Butler, Kevin and Thomas, Kurt and et al.},
	institution = {SNF and EC},
	isbn = {978-1-939133-31-1},
	language = {en},
	note = {31st USENIX Security Symposium (USENIX Security 22); Conference Location: Boston, MA, USA; Conference Date: August 10--12, 2022},
	pages = {1291 - 1306},
	publisher = {Advanced Computing Systems Association},
	size = {16 p. accepted version},
	title = {LTrack: Stealthy Tracking of Mobile Phones in LTE},
	type = {Conference Paper},
	year = {2022},
	Bdsk-Url-1 = {https://doi.org/10.3929/ethz-b-000517466}}


@misc{oslo_self-drving-trial,
	Howpublished = {\url{https://www.intelligenttransport.com/transport-news/112949/oslo-self-driving-trial/}},
	Lastvisited = {'15.01.2021},
	Note = {[Online; Accessed 15. January 2021]},
	Title = {{Self-driving Trial Oslo}}}

@misc{uk_self-drving-trial,
	Howpublished = {\url{https://www.newswise.com/articles/work-begins-on-autonomous-vehicle-trial-route}},
	Lastvisited = {'15.01.2021},
	Note = {[Online; Accessed 15. January 2021]},
	Title = {{Self-driving Trial UK}}}

@misc{Wiggle,
  doi = {10.48550/ARXIV.2209.00080},
  url = {https://arxiv.org/abs/2209.00080},
  author = {Dickey, Connor and Smith, Christopher and Johnson, Quentin and Li, Jingcheng and Xu, Ziqi and Lazos, Loukas and Li, Ming},
  keywords = {Cryptography and Security (cs.CR), FOS: Computer and information sciences, FOS: Computer and information sciences},
  title = {Wiggle: Physical Challenge-Response Verification of Vehicle Platooning},
  publisher = {arXiv},
  year = {2022},
  copyright = {arXiv.org perpetual, non-exclusive license}
}



@inproceedings{rasmussen2007secnav,
	Author = {Rasmussen, Kasper Bonne and Capkun, Srdjan and Cagalj, Mario},
	Booktitle = {Proceedings of the 13th annual ACM international conference on Mobile computing and networking},
	Organization = {ACM},
	Pages = {310--313},
	Title = {Secnav: secure broadcast localization and time synchronization in wireless networks},
	Year = {2007}}

@article{capkun2010integrity,
	Author = {Capkun, Srdjan and Cagalj, Mario and Karame, Ghassan and Tippenhauer, Nils Ole},
	Journal = {IEEE Transactions on Mobile Computing},
	Number = {11},
	Pages = {1608--1621},
	Publisher = {IEEE},
	Title = {Integrity regions: Authentication through presence in wireless networks},
	Volume = {9},
	Year = {2010}}

@inproceedings{brelurut2015survey,
	Author = {Brelurut, Agn{\`e}s and Gerault, David and Lafourcade, Pascal},
	Booktitle = {International Symposium on Foundations and Practice of Security},
	Organization = {Springer},
	Pages = {29--49},
	Title = {Survey of distance bounding protocols and threats},
	Year = {2015}}

@inproceedings{ranganathan2015proximity,
	Author = {Ranganathan, Aanjhan and Danev, Boris and Capkun, Srdjan},
	Booktitle = {Proceedings of the 31st Annual Computer Security Applications Conference},
	Organization = {ACM},
	Pages = {271--280},
	Title = {Proximity verification for contactless access control and authentication systems},
	Year = {2015}}

@article{achlioptas2003database,
	Author = {Achlioptas, Dimitris},
	Journal = {Journal of computer and System Sciences},
	Number = {4},
	Pages = {671--687},
	Publisher = {Elsevier},
	Title = {Database-friendly random projections: Johnson-Lindenstrauss with binary coins},
	Volume = {66},
	Year = {2003}}

@inproceedings{rasmussen2010realization,
	Author = {Rasmussen, Kasper Bonne and Capkun, Srdjan},
	Booktitle = {USENIX Security Symposium},
	Pages = {389--402},
	Title = {Realization of RF Distance Bounding.},
	Year = {2010}}

@inproceedings{tippenhauer2011requirements,
	Author = {Tippenhauer, Nils Ole and P{\"o}pper, Christina and Rasmussen, Kasper Bonne and Capkun, Srdjan},
	Booktitle = {Proceedings of the 18th ACM conference on Computer and communications security},
	Organization = {ACM},
	Pages = {75--86},
	Title = {On the requirements for successful GPS spoofing attacks},
	Year = {2011}}

@article{ganeriwal2008secure,
	Author = {Ganeriwal, Saurabh and P{\"o}pper, Christina and {\v{C}}apkun, Srdjan and Srivastava, Mani B},
	Journal = {ACM Transactions on Information and System Security (TISSEC)},
	Number = {4},
	Pages = {23},
	Publisher = {ACM},
	Title = {Secure time synchronization in sensor networks},
	Volume = {11},
	Year = {2008}}

@inproceedings{vasisht2016decimeter,
	Author = {Vasisht, Deepak and Kumar, Swarun and Katabi, Dina},
	Booktitle = {NSDI},
	Pages = {165--178},
	Title = {Decimeter-Level Localization with a Single WiFi Access Point.},
	Volume = {16},
	Year = {2016}}

@inproceedings{bahl2000radar,
	Author = {Bahl, Paramvir and Padmanabhan, Venkata N},
	Booktitle = {INFOCOM 2000. Nineteenth Annual Joint Conference of the IEEE Computer and Communications Societies. Proceedings. IEEE},
	Organization = {Ieee},
	Pages = {775--784},
	Title = {RADAR: An in-building RF-based user location and tracking system},
	Volume = {2},
	Year = {2000}}



@misc{hack_key,
	Howpublished = {\url{https://www.wired.com/2017/04/just-pair-11-radio-gadgets-can-steal-car/}},
	Lastvisited = {'10.11.18},
	Note = {[Online; Accessed November 10th 2018]},
	Title = {Radio Attack Lets Hackers Steal Cars With Just \$20 Worth of Gear.}}

@misc{nytimes,
	Howpublished = {\url{https://www.nytimes.com/2015/04/16/style/keeping-your-car-safe-from-electronic-thieves.html}},
	Lastvisited = {'10.11.18},
	Note = {[Online; Accessed November 10th 2018]},
	Title = {"Keeping Your Care Safe From Electronic Thieves."}}

@misc{SecuKey,
	Howpublished = {\url{www.secukey.org}},
	Lastvisited = {'20.12.20},
	Note = {[Online; Accessed December 20th 2020]},
	Title = {"SecuKey"}}



@book{lapidoth2017foundation,
	Author = {Lapidoth, Amos},
	Publisher = {Cambridge University Press},
	Title = {A foundation in digital communication},
	Year = {2017}}

@book{cover2012elements,
	Author = {Cover, Thomas M and Thomas, Joy A},
	Publisher = {John Wiley \& Sons},
	Title = {Elements of information theory},
	Year = {2012}}

@misc{bbcdrones,
	Author = {Mary-Ann Russon},
	Howpublished = {\url{http://www.bbc.com/news/business-43906846}},
	Month = May,
	Title = {{Drones to the rescue!}},
	Year = {2018}}

@misc{postdrones,
	Author = {Swiss Post},
	Howpublished = {\url{https://www.post.ch/en/about-us/company/media/press-releases/2017/swiss-post-drone-to-fly-laboratory-samples-for-ticino-hospitals}},
	Month = May,
	Title = {{Drones as transportation vehicle}},
	Year = {2018}}

@inproceedings{Polypoint_MAC,
	Author = {Kempke, Benjamin and Pannuto, Pat and Dutta, Prabal},
	Booktitle = {ACM HotWireless},
	Numpages = {5},
	Pages = {16--20},
	Title = {{PolyPoint: Guiding Indoor Quadrotors with Ultra-Wideband Localization}},
	Year = {2015}}


@inproceedings{SurePoint,
	Author = {Kempke, Benjamin and Pannuto, Pat and Dutta, Prabal},
	Booktitle = {ACM SenSys},
	Numpages = {2},
	Pages = {318--319},
	Title = {{SurePoint: Exploiting Ultra Wideband Flooding and Diversity to Provide Robust, Scalable, High-Fidelity Indoor Localization}},
	Year = {2016}}

@article{Nils_Integrity,
	Author = {Tippenhauer, Nils Ole and Rasmussen, Kasper Bonne and \v{C}apkun, Srdjan},
	Journal = {Computer Networks},
	Number = {P1},
	Numpages = {8},
	Pages = {31--38},
	Title = {{Physical-layer Integrity for Wireless Messages}},
	Volume = {109},
	Year = {2016}}

@inproceedings{Nils_WLAN_attack,
	Author = {Tippenhauer, Nils Ole and Rasmussen, Kasper Bonne and P\"opper, Christina and \v{C}apkun, Srdjan},
	Booktitle = {ACM/Usenix MobiSys},
	Title = {{Attacks on Public WLAN-based Positioning}},
	Year = {2009}}

@inbook{UWB_Book_2,
	Author = {Nguyen, Cam and Miao, Meng},
	Booktitle = {Design of CMOS RFIC Ultra-Wideband Impulse Transmitters and Receivers},
	Pages = {7--24},
	Publisher = {Springer},
	Title = {{Fundamentals of UWB Impulse Systems}},
	Year = {2017}}

@misc{timedomain,
	Author = {{Humatics}},
	Howpublished = {\url{http://www.timedomain.com/products/pulson-440/}},
	Lastvisited = {23.10.17},
	Note = {[Online; Accessed 23. October 2017]},
	Title = {{Time Domain's PulsON} ("P440")}}

@misc{zebra,
	Author = {{Zebra Technologies}},
	Howpublished = {\url{https://www.zebra.com/us/en/solutions/location-solutions/enabling-technologies/dart-uwb.html}},
	Lastvisited = {22.10.18},
	Note = {[Online; Accessed 22. October 2018]},
	Title = {"Sapphire Dart Ultra Wideband (UWB) real time locating system 2010."}}

@article{leadingedge1,
	Author = {Guvenc, Ismail and Sahinoglu, Zafer and Orlik, Philip and Arslan, Huseyin},
	Doi = {10.1007/s11277-008-9549-3},
	Journal = {Wireless Personal Communications},
	Month = {03},
	Pages = {585-603},
	Title = {Searchback Algorithms for TOA Estimation in Non-coherent Low-rate IR-UWB Systems},
	Volume = {48},
	Year = {2009},
	Bdsk-Url-1 = {https://doi.org/10.1007/s11277-008-9549-3}}

@article{leadingedge2,
	Author = {Dardari, Davide and Conti, Andrea and Ferner, Ulric and Giorgetti, Andrea and Win, Moe},
	Doi = {10.1109/JPROC.2008.2008846},
	Journal = {Proceedings of the IEEE},
	Month = {03},
	Pages = {404 - 426},
	Title = {Ranging With Ultrawide Bandwidth Signals in Multipath Environments},
	Volume = {97},
	Year = {2009},
	Bdsk-Url-1 = {https://doi.org/10.1109/JPROC.2008.2008846}}

	@article{leadingedge3,
	Author = {Sharp, Ian and Yu, Kegen and Guo, Y Jay},
	Journal = {IET communications},
	Number = {10},
	Pages = {1616--1627},
	Publisher = {IET},
	Title = {Peak and leading edge detection for time-of-arrival estimation in band-limited positioning systems},
	Volume = {3},
	Year = {2009}}



@inproceedings{Cristina_ESORICS,
	Author = {P{\"o}pper, Christina and Tippenhauer, Nils Ole and Danev, Boris and \v{C}apkun, Srdjan},
	Booktitle = {Computer Security -- ESORICS 2011},
	Editor = {Atluri, Vijay and Diaz, Claudia},
	Pages = {40--59},
	Publisher = {Springer},
	Title = {{Investigation of Signal and Message Manipulations on the Wireless Channel}},
	Year = {2011}}

@article{Taponecco_Overshadow,
	Author = {L. Taponecco and P. Perazzo and A. A. D'Amico and G. Dini},
	Journal = {IEEE Communications Letters},
	Number = {2},
	Pages = {257-260},
	Title = {{On the Feasibility of Overshadow Enlargement Attack on IEEE 802.15.4a Distance Bounding}},
	Volume = {18},
	Year = {2014}}

	@article{Multiple_leading_edge,
	Author = {A. Compagno and M. Conti and A. A. D'Amico and G. Dini and P. Perazzo and L. Taponecco},
	Journal = {IEEE Transactions on Information Forensics and Security},
	Number = {7},
	Pages = {1565-1577},
	Title = {{Modeling Enlargement Attacks Against UWB Distance Bounding Protocols}},
	Volume = {11},
	Year = {2016}}


@article{zandbergen2009accuracy,
	Author = {Zandbergen, Paul A},
	Journal = {Blackwell Transactions in GIS},
	Number = {s1},
	Publisher = {Blackwell Publishing Ltd},
	Title = {{Accuracy of iPhone locations: A Comparison of Assisted GPS, WiFi and Cellular Positioning}},
	Volume = {13},
	Year = {2009}}

@inproceedings{Channel_model_final_report,
	Author = {Andreas F. Molisch and Kannan Balakrishnan and Chia-chin Chong and Shahriar Emami and Andrew Fort and Johan Karedal and Juergen Kunisch and Hans Schantz and Ulrich Schuster and Kai Siwiak},
	Booktitle = {Converging: Technology, work and learning. Australian Government Printing Service},
	Howpublished = {\url{http://www.ieee802.org/15/pub/04/15-04-0662-02-004a-channel-model-final-report-r1.pdf}},
	Publisher = {[Online; Accessed 4. November 2018]},
	Title = {{IEEE 802.15.4a channel model - final report}},
	Year = {2004}}


@ARTICLE{Molisch_paper_channel_UWB,
  author={Molisch, A.F.},
  journal={IEEE Transactions on Vehicular Technology}, 
  title={Ultrawideband propagation channels-theory, measurement, and modeling}, 
  year={2005},
  volume={54},
  number={5},
  pages={1528-1545},
  doi={10.1109/TVT.2005.856194}}



@inproceedings{ChannelCondition_material,
	Author = {A. Muqaibel and A. Safaai-Jazi and A. Bayram and S. M. Riad},
	Booktitle = {IEEE Antennas and Propagation Society International Symposium},
	Pages = {623-626},
	Title = {Ultra wideband material characterization for indoor propagation},
	Volume = {4},
	Year = {2003}}

@article{enlargement_miscontrol,
	Articleno = {27},
	Author = {Perazzo, Pericle and Taponecco, Lorenzo and D'amico, Antonio A. and Dini, Gianluca},
	Issue_Date = {November 2016},
	Journal = {ACM Transactions on Sensor Networks},
	Number = {4},
	Numpages = {32},
	Pages = {27:1--27:32},
	Publisher = {ACM},
	Title = {{Secure Positioning in Wireless Sensor Networks Through Enlargement Miscontrol Detection}},
	Volume = {12},
	Year = {2016}}

@article{Non-Coherent-receiver-explain,
	Author = {K. Witrisal and G. Leus and G. J. M. Janssen and M. Pausini and F. Troesch and T. Zasowski and J. Romme},
	Journal = {IEEE Signal Processing Magazine},
	Number = {4},
	Pages = {48-66},
	Title = {Noncoherent ultra-wideband systems},
	Volume = {26},
	Year = {2009}}

@article{Molisch_channel_model_CM_5,
	Author = {A. F. Molisch and D. Cassioli and C. Chong and S. Emami and A. Fort and B. Kannan and J. Karedal and J. Kunisch and H. G. Schantz and K. Siwiak and M. Z. Win},
	Journal = {IEEE Transactions on Antennas and Propagation},
	Number = {11},
	Pages = {3151-3166},
	Title = {{A Comprehensive Standardized Model for Ultrawideband Propagation Channels}},
	Volume = {54},
	Year = {2006}}

@Inbook{Valero2022,
author="Valero, Jos{\'e} Mar{\'i}a Jorquera
and S{\'a}nchez, Pedro Miguel S{\'a}nchez
and Lekidis, Alexios
and Martins, Pedro
and Diogo, Pedro
and P{\'e}rez, Manuel Gil
and Celdr{\'a}n, Alberto Huertas
and P{\'e}rez, Gregorio Mart{\'i}nez",
editor="Abd El-Latif, Ahmed A.
and Abd-El-Atty, Bassem
and Venegas-Andraca, Salvador E.
and Mazurczyk, Wojciech
and Gupta, Brij B.",
title="Trusted Execution Environment-Enabled Platform for 5G Security and Privacy Enhancement",
bookTitle="Security and Privacy Preserving for IoT and 5G Networks: Techniques, Challenges, and New Directions",
year="2022",
publisher="Springer International Publishing",
address="Cham",
pages="203--223",
abstract="With the deployment of 5G networks and the beginning of the design of beyond 5G communications, new critical requirements are emerging in terms of performance, security, and trust for leveraged technologies, such as Software Defined Networking (SDN) and Network Function Virtualization (NFV). One of the requirements at the security and trust level is that when delegating critical tasks and data to the infrastructure deployed in an external domain, the client needs guarantees that the execution has been carried out securely, without data breaches or compromises during computing tasks. To meet this need, this chapter proposes a framework that uses Trusted Execution Environments (TEEs), processing environments isolated from the rest of the system to guarantee the security of the data and tasks processed in them, in order to improve the security of 5G environments. This framework enables the deployment of TEE as a cloud service, also denoted as TEE-as-a-Service or TEEaaS, allowing customers to take advantage of its benefits without having to deal with the configuration of the environment and hardware. Furthermore, this chapter also discusses current trends as well as future challenges related to the deployment of TEEs in 5G environments, providing key aspects for future solutions in the area.",
isbn="978-3-030-85428-7",
doi="10.1007/978-3-030-85428-7_9",
url="https://doi.org/10.1007/978-3-030-85428-7_9"
}







@article{aanjhan_magazine,
	Author = {A. Ranganathan and S. Capkun},
	Journal = {IEEE Security Privacy},
	Number = {3},
	Pages = {52-58},
	Title = {Are We Really Close? Verifying Proximity in Wireless Systems},
	Volume = {15},
	Year = {2017}}

@inproceedings{Aanjhan_Contactless,
	Acmid = {2818004},
	Author = {Ranganathan, Aanjhan and Danev, Boris and Capkun, Srdjan},
	Booktitle = {Proceedings of the 31st Annual Computer Security Applications Conference},
	Doi = {10.1145/2818000.2818004},
	Isbn = {978-1-4503-3682-6},
	Numpages = {10},
	Pages = {271--280},
	Publisher = {ACM},
	Series = {ACSAC 2015},
	Title = {Proximity Verification for Contactless Access Control and Authentication Systems},
	Url = {http://doi.acm.org/10.1145/2818000.2818004},
	Year = {2015},
	Bdsk-Url-1 = {http://doi.acm.org/10.1145/2818000.2818004},
	Bdsk-Url-2 = {https://doi.org/10.1145/2818000.2818004}}

@inproceedings{Aanjhan_chirp,
	Author = {Ranganathan, Aanjhan and Danev, Boris and Francillon, Aur{\'e}lien and Capkun, Srdjan},
	Booktitle = {Proceedings of the fifth ACM conference on Security and Privacy in Wireless and Mobile Networks},
	Organization = {ACM},
	Pages = {15--26},
	Title = {Physical-layer attacks on chirp-based ranging systems},
	Year = {2012}}

@inproceedings{EPFL_GPS,
	Author = {P. Papadimitratos and A. Jovanovic},
	Booktitle = {MILCOM 2008 - 2008 IEEE Military Communications Conference},
	Doi = {10.1109/MILCOM.2008.4753512},
	Issn = {2155-7578},
	Pages = {1-7},
	Title = {GNSS-based Positioning: Attacks and countermeasures},
	Year = {2008},
	Bdsk-Url-1 = {https://doi.org/10.1109/MILCOM.2008.4753512}}

@book{UWB_BOOK_Bensky,
	Address = {Norwood, MA, USA},
	Author = {Bensky, Alan},
	Isbn = {1596931302, 9781596931305},
	Publisher = {Artech House, Inc.},
	Title = {Wireless Positioning Technologies and Applications},
	Year = {2007}}



@inproceedings{HildurPhase,
	Author = {{\'O}lafsd{\'o}ttir, Hildur and Ranganathan, Aanjhan and \v{C}apkun, Srdjan},
	Booktitle = {International Conference on Cryptographic Hardware and Embedded Systems},
	Organization = {Springer},
	Pages = {490--509},
	Title = {On the Security of Carrier Phase-based Ranging},
	Year = {2017}}

@inproceedings{WifiBlueetothRelayAttack,
	Author = {H. T. T. Truong and Xiang Gao and B. Shrestha and N. Saxena and N. Asokan and P. Nurmi},
	Booktitle = {2014 IEEE International Conference on Pervasive Computing and Communications (PerCom)},
	Doi = {10.1109/PerCom.2014.6813957},
	Pages = {163-171},
	Title = {Comparing and fusing different sensor modalities for relay attack resistance in Zero-Interaction Authentication},
	Year = {2014},
	Bdsk-Url-1 = {https://doi.org/10.1109/PerCom.2014.6813957}}

@misc{ATMEL,
	Howpublished = {\url{http://www.atmel.com/Images/Atmel-8443-RTB-Evaluation-Application-Software-Users-Guide_Application-Note_AVR2152.pdf}},
	Lastvisited = {'23.10.17},
	Note = {[Online; Accessed 23. October 2017]},
	Title = {Atmel phase difference measurement}}



@misc{deca,
	Howpublished = {\url{https://www.decawave.com/products/dw1000}},
	Lastvisited = {'23.10.17},
	Note = {[Online; Accessed 23. October 2017]},
	Title = {{DecaWave} ``DW1000 product description and applications"}}

@inproceedings{fontana2007observations,
	Author = {Fontana, Robert J and Richley, Edward A},
	Booktitle = {Ultra-Wideband, 2007. ICUWB 2007. IEEE International Conference on},
	Organization = {IEEE},
	Pages = {334--338},
	Title = {Observations on low data rate, short pulse UWB systems},
	Year = {2007}}

@inproceedings{Reid_DB_terroristFraud,
	Acmid = {1229314},
	Author = {Reid, Jason and Nieto, Juan M. Gonzalez and Tang, Tee and Senadji, Bouchra},
	Booktitle = {Proceedings of the 2Nd ACM Symposium on Information, Computer and Communications Security},
	Doi = {10.1145/1229285.1229314},
	Pages = {204--213},
	Publisher = {ACM},
	Series = {ASIACCS '07},
	Title = {Detecting Relay Attacks with Timing-based Protocols},
	Url = {http://doi.acm.org/10.1145/1229285.1229314},
	Year = {2007},
	Bdsk-Url-1 = {http://doi.acm.org/10.1145/1229285.1229314},
	Bdsk-Url-2 = {https://doi.org/10.1145/1229285.1229314}}



@misc{iBeaconRSS,
	Howpublished = {\url{https://developer.apple.com/ibeacon/Getting-Started-with-iBeacon.pdf}},
	Lastvisited = {'05.01.21},
	Note = {[Online; Accessed 05 January 2021]},
	Title = {"Getting Started with iBeacon"}}

@inproceedings{hu2003packet,
	Author = {Hu, Y-C and Perrig, Adrian and Johnson, David B},
	Booktitle = {INFOCOM 2003},
	Organization = {IEEE},
	Pages = {1976--1986},
	Title = {Packet leashes: a defense against wormhole attacks in wireless networks},
	Volume = {3},
	Year = {2003}}

@inproceedings{sastry2003secure,
	Author = {Sastry, Naveen and Shankar, Umesh and Wagner, David},
	Booktitle = {Proceedings of the 2nd ACM workshop on Wireless security},
	Organization = {ACM},
	Pages = {1--10},
	Title = {Secure verification of location claims},
	Year = {2003}}

@article{capkun2006secure,
	Author = {Capkun, Srdjan and Hubaux, J-P},
	Journal = {IEEE Journal on Selected Areas in Communications},
	Number = {2},
	Pages = {221--232},
	Publisher = {IEEE},
	Title = {Secure positioning in wireless networks},
	Volume = {24},
	Year = {2006}}



@article{FMCW_attack,
	Author = {Nashimoto, Shoei and Suzuki, Daisuke and Miura, Noriyuki and Machida, Tatsuya and Matsuda, Kohei and Nagata, Makoto},
	Da = {2021/01/07},
	Date-Added = {2021-01-18 21:00:12 +0100},
	Date-Modified = {2021-01-18 21:01:40 +0100},
	Doi = {10.1007/s13389-020-00252-5},
	Id = {Nashimoto2021},
	Isbn = {2190-8516},
	Journal = {Journal of Cryptographic Engineering},
	Rating = {1},
	Read = {0},
	Title = {Low-cost distance-spoofing attack on FMCW radar and its feasibility study on countermeasure},
	Ty = {JOUR},
	Url = {https://doi.org/10.1007/s13389-020-00252-5},
	Year = {2021},
	Bdsk-Url-1 = {https://doi.org/10.1007/s13389-020-00252-5}}

@article{Phase-Miesen,
	Author = {R. Miesen and A. Parr and Jochen Schleu and M. Vossiek},
	Journal = {2013 IEEE International Conference on RFID-Technologies and Applications (RFID-TA)},
	Pages = {1-6},
	Title = {360 degree carrier phase measurement for UHF RFID local positioning},
	Year = {2013}}



@inproceedings{ChannelEstimationAttack,
	Author = {Eberz, Simon	and Strohmeier, Martin	and Wilhelm, Matthias	and Martinovic, Ivan},
	Booktitle = {Computer Security -- ESORICS 2012"},
	Publisher = {"Springer Berlin Heidelberg"},
	Title = {A Practical Man-In-The-Middle Attack on Signal-Based Key Generation Protocols"}}
 




 
@inproceedings{MLApproachin5G,
	Author = {Magnus Malmström, Isaac Skog, Sara Modarres Razavi, Yuxin Zhao and Fredrik},
	BookTitle = {2019 16th Workshop on Positioning, Navigation and Communications (WPNC)},
	Pages = {1-6},
	Title = {5G Positioning: A Machine Learning Approach},
	Year = {2019}}


 @article{DLinmmwave5G,
	Author = {Joao Gante, Gabriel Falcao and Leonel Sousa},
	Journal = {Neural Processing Letters Volume 50},
	Pages = {487-514},
	Title = {Deep Learning Architectures for Accurate Millimeter Wave Positioning in 5G},
	Year = {2020}}
